\documentclass[12pt,paper]{article}
\usepackage{jheppub}
\usepackage[usenames,dvipsnames,table]{xcolor}

\usepackage{physics}
\usepackage{caption,subcaption}
\graphicspath{{figures/}}

\title{Islands in Multiverse Models}

\author[a]{Sergio E. Aguilar-Gutierrez,}
\author[a,b]{Aidan Chatwin-Davies,}
\author[a]{Thomas Hertog,}
\author[c,d]{\\Natalia Pinzani-Fokeeva,}
\author[a]{and Brandon Robinson}

\affiliation[a]{Institute for Theoretical Physics, KU Leuven, \\Celestijnenlaan 200D, B-3001 Leuven, Belgium}
\affiliation[b]{ Department of Physics and Astronomy, University of British Columbia, \\ 6224 Agricultural Road, Vancouver, BC, V6T 1Z1}
\affiliation[c]{Center for Theoretical Physics, Massachusetts Institute of Technology,\\ Cambridge, MA 02139, USA}
\affiliation[d]{Dipartimento di Fisica e Astronomia, Universit\'a di Firenze, \\ Via G. Sansone 1, I-50019, Sesto Fiorentino, Firenze, Italy}

\newcommand{\Eq}[1]{Eq.~(\ref{#1})}
\newcommand{\Eqs}[2]{Eqs.~(\ref{#1}) and (\ref{#2})}
\newcommand{\Sec}[1]{Sec.~\ref{#1}}

\newcommand{\Fig}[1]{Fig.~\ref{#1}}
\newcommand{\App}[1]{App.~\ref{#1}}
\renewcommand{\Ref}[1]{Ref.~\cite{#1}}
\newcommand{\Refs}[1]{Refs.~\cite{#1}}

\newcommand{\mbf}[1]{\mathbf{#1}}
\newcommand{\mrm}[1]{\mathrm{#1}}

\newcommand{\dee}{\mathrm{d}}

\def\Daoi{D^{\ge 1}}
\def\p0{\phi_0}
\def\cF{{\cal F}}
\def\aoi{D^{\ge 1}}
\def\lambdabar{\protect\@lambdabar}
\def\@lambdabar{%
\relax
\bgroup
\def\@tempa{\hbox{\raise.73\ht0
\hbox to0pt{\kern.25\wd0\vrule width.5\wd0
height.1pt depth.1pt\hss}\box0}}%
\mathchoice{\setbox0\hbox{$\displaystyle\lambda$}\@tempa}%
{\setbox0\hbox{$\textstyle\lambda$}\@tempa}%
{\setbox0\hbox{$\scriptstyle\lambda$}\@tempa}%
{\setbox0\hbox{$\scriptscriptstyle\lambda$}\@tempa}%
\egroup
}

\emailAdd{sergio.ernesto.aguilar@gmail.com}
\emailAdd{achatwin@phas.ubc.ca}
\emailAdd{thomas.hertog@kuleuven.be}
\emailAdd{n.pinzanifokeeva@gmail.com}
\emailAdd{brandon.robinson@kuleuven.be}

\begin{document}

\abstract{
We consider multiverse models in two-dimensional linear dilaton-gravity theories as toy models of false vacuum eternal inflation. Coupling conformal matter we calculate the Von Neumann entropy of subregions. When these are sufficiently large we find that an island develops covering most of the rest of the multiverse, leading to a Page-like transition. This resonates with a description of multiverse models in semiclassical quantum cosmology, where a measure for local predictions is given by saddle point geometries which coarse-grain over any structure associated with eternal inflation beyond one's patch.
}

\maketitle

\clearpage

\section{Introduction}

The multiverse suffers from an information loss problem akin to that of black holes: the so-called ``measure problem'' \cite{Linde:1993xx}. This arises in cosmological models that assume a classical near de Sitter (dS) background, in which quantum fluctuations produce physically distinct patches where inflation locally ends and a more interesting cosmological evolution can ensue. The archetypal example is false vacuum-driven eternal inflation with multiple decay channels. As in the case of black holes, the ``problem'' concerns a breakdown of predictivity. The formation of a mosaic of bubbles or patches with different physical properties, say different statistical features of the Cosmic Microwave Background (CMB), means the theory fails to predict what we should observe. 

Semiclassical quantum cosmology (QC) in low-energy gravity offers a very different description of the multiverse that is seemingly at odds with the view of a cosmic patchwork of bubbles. The global fine-grained mosaic of bubbles in the description above is replaced in semiclassical QC with a small number of distinct saddle point geometries. The latter are associated with {\it coarse-grained} descriptions of the universe.\footnote{This is terminology from decoherent histories quantum mechanics. In this context, by ``fine-graining'' we mean retaining information on the largest scales whereas ``coarse-graining'' does not. Indeed the specific coarse-grained saddle geometries that will be of interest to us later will contain fine-grained information in a local region.}  Specifically, each individual saddle geometry contains information about a limited cosmic patch or bubble only, while coarse-graining, or averaging, over any putative mosaic structure on much larger scales. It has been argued that this semiclassical description resolves the information loss problem associated with multiverse cosmology \cite{Hartle:2010dq}.  The semiclassical theory encodes any ``global'' information that is relevant to the prediction of local observations of a given observer not as a mosaic structure, but as distinct past (saddle point) histories of a given observer, with the relative weighting of saddle points specifying a measure.

The semiclassical recovery of information in multiverse cosmology bears striking similarities to the recent low-energy gravity description of black hole evaporation as a unitary process \cite{Penington:2019npb, Almheiri:2019psf, Penington:2019kki, Almheiri:2019qdq, Marolf:2020xie, Almheiri:2020cfm, Marolf:2020rpm}. In both cases, the semiclassical low-energy framework appears to capture the essential quantum physics without an explicit knowledge of the microscopic quantum state. Equally striking, semiclassical reasoning appears to cast doubt on the assumption that a definite spacetime background with independent degrees of freedom exists well beyond horizons, let alone indefinitely, in a manner that is independent of the observable of interest. Instead, an additional saddle appears when a given observer aims to perform some of the extraordinarily complicated measurements needed to recover a significant amount of information. For example, the semiclassical calculation of the fine-grained Von Neumann entropy of Hawking radiation which reproduces the ``Page curve'’ \cite{Page:1993wv, Page:2013dx}, long regarded as a key signature of unitary evolution, involves additional saddles: replica wormholes \cite{Penington:2019kki, Almheiri:2019qdq, Marolf:2020xie, Almheiri:2020cfm, Marolf:2020rpm}.

A complementary and calculationally tractable description of the semiclassical purification process is provided by the ``island rule.'' According to this, the Von Neumann entropy of Hawking radiation collected in a region $R$ can be obtained by extremizing the generalized entropy over possible configurations $R\cup I$, where $I$ is an additional island region, and then taking the resulting global minimum,
\begin{equation} \label{eq:islands}
    S(\rho_R) = \min \underset{I}{\mrm{ext}} ~ S_{\mrm{gen}}(R \cup I),
\end{equation}
where 
\begin{equation} \label{eq:gen-entropy}
    S_{\mrm{gen}}(X) = S_{\mrm{semi-cl}}(X)+\frac{{\rm Area}(\partial X)}{4 G_N}.
\end{equation}
Here, $S_{\mrm{semi-cl}}(X)$ is the Von Neumann entropy of quantum fields of region $X$ of a classical background geometry, and ${\rm Area}(\partial X)$ is the gravitational area term of the boundary $\partial X$. While originally motivated on the basis of considerations of holographic entanglement entropy \cite{Ryu_2006, Hubeny:2007xt, Faulkner:2013ana, Engelhardt:2014gca}, the island rule \Eq{eq:islands} in the context of black holes was later found to be consonant with an analysis based on the semiclassical gravitational path integral \cite{Penington:2019kki, Almheiri:2019qdq}. 

Moving back to cosmology, the island prescription opens up a new semiclassical angle to study the multiverse. This is interesting, for an oft-voiced criticism against the semiclassical quantum cosmology resolution of the measure problem has been that the saddle point approximation of the wavefunction of the universe simply {\it misses} relevant information in the global fine-grained patchwork that eternal inflation supposedly generates. If, however, large islands were to develop in multiverse configurations whenever one calculates sufficiently refined observables, then this would suggest that the coarse-graining inherent in the semiclassical theory is not a bug but a feature, and an interesting one indeed. The goal of this paper is to explore precisely this possibility. We do so in two-dimensional toy model multiverse cosmologies where explicit computations of the Von Neumann entropy of matter fields are possible, and we then relate our findings in these models to the more general semiclassical QC description of eternal inflation. 

We pursue our analysis in the Jackiw-Teitelboim (JT) theory of two-dimensional linear dilaton gravity \cite{Teitelboim:1983ux, Jackiw:1984je}. JT gravity has seen a recent resurgence in interest as a simple solvable theory of quantum gravity \cite{Almheiri:2014cka, Maldacena:2016upp} and given its implications in low dimensional holography (see e.g. \cite{Saad:2019lba}). For our purposes, we shall be primarily interested in the de Sitter version of JT gravity \cite{Maldacena:2019cbz,Cotler:2019nbi}, which has featured in earlier studies of islands in low-dimensional cosmological toy models \cite{Chen:2020tes,Hartman:2020khs,Balasubramanian:2020xqf,Sybesma:2020fxg, Aalsma:2021bit,Manu:2020tty}.
We construct a first toy model multiverse by analytic continuation of the dS$_2$ solution.
Then, inspired by \cite{Hartman:2020khs}, we generalize the model by allowing for regions to be excised and replaced with bubbles of zero- or negative-curvature spacetime, and we couple conformal matter to the background metric.
The result is a low-dimensional model for the global mosaic spacetimes featuring in traditional (classical) studies of eternal inflation.
We use the value of the dilaton to characterize regions of spacetime with different physical properties, identifying regions of weak gravity and of strong gravity along the way.
We then consider interval subregions $R$ located in weakly gravitating regions of the background. Using the island formula \eqref{eq:islands}, we compute the Von Neumann entropy associated to $R$ and study its dependence on properties of $R$ and properties of the global spacetime.

In all cases that we analyze, we are able to show that for a sufficiently large region $R$, and at sufficiently late times, an island develops. Consequently, while initially the Von Neumann entropy of the region grows with its size, a Page-like transition occurs at a critical point beyond which a configuration with a non-trivial island minimizes the generalized entropy \eqref{eq:gen-entropy}. 

Further, we find rather universally that islands, when they exist, cover nearly all of the multiverse structure outside $R$. This agrees with the results of a recent  work \cite{Langhoff:2021uct} which considers the formation of islands in a higher dimensional multiverse setting using the ``island finder'' prescription \cite{Bousso:2021sji}. Both sets of results lend support to the intuition emanating from the semiclassical QC description of the multiverse that distant regions may not carry independent degrees of freedom, and thus that the huge coarse-graining which the semiclassical theory encodes may be appropriate to derive well-defined predictions for local observations. 

The precise point at which the Page-like transition occurs depends on the details of the multiverse configuration. Nonetheless, reading the semiclassical QC description the other way around, we are led to conjecture that, quite generally, islands should form at the threshold of the regime of eternal inflation that surrounds the weakly or non-gravitating patch containing $R$, provided of course one considers an appropriate observable. The picture that arises is that of an ``inside out'' version of black holes, in which the definite classical spacetime around us corresponds to an oasis surrounded by (quantum) uncertainty \cite{Hartle:2010dq,Hartle:2016tpo}.

The organization of this paper is as follows. We begin in \Sec{subsec:JT} with a brief review of de Sitter, flat, and anti de Sitter versions of JT gravity, and a discussion of how to glue these solutions together to construct JT multiverse models. We also review in \Sec{sec:CFTentropy} results for the generalized entropy of an interval in a probe conformal field theory (CFT) with large central charge coupled to JT gravity.  In \Sec{Sec:islands}, we analyze the generalized entropy of an interval region in a single de Sitter or flat patch of the JT multiverse, and demonstrate the late time, large interval entropy preferring the formation of a large single island. We comment on other configurations including multiple small islands and intervals spanning several patches in \App{app:approx-checks}.  In \Sec{Sec:quantum-cosmology}, we develop the analogy between the qualitative general lessons from our investigations of the JT multiverse and a semiclassical quantum cosmology description of higher dimensional inflationary multiverses. We conclude with an extensive discussion of our results in \Sec{Sec:conclusion}, and point to some future directions.

\subsection{Erratum}

We did not properly account for contributions from the Weyl anomaly and the Casimir energy in Secs.~\ref{sec:CFTentropy} and \ref{Sec:islands}.
They combine to give a source for the metric equation of motion; see, e.g., \cite{Balasubramanian:2020xqf}.
These contributions cancel when $n=1$.\footnote{We thank Edgar Shaghoulian for pointing this out to us.}
As a consequence, only the calculations and results in \Sec{Sec:islands} that operate in the $\phi_r/G_N \gg c$ limit---where we can neglect backreaction with certain caveats---are reliable.
We discuss further details in the addendum in \App{app:addendum}.
However, the following items which assumed $\phi_r/G_N \ll c$ should not be considered reliable:
\begin{itemize}
\item The paragraph containing Eqs.\eqref{eq:approx_Sgen_end} and \eqref{eq:Page-approx-2}.
\item The numerical Page curves in Figs.~\ref{FigPagedS2n}-\ref{Figtime}, \ref{fig:flat-bubble-Page}-\ref{fig:AdS-bubble-Page}, \ref{fig:FFdS2n}-\ref{fig:crunching0}, \ref{fig:2islands}.
\item The sentence containing  \Eq{eq:Sisland-flat2} and the one that immediately follows it.
\end{itemize}

\section{Jackiw–Teitelboim ``cosmology''}
\label{Sec:JT}

In this section,  first we discuss our toy model of cosmology in  \Sec{subsec:JT} and then, in \Sec{sec:CFTentropy}, we prepare the formulae for the computation of the generalized entropy of subregions. We borrow techniques developed in \Refs{Chen:2020tes,Hartman:2020khs} and expand them to build two-dimensional de Sitter solutions with multiple flat, crunching, and expanding bubbles.

\subsection{Bubbles in de Sitter JT gravity}
\label{subsec:JT}

Our starting point is  the de Sitter version of JT gravity, extensively studied in Refs.~\cite{Anninos:2017hhn,Anninos:2018svg,Maldacena:2019cbz,Cotler:2019nbi}.
 It is a theory of two-dimensional spacetime with positive curvature coupled to a dilaton field, $\phi$.
The action is given by
\begin{equation}\label{eq:JT-action-g-only}
I_{dS\text{-}JT}[g_{\mu\nu},\phi] = \frac{\phi_0}{16\pi G_N} \int_{\mathcal{M}} d^2x\sqrt{-g} \mathcal{R} + \frac{1}{16\pi G_N} \int_{\mathcal{M}} d^2x \sqrt{-g}\phi (\mathcal{R}-2) +I_{GHY}[g_{\mu\nu},\phi],
\end{equation}
where $\mathcal{R}$ is the bulk scalar curvature, $I_{GHY}$ is the Gibbons-Hawking-York counterterm, 
\begin{equation}\label{eq:JT-g-counterterm-action}
I_{GHY}[g_{\mu\nu},\phi] = \frac{\phi_0}{8\pi G_{N}}\int_{\partial\mathcal{M}} K + \frac{1}{8\pi G_N} \int_{\partial\mathcal{M}}\tilde\phi (K - 1),
\end{equation}
$\tilde\phi$ is the boundary value of the dynamical dilaton, and $K$ is the trace of the extrinsic curvature of the boundary $\partial\mathcal{M}$ of a manifold $\mathcal{M}$. In writing \Eq{eq:JT-action-g-only}, we have also included a topological term proportional to $\phi_0$, a positive constant. Moreover, we have set the length scale of the cosmological constant to one.

\begin{figure}[t]
    \centering
    \includegraphics[scale=0.2]{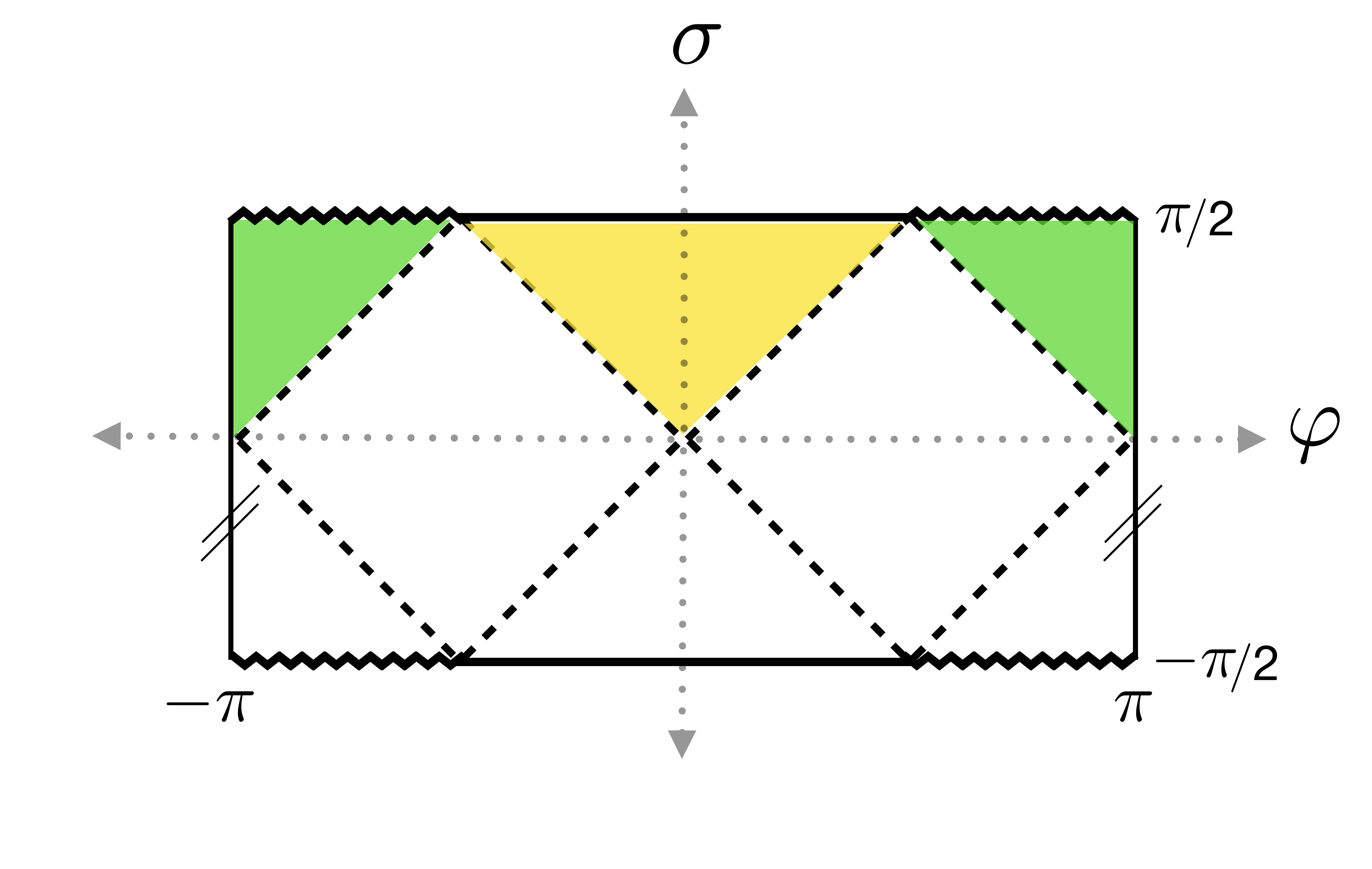}
    \caption{Penrose diagram for $\mrm{dS}_2$. The horizontal line $\sigma = \pi/2$ corresponds to $\mathcal{I}^+$. The expanding patch (the past domain of dependence of the part of $\mathcal{I}^+$ where $\phi$ diverges to $+\infty$) is shaded in yellow, and the crunching patch (the past domain of dependence of the part of $\mathcal{I}^+$ where $\phi$ diverges to $-\infty$) is shaded in green.}
    \label{fig:dS2-penrose}
\end{figure}

Varying $I_{dS\text{-}JT}$ with respect to $\phi$ enforces $\mathcal{R} = 2$; i.e., the spacetime is fixed to be locally $\mrm{dS}_2$.
Consider first the exactly $\mrm{dS}_2$ solution.
We may write its line element in terms of compact global coordinates $(\sigma, \varphi)$ as
\begin{equation}\label{eq:dS2-line-element}
    \dee s^2 = \sec^2 \sigma \left( -\dee \sigma^2 + \dee \varphi^2 \right),
\end{equation}
where the timelike coordinate $\sigma$ takes values in $(-\pi/2,\pi/2)$ and the spacelike coordinate $\varphi \in (-\pi, \pi)$ is periodically identified at its endpoints.
These coordinates cover the whole $\mrm{dS}_2$ manifold and are useful for depicting its conformal structure in a Penrose diagram, as shown in \Fig{fig:dS2-penrose}.
Varying $I_{dS\text{-}JT}$ with respect to $g_{\mu\nu}$ gives the metric equation of motion
\begin{equation}\label{eq:JT-g-only-metric-eom}
(g_{\mu\nu} \nabla^2 - \nabla_\mu \nabla_\nu + g_{\mu\nu})\phi = 0,
\end{equation}
whose solution for the line element \eqref{eq:dS2-line-element} is given by
\begin{equation}\label{eq:dS2-dilaton}
    \phi(\sigma,\varphi) = \phi_r \frac{\cos \varphi}{\cos \sigma},
\end{equation}
where $\phi_r > 0$, which satisfies $\tilde{\phi}=\phi_r/\varepsilon$ for a small UV cutoff $\varepsilon$.

Although the spacetime metric has $\mathcal{R} = 2$ everywhere, we can think of this $\mrm{dS}_2$ solution  as a simple, low-dimensional model of a cosmological spacetime that has two types of regions with different physical properties.
These two types of regions are distinguished by the dilaton's behavior in the asymptotic future \cite{Chen:2020tes,Hartman:2020khs}.
Namely, $\mathcal{I}^+$ can be partitioned into an interval where $\phi \rightarrow +\infty$ and an interval where $\phi \rightarrow -\infty$.
The two types of region in question are then identified with these intervals' past domains of dependence.
In previous literature, the past domain of dependence of the part of $\mathcal{I}^+$ where $\phi \rightarrow +\infty$ has been called an ``expanding patch'', and the past domain of dependence of its complement has been called a ``crunching patch.''
The intuition for this termninology comes from viewing JT gravity as descending from a higher dimensional theory, which we briefly review here for completeness;  see, e.g., \cite{Maldacena:2019cbz, Cotler:2019nbi} for more details.
However, we emphasize that we will always treat the  two-dimensional de Sitter JT gravity theory as a standalone toy model of cosmology.

Starting from the de Sitter-Schwarzschild black hole solution to four-dimensional Einstein gravity,
\begin{equation}
 \dee s^2 = -f(r)\dee t^2 +f(r)^{-1}\dee r^2 +r^2\dee\Omega_2^2 \qquad \text{with} \qquad f(r) = 1- 2M/r - r^2/L_{4d}^2,
\end{equation}
one can obtain  de Sitter JT gravity via dimensional reduction of the near horizon geometry.
The procedure follows by taking the limit where the zeros ($0<r_-<r_+$) of $f(r)$ degenerate ($r_- = r_+ = \hat{r}$), i.e. the zero temperature limit, which produces the Nariai dS$_2\times\,\text{S}^2$ spacetime geometry:
\begin{equation} \label{eq:JT-expansion}
 \dee s^2 = \hat{r}^2\dee s_{\mrm{dS}_2}^2 +\hat{r}^2(1+\delta )^2\dee\Omega_2^2 \qquad \text{with} \qquad 1+\delta = r/\hat{r}.
\end{equation}
Expanding the four-dimensional theory perturbatively to zeroth order in $\delta$ and dimensionally reducing on the transverse $\mrm{S}^2$ gives the topological terms in Eqs.~\eqref{eq:JT-action-g-only} and \eqref{eq:JT-g-counterterm-action} with $\phi_0 = L_{4d}^2 \hat{r}^2/4G_{4d}$.
By including the leading deformation at $O(\delta)$, the same dimensional reduction yields the full JT gravity action in \Eq{eq:JT-action-g-only} with dynamical dilaton $\phi = 2\phi_0\delta$.
Locations where the dilaton becomes negative therefore correspond to the black hole interior in the higher dimensional picture, and positive values of the dilaton correspond to the black hole exterior, where spacetime expands eternally as $\phi \rightarrow +\infty$.
While $\phi$ cannot be less than $-1$ according to \Eq{eq:JT-expansion}, $\phi \rightarrow -\infty$ in the two-dimensional model is commonly viewed as signalling the  eventual black hole singularity in the higher dimensional theory.
This motivates the nomenclature ``expanding patch'' and ``crunching patch,'' which we will continue to use throughout this work.
More relevantly, we call spacetime regions in which $\phi \rightarrow + \infty$ regions of weak gravity, and regions where $\phi \rightarrow - \infty$ regions of strong gravity in our model.
Much like the terms ``expanding'' and ``crunching,'' this identification is inspired by the higher dimensional theory; the black hole singularity is clearly a region of strong gravity, while the four-dimensional Newton's constant is small when the dilaton is large.
However, we take this identification to be intrinsic to the two-dimensional model itself and independent of any specific choice of parameters.

\begin{figure}[t]
    \centering
    \includegraphics[scale=0.22]{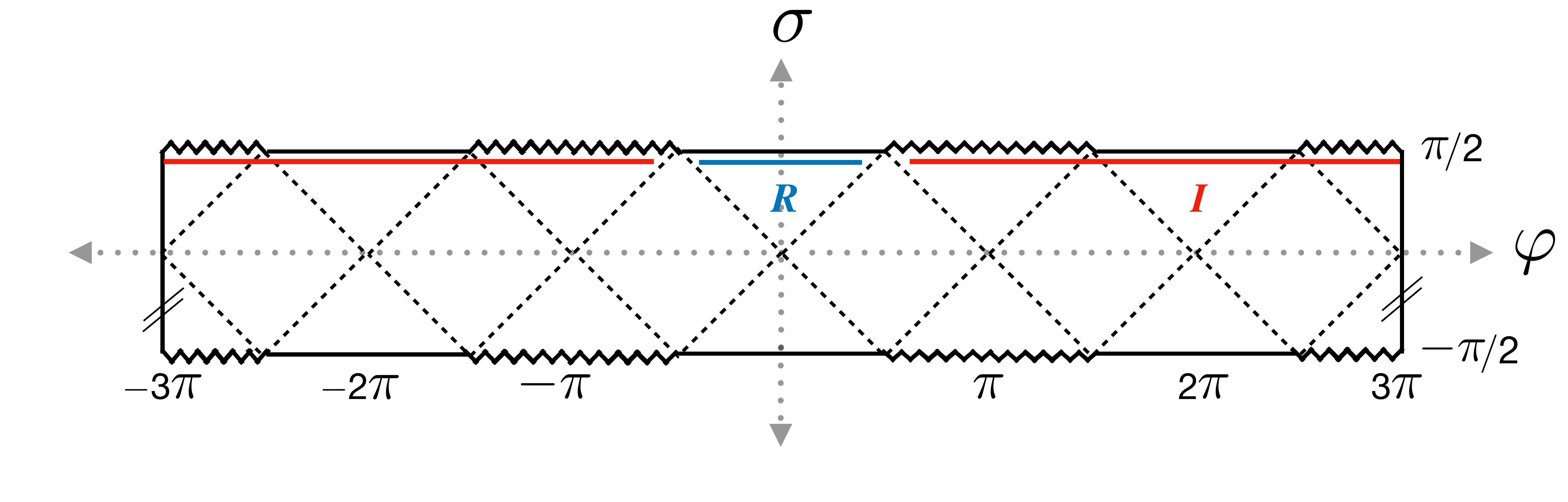}
    \caption{Penrose diagram for extended $\mrm{dS}_2^n$. The case with $n=3$ is drawn here for illustration. The region $R$ lies in the expanding patch that is centered about $\varphi=0$. We take as an ansatz an island, $I$, whose endpoints $(\sigma_I, \pm \varphi_I)$ are in the crunching patches that are adjacent to $R$.}
    \label{fig:dS2n-penrose}
\end{figure}

We can push this low-dimensional cosmological model further by making two additional observations.
First, one can analytically extend the spacetime by allowing the angular coordinate $\varphi$ to be $2\pi n$-periodic for natural numbers $n \geq 1$.
This results in a larger spacetime  where the line element is still given by \Eq{eq:dS2-line-element} and on which the dilaton is still given by \Eq{eq:dS2-dilaton}, but now we allow $\varphi$ to take values in $(-n\pi, n\pi)$.
The Penrose diagram of such an extension consists of $n$ copies of the diagram in \Fig{fig:dS2-penrose} that are glued together before being periodically identified along the leftmost and rightmost sides, as illustrated in \Fig{fig:dS2n-penrose} with $n=3$.
In terms of an embedding of $\mrm{dS}_2$ as a hyperboloid in $\mathbb{R}^{1,2}$, such an extension covers the hyperboloid $n$ times.
We will denote this $n$-fold extension of $\mrm{dS}_2$ by $\mrm{dS}_2^n$, and the decompactified limit is obtained by formally taking $n \rightarrow +\infty$.\footnote{While an $n$-fold extension of $\mrm{dS}_2$ is a  well-defined classical solution that can function as a background for a quantum field theory, subtleties arise if one tries to define a quantum state for the gravitational sector. We will discuss this point in \Sec{Sec:conclusion}.}

\begin{figure}
    \centering
    \includegraphics[scale=0.15]{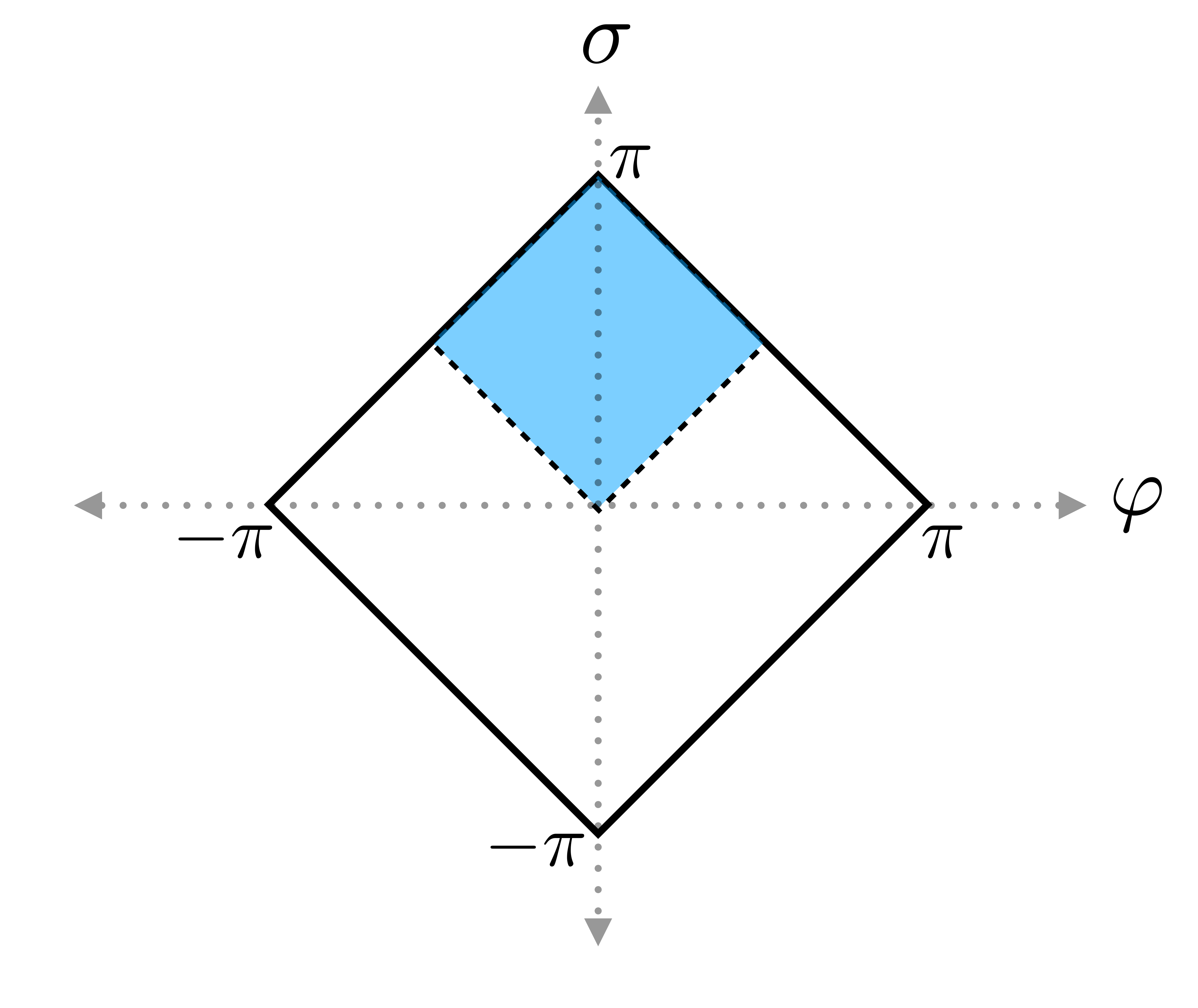} ~ \includegraphics[scale=0.15]{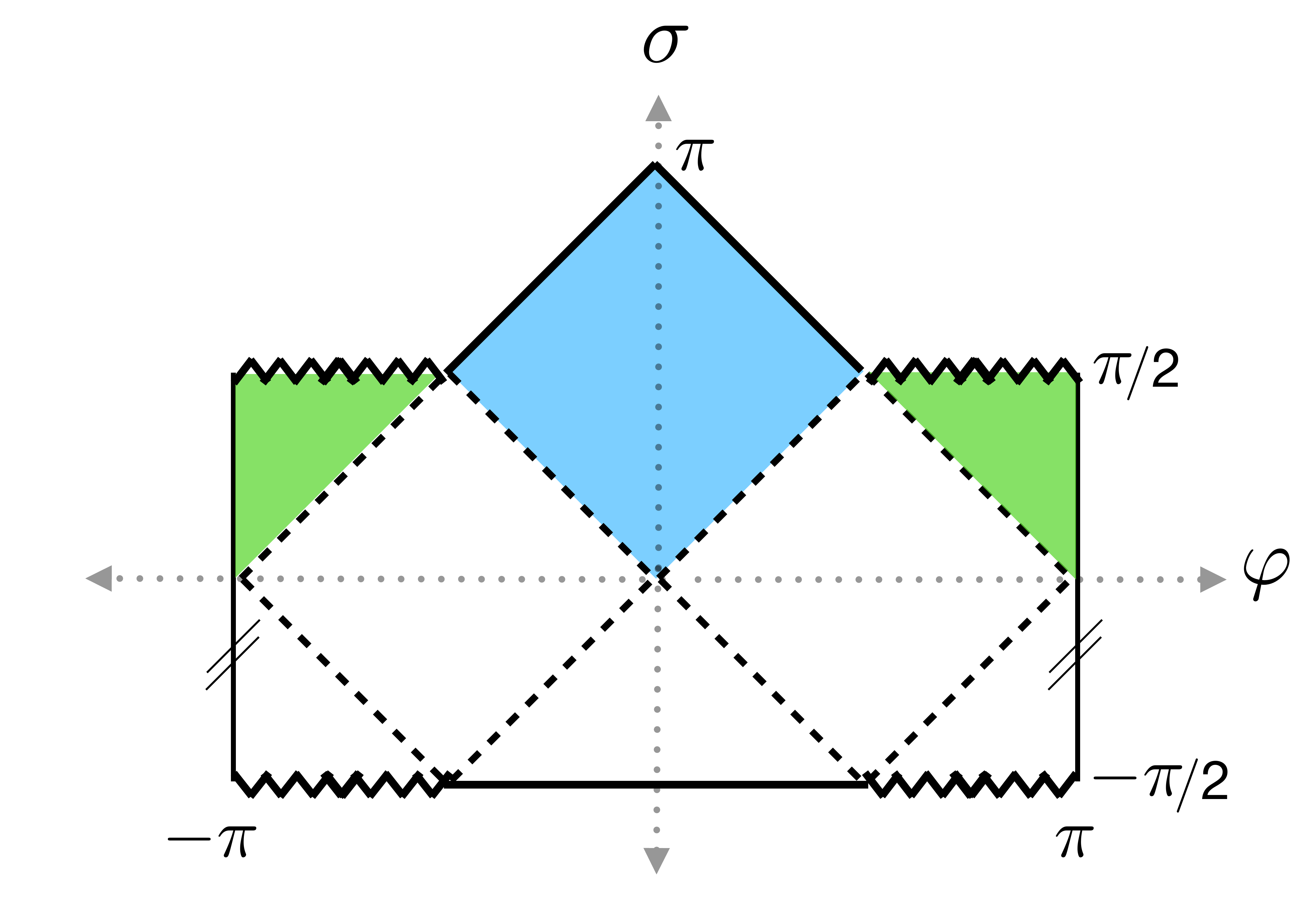}
    \caption{(Left) Penrose diagram for $\mathbb{R}^{1,1}$. (Right) Penrose diagram for $\mrm{dS}_2$ where the expanding patch has been replaced with a bubble of flat spacetime. The potion of full $\mathbb{R}^{1,1}$ that this bubble corresponds to is shaded in the left diagram.}
    \label{fig:flat-penrose}
\end{figure}

Second, as was pointed out in \Ref{Hartman:2020khs}, one can excise an expanding patch or a crunching patch and replace it with a patch of flat spacetime.
A flat version of JT gravity is obtained by replacing the integrand of the second term in \Eq{eq:JT-action-g-only} with $\sqrt{-g} \left( \phi \mathcal{R} - 2 \right)$, leading to\footnote{Notice that there are other possible choices in place of \Eq{eq:JT-action-g-only-flat} that would give flat spacetime solutions in two dimensions. We followed the conventions of \cite{Hartman:2020khs},  which result in the dilaton diverging to $+ \infty$ toward the future.}
\begin{equation}\label{eq:JT-action-g-only-flat}
I_{flat\text{-}JT}[g_{\mu\nu},\phi] = \frac{\phi_0}{16\pi G_N} \int_{\mathcal{M}} d^2x\sqrt{-g} \mathcal{R} + \frac{1}{16\pi G_N} \int_{\mathcal{M}} d^2x \sqrt{-g} (\phi\mathcal{R}-2) +I_{GHY}[g_{\mu\nu},\phi].
\end{equation}
The resulting dilaton equation of motion is $\mathcal{R} = 0$, and the metric equation of motion is
\begin{equation}\label{eq:JT-g-only-metric-eom-flat}
(g_{\mu\nu} \nabla^2 - \nabla_\mu \nabla_\nu)\phi + g_{\mu\nu} = 0.
\end{equation}
In terms of the usual planar coordinates $(t,x)$ for which $\dee s^2 = -\dee t^2 + \dee x^2$, the general solution for the dilaton is $\phi(t,x) = \frac{1}{2}(t^2 - x^2) + At + Bx + C$ for constants $A$, $B$, and $C$.
Let us instead choose coordinates $(\sigma, \varphi)$ by defining  
\begin{equation}
\begin{aligned}\label{eq:tx}
    t &= \tan\left( \frac{\sigma+\varphi}{2} \right) + \tan \left( \frac{\sigma-\varphi}{2} \right), \\
    x &= \tan\left( \frac{\sigma+\varphi}{2} \right) - \tan \left( \frac{\sigma-\varphi}{2} \right).
\end{aligned}
\end{equation}
The range of these coordinates is $|\sigma \pm \varphi| < \pi$, and the line element reads
\begin{equation}\label{eq:mink2-line-element}
    \dee s^2 = \frac{-\dee \sigma^2 + \dee \varphi^2}{\tfrac{1}{4}(\cos\sigma+\cos\varphi)^2}=\frac{-\dee \sigma^2 + \dee \varphi^2}{\cos^2(\frac{\sigma+\varphi}{2})\cos^2(\frac{\sigma-\varphi}{2})}.
\end{equation}
If we set the integration constants $A = B = 0$ and $C = \phi_r$, the dilaton reads
\begin{equation}\label{eq:mink2-dilaton}
    \phi(\sigma,\varphi) = \phi_r + 2\tan\left( \frac{\sigma+\varphi}{2} \right) \tan \left( \frac{\sigma-\varphi}{2} \right),
\end{equation}
and we can continuously join the flat solution in Eqs.~\eqref{eq:mink2-line-element} and \eqref{eq:mink2-dilaton} to the $\mrm{dS}_2$ solution in Eqs.~\eqref{eq:dS2-line-element} and \eqref{eq:dS2-dilaton} along the line segments $\sigma = |\varphi|$; see \Fig{fig:flat-penrose}.
The dilaton's first derivatives will be discontinuous whenever $\phi_r \neq 1$, which signals that the interface must carry some tension.
We will return to this point in the next subsection. 
Then, by extension, it follows that for the right choice of integration constants (as well as an appropriate offset for $\varphi$), one can substitute a flat patch as defined by \Eqs{eq:mink2-line-element}{eq:mink2-dilaton} for any expanding or crunching patch in an extended $\mrm{dS}_2^n$ manifold.
In this way, we can build up a  model which we call a  ``JT multiverse'' that consists of a pattern of expanding, crunching, and flat patches that can be arbitrarily long.

\begin{figure}
    \centering
    \includegraphics[scale=0.15]{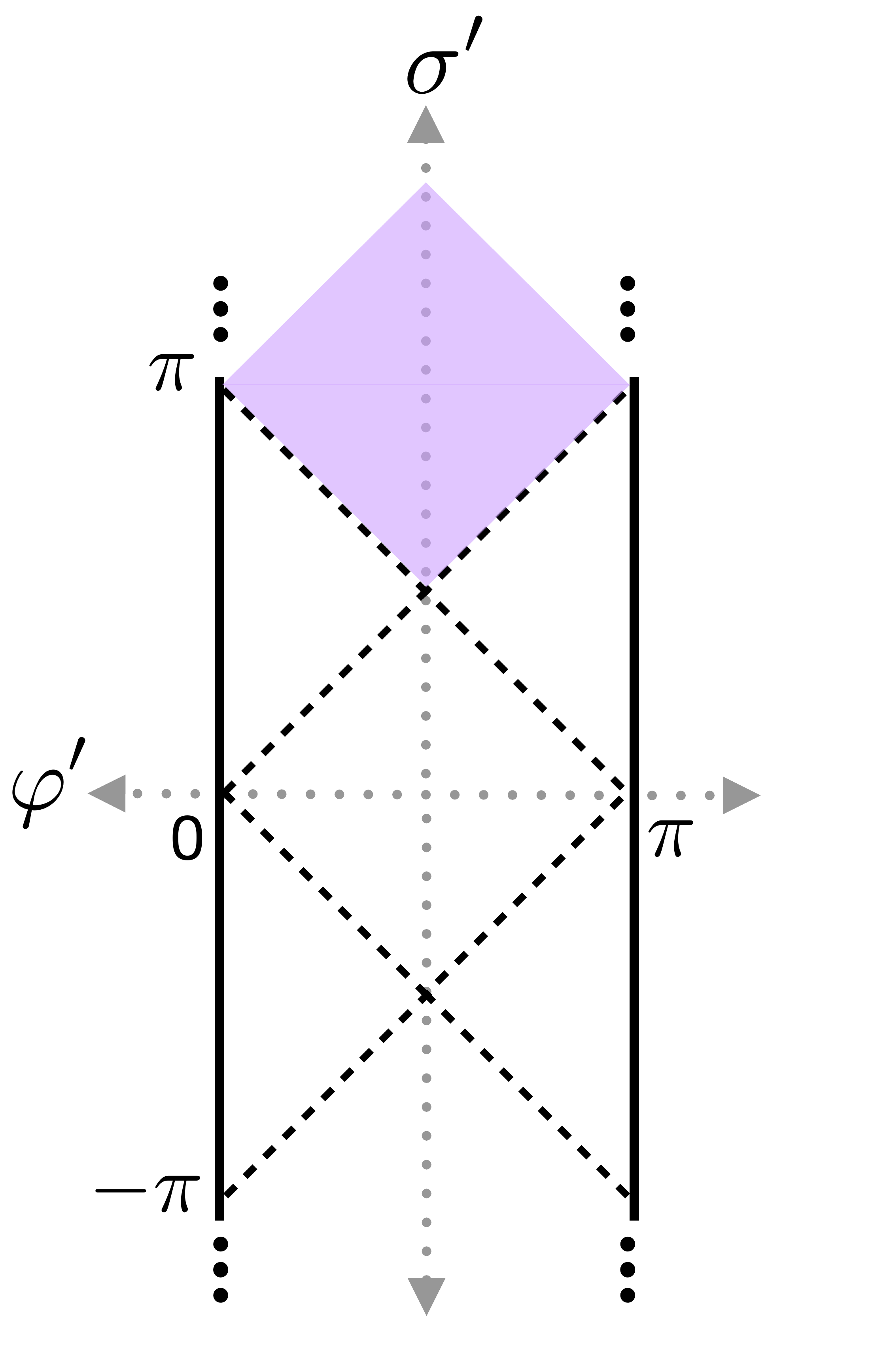} ~ \includegraphics[scale=0.15]{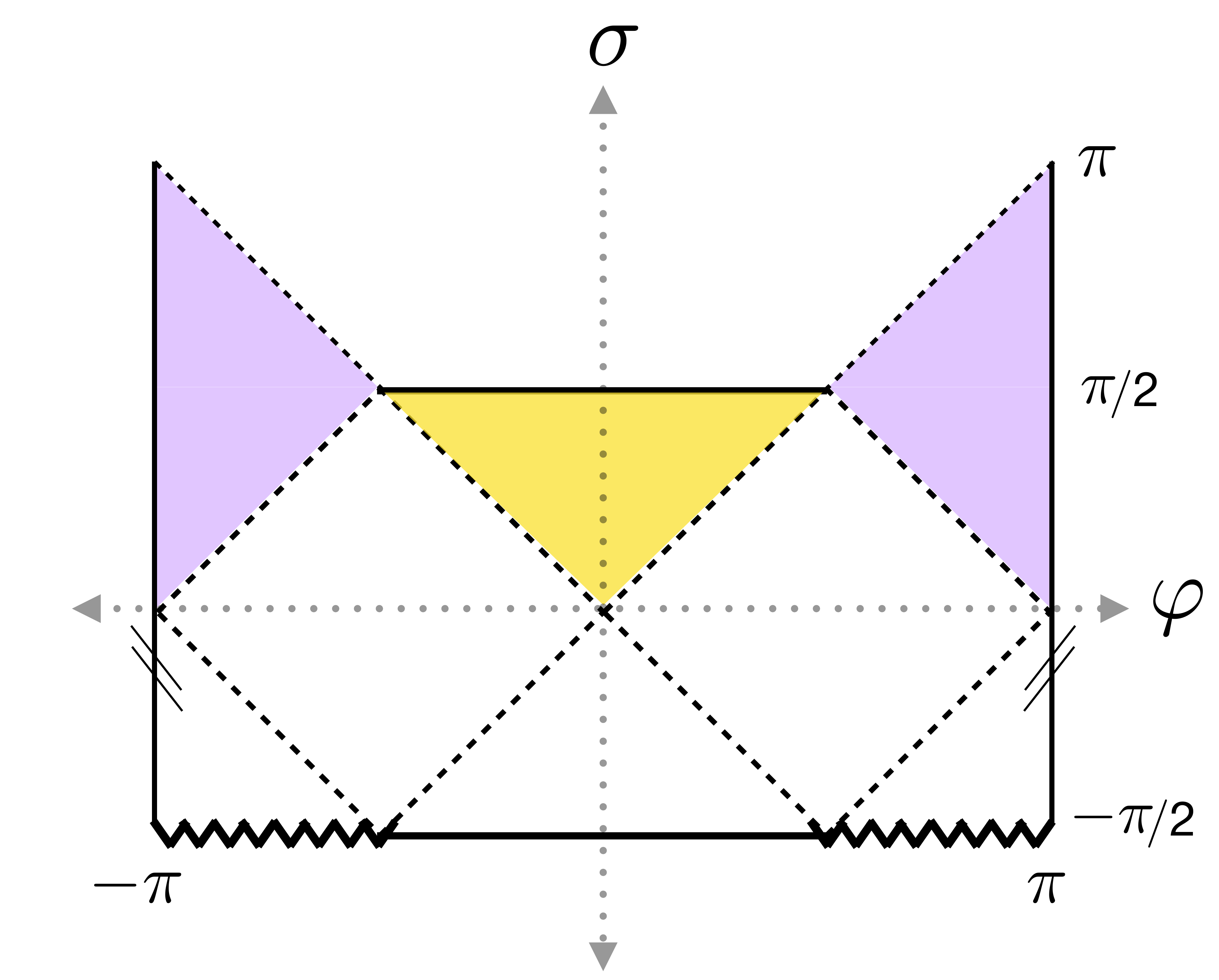}
    \caption{(Left) Penrose diagram for global $\mrm{AdS}_2$. (Right) Penrose diagram for $\mrm{dS}_2$ where the crunching patch has been replaced with a bubble of $\mrm{AdS}_2$. The potion of the $\mrm{AdS}_2$ manifold that this bubble corresponds to is shaded in the left diagram. Although not illustrated here, the diamond centred about $(\sigma',\varphi') = (0,\pi/2)$ could be used to replace the expanding patch of $\mrm{dS}_2$.}
    \label{fig:AdS-Penrose}
\end{figure}

It is also possible to patch in a portion of a two-dimensional anti de Sitter ($\mrm{AdS}_2$) spacetime in lieu of an expanding or crunching $\mrm{dS}_2$ patch.\footnote{The extension of $\varphi$'s range and the inclusion of $\mrm{AdS}_2$ bubbles are both departures from the model proposed in \Ref{Hartman:2020khs}.}
Upon replacing $(\mathcal{R}-2)$ with $(\mathcal{R}+2)$ in \Eq{eq:JT-action-g-only}, the usual story for $\mathcal{R} = -2$ follows \cite{Maldacena:2016upp,Sarosi:2017ykf}: in terms of Poincar\'e coordinates $(t,z)$, the $\mrm{AdS}_2$ line element reads $\dee s^2 = z^{-2}(-\dee t^2 + \dee z^2)$ and the general solution for the dilaton is $\phi = (A + Bt + C(t^2+z^2))/z$.
We will instead work with global coordinates $(\sigma',\varphi')$, where $t \pm z = \tan((\sigma' \pm \varphi')/2)$.
The $\mrm{AdS}_2$ line element then reads
\begin{equation}
    \dee s^2 = \csc^2\varphi'(- \dee \sigma'\,^2 + \dee \varphi'\,^2),
\end{equation}
where $\sigma' \in \mathbb{R}$ and $\varphi' \in (0, \pi)$; see \Fig{fig:AdS-Penrose} for a Penrose diagram.
For the dilaton, we set $A = C = 0$ and $B = \phi_r$ to obtain
\begin{equation}
    \phi = \phi_r \frac{\sin\sigma'}{\sin\varphi'},
\end{equation}
where $\phi_r$ is the same as in \Eq{eq:dS2-dilaton} so that we may perform a continuous gluing.
If we shift the global coordinates by defining $\varphi = \varphi'-\pi/2$ and $\sigma = \sigma' - \pi/2$ (resp. $\sigma = \sigma' + \pi/2$), then we can glue a diamond with $\sigma > |\varphi|$ into a crunching patch (resp. expanding patch).
In terms of these shifted coordinates, the $\mrm{AdS}_2$ line element reads
\begin{equation} \label{eq:AdS2-line-element}
    \dee s^2 = \sec^2\varphi(- \dee \sigma^2 + \dee \varphi^2),
\end{equation}
and the dilaton is given by
\begin{equation} \label{eq:AdS2-dilaton}
    \phi = \pm \phi_r \frac{\cos \sigma}{\cos \varphi},
\end{equation}
where we must take the positive sign when substituting for an expanding patch and the negative sign when substituting for a crunching patch.

A bubble of $\mrm{AdS}_2$ behaves somewhat similarly to a crunching patch regarding whether or not an island forms, and so we will not focus too much on such bubbles.
Nevertheless, it is interesting and satisfying that one can construct toy JT multiverses that contain both flat and negative curvature bubbles in an ambient positive curvature spacetime.
As such, these JT multiverses are low-dimensional models for the sorts of mosaic universes predicted by traditional eternal inflation.
In such universes, instantons can nucleate bubbles that have different values of the cosmological constant and different physical properties within an ambient, eternally inflating spacetime that has a positive cosmological constant.
In the JT multiverses considered here, the background curvature and asymptotic behavior of the dilaton are proxies for different, distinguishable cosmological properties.

\subsection{CFTs and generalized entropy in JT gravity} \label{sec:CFTentropy}

Here we consider deforming the action of de Sitter and flat JT gravity in \Eq{eq:JT-action-g-only} and \Eq{eq:JT-action-g-only-flat}, respectively, by coupling to the background metric $g_{\mu\nu}$ a two-dimensional CFT
with field content collectively denoted by $\psi$, where $I_{CFT}[g_{\mu\nu},\psi]$ is the action of the Lorentzian CFT describing the matter sector. Crucially, we assume as usual that the CFT does not couple to the dilaton, and so there is no backreaction to take us away from the  background solution of the dilaton equation of motion. Further, we assume that the central charge of the CFT is very large, $c\gg 1$, such that we can consistently treat the gravitational sector in the presence of matter at the semiclassical level and neglect fluctuations in the boundary mode of the dynamical dilaton.\footnote{Given that our perspective on JT gravity coupled to a probe CFT is purely two-dimensional, we may freely dial $\phi_0$ and $\phi_r$ so long as $c\gg 1$.  However, if we were to consider our model as embedded in a higher dimensional theory, e.g. a feature necessary in the analysis of \cite{Aalsma:2021bit}, then the parameters of the lower dimensional theory would have to lie in a hierarchy $1\ll c\ll \phi_r/G_N \ll \phi_0/G_N$ in order to work in a semiclassical regime where gravity is weak and the matter sector is a probe of the classical background.} Finally, we require that the CFT is in a global vacuum and therefore that the stress tensor has vanishing one-point function $\langle T_{\mu\nu}\rangle=0$ in the geometries that we consider below. With these assumptions, the metric equations of motion in \Eqs{eq:JT-g-only-metric-eom}{eq:JT-g-only-metric-eom-flat} are left unmodified. 

While requiring $\langle T_{\mu\nu}\rangle =0$, we need to be careful about trace anomaly contributions,
\begin{align}
\langle T^\mu{}_\mu\rangle = \frac{c}{12} \mathcal{R},
\end{align}
arising in regions of our JT multiverse solutions with non vanishing $\mathcal{R}$.
Such a term can enter as a source for the  dilaton as can be seen, for example, by computing the trace of \Eq{eq:JT-g-only-metric-eom},
\begin{align}
(\nabla^2 + 2)\phi = 8\pi G_{N} \langle T^\mu{}_\mu\rangle.
\end{align}
However, including boundary contributions, the integrated trace anomaly of a two-dimensional CFT takes the form 
\begin{align}
   \int d^2x\sqrt{-g} \langle T^\mu{}_\mu \rangle = \frac{c}{24\pi}\int_{\mathcal{M}} d^2x \sqrt{-g}\mathcal{R} + \frac{c}{12\pi} \int_{\partial\mathcal{M}} K.
\end{align} 
Therefore,
at the level of the action, we can redefine the constant value of the dilaton $\phi_0 \to \phi^\prime_0 = \phi_0 + 2c G_N/3$ to remove the source.
Thus, without loss of generality, we will assume the dilaton obeys source-free metric equations of motion in what follows.

Ultimately, we will be interested in computing the Von Neumann entropy $S(\rho_R)$ associated to a subregion $R$ in our JT multiverse plus CFT model.
According to the island formula \Eq{eq:islands}, we will therefore need to compute the generalized entropy  for different configurations of $R$ and $I$. From \Eq{eq:gen-entropy}, we obtain
\begin{align}\label{eq:gen}
    S_{\rm gen}(R\cup I) =  S_{\rm CFT}(R\cup I)+\frac{\rm{Area}(\partial I)}{4G_N}  - S_{\rm{ct}}(\partial I),
\end{align}
where $S_{\rm CFT}$ is the semiclassical entropy of CFT fields---that is, the entropy of the quantum  fields on a fixed background geometry evaluated according to the conventional techniques of quantum field theory in curved spacetime.
The second term in \Eq{eq:gen} is the gravitational contribution to generalized entropy coming from the boundary of the island.
In JT gravity, the ``area'' of the boundary of the island is just  $\phi_0 + \phi$, evaluated at and summed over all of the island's endpoints.
We omit an area term due to the boundary of $R$; in principle we could include this contribution, but it would not change any of our conclusions, as we will see shortly.
In writing \Eq{eq:gen}, we have included  $S_{\mrm{ct}}$,  a counterterm originating from the gravitational contribution that renormalizes the UV divergence in $S_{\rm CFT}$ coming from the boundary of $I$.

To compute the semiclassical entropy of fields, owing to the simplicity of our model, we can use standard universal results of Von Neumann entropy of a two-dimensional CFT in Minkowski vacuum \cite{Holzhey:1994we,Calabrese:2004eu,Calabrese:2009qy}. 
For a subregion taken to be a single interval of proper length $\ell$, it is given by
\begin{align}\label{eq:2d-EE-no-island-1}
    S_{\rm CFT} = \frac{c}{6} \log \frac{\ell^2}{\epsilon_{\rm uv}^2} + s_0,
\end{align}
where $\epsilon_{\rm uv}\ll \ell$ is a UV regulator and $s_0$ is a scheme-dependent constant. In our case, we are working under the assumption that all CFT fields are in a vacuum state of a JT multiverse geometry written in $(\sigma,\varphi)$ coordinates.  Therefore, we need to translate our global coordinates to those in which the CFT is in a Minkowski vacuum. 

For all values of $\mathcal{R}$, we can put the background metric in the form
\begin{align}\label{eq:generic-extended-metric-1}
    \dee s^2 = \frac{1}{\omega^2(\sigma,\varphi)}(-\dee\sigma^2+\dee\varphi^2),
\end{align}
where
\begin{equation}\label{eq:conformal-factors}
    \omega(\sigma,\varphi) = \left\{
    \begin{array}{ll}
    \cos \sigma & (\sigma,\varphi) ~ \text{in a} ~ \mathcal{R} = 2 ~ \text{patch} \\[2mm]
    \tfrac{1}{2} (\cos \sigma + \cos \varphi) & (\sigma,\varphi) ~ \text{in a} ~ \mathcal{R} = 0 ~ \text{patch} \\[2mm]
    \cos \varphi & (\sigma,\varphi) ~ \text{in a} ~ \mathcal{R} = -2 ~ \text{patch.}
    \end{array} \right.
\end{equation}
By rescaling $\sigma = n\tilde\sigma$ and $\varphi=n\tilde\varphi$, such that the spatial coordinate takes values $\tilde\varphi \in (-\pi,\pi)$, we may perform the coordinate transformation
\begin{align}
    z = e^{-i(\tilde\sigma+\tilde\varphi)},\hspace{1.25cm} \bar{z} = e^{-i(\tilde\sigma-\tilde\varphi)},\label{xxbar}
\end{align}
and the metric
 \Eq{eq:generic-extended-metric-1}  becomes
\begin{align}\label{eq:generic-extended-metric-2}
 \dee s^2 = \frac{n^2\dee z \dee\bar z}{\omega^2(n\tilde\sigma,n\tilde\varphi) e^{-2i\tilde\sigma}}=:\frac{\dee z \dee \bar{z}}{\Omega^2(z,\bar{z})}. 
\end{align}
With respect to $(z,\bar{z})$ coordinates, the CFT is in a Minkowski vacuum up to a Weyl rescaling.

Thus, given the entangling region be an interval with endpoints at $(z_1,\bar{z}_1)$ and $(z_2,\bar{z}_2)$, using the metric \Eq{eq:generic-extended-metric-2} in the universal formula \Eq{eq:2d-EE-no-island-1}, we find
\begin{align}\label{eq:2d-EE-no-island-2}
    S_{\rm CFT} = \frac{c}{6}\log\left[\frac{z_{12}\bar{z}_{12}}{\epsilon_{\rm uv}^2 \Omega_1\Omega_2}\right],
\end{align}
where $z_{ij} := z_i - z_j$ (resp. $\bar{z}_{ij}$) and $\Omega_i := \Omega(z_i,\bar{z}_i)$. Following from the coordinate transformations above for a CFT on the background described in \Eq{eq:generic-extended-metric-1}, we find the following expression for the single interval Von Neumann entropy
\begin{align}
    S_{\rm CFT}=\frac{c}{6}\log \left[\frac{2n^2(\cos(\sigma_{ij}/n) - \cos(\varphi_{ij}/n))}{\epsilon_{\rm uv}^2\omega_1\omega_2 }\right] + s_0,\label{eq:2d-EE-no-island-3}
\end{align}
where we adopt the notation $\sigma_{ij},\,\varphi_{ij}$ from above for $z_{ij}$ and $\omega_i := \omega(\sigma_i,\varphi_i)$. \Eq{eq:2d-EE-no-island-3} will prove useful for comparisons in the following sections in our search for islands in dS$_2^n$.

Since the matter sector we are considering is a two-dimensional CFT, the regularizing term $S_{\rm ct}(\partial I)$ takes a simple form.  That is, if we consider an island configuration of a system of disjoint intervals $I = \bigsqcup_j I_j$ with $k$ endpoints, then
\begin{align}
    S_{\rm ct} = k\frac{c}{6}\log \frac{\epsilon_{\rm rg}}{\epsilon_{\rm uv}},
\end{align}
where $\epsilon_{\rm rg}\gg \epsilon_{\rm uv}$ is an arbitrary renormalization scale.  The appearance of this scale can be thought of as due to contact terms in the non-minimally coupled CFT which contributes to the RG flow of $1/G_N$  \cite{Bousso:2015mna}.\footnote{This new scale, $\epsilon_{\rm rg}$, can be absorbed into $G_N$, but we will keep it explicit throughout the following sections.} As we will use in the subsequent section for single island configurations, and for multiple disjoint islands in the appendix, the net effect of $S_{\rm ct}$ on the rest of the non-geometrical part of the generalized entropy, i.e. $S_{\rm CFT}$, will be to renormalize $\epsilon_{\rm uv}^2 \to \epsilon_{\rm rg} \epsilon_{\rm uv}$. 

 Before moving on, there are a few remaining subtleties that we must address.  The above review of generalized entropy for large $c$ CFTs in JT gravity implicitly assumed a smooth gluing of the interfaces between different patches.  However, since we consider configurations with patches of different $\mathcal{R}$ glued together below, we should address the possible shortcomings of our approach.

First, we will assume below that it is sufficient to consider transparent boundary conditions for the CFT matter at the interface between patches.  That is, from the perspective of the CFT, the interface is trivial.  However, in the case that the interface carries some non-trivial tension due to $\phi_r \neq 1$, it is not clear {\textit{a priori}} that this assumption holds insofar as the presence of such an interface could break conformal symmetry by interface couplings between the bulk conformal matter and interface-localized degrees of freedom. For the following analysis, we can either assume that no interface couplings appear, such that the probe CFT is completely decoupled, or that the conformal symmetry enjoyed by the probe CFT is manifest, at least approximately, in regions far from a non-trivial interface.  With either of these assumptions, it is possible to apply the above results for the Von Neumann entropy of the CFT (reliably in regions far from a non-trivial interface) in all cases. 

This brings us to the last point that we need to address regarding the configurations of the entangling region $R$ and the islands $I$.  It is well known that in two-dimensional CFTs on a background with a non-empty boundary the Von Neumann entropy for a region $R$ that has non-trivial intersection with the boundary is not simply given by \Eq{eq:2d-EE-no-island-1} but rather picks up an additional universal $\log(g)$ term \cite{Affleck:1991tk, Calabrese:2009qy}.  The same $\log(g)$ could ostensibly appear in the generalized entropy if there exists a non-empty intersection between a non-trivial interface between patches in dS$_2^n$ and $R\cup I$.  However, since the $g$-function is not extensive in the size of the region, neglecting its effects will not change the results of our analysis in any meaningful way.

\section{Islands in JT multiverses}
\label{Sec:islands}

In this section, we use the island formula \eqref{eq:islands} to compute the Von Neumann entropy associated to a spacelike interval $R$ in the JT multiverses coupled to a CFT described above.
In particular, we consider regions $R$ that are confined to a single patch, and we look for islands $I$ that are supported outside of $R$'s patch.
We first consider the case of $\mrm{dS}_2^n$, followed by the case where we include flat and negatively curved bubbles.

\subsection{Extended dS$_{2}$}

Consider an $n$-fold extension of $\mrm{dS}_2$ with a line element and dilaton given by \Eqs{eq:dS2-line-element}{eq:dS2-dilaton} respectively, and where the coordinate $\varphi$ runs from $-n \pi$ to $n \pi$.
Let $R$ be a spacelike interval with endpoints $(\sigma_R, \varphi_R)$ and $(\sigma_R, -\varphi_R)$,\footnote{More accurately, the endpoints of $R$ define a causal diamond to which the entropy of $R$ is associated.} where we take $0 < \varphi_R \leq \sigma_R$ so that $R$ is contained within a single expanding patch, as shown in \Fig{fig:dS2n-penrose}. For this configuration, let us compute the von Neumann entropy of the reduced state on $R$, per the island formula.
We must therefore look for extrema of the generalized entropy $S_\mrm{gen}(R\cup I)$ with respect to the inclusion of island regions, $I$, and identify the extremum that gives the smallest generalized entropy.

One extremum is of course the trivial island, $I = \varnothing$.
In this case, the entropy of $R$ reduces to
\begin{equation} \label{eq:trivial-island}
    S_\mrm{gen}(R) \equiv S_\mrm{CFT}(R) = \frac{c}{3}\log \left[ {\frac{2n\sin(\varphi_R/n)}{\epsilon_{\mrm{uv}}\cos \sigma_R}} \right],
\end{equation}
where we have used \Eq{eq:2d-EE-no-island-3} with $\omega = \cos \sigma$, and here and henceforth we drop the non-universal constant $s_0$.
Following \cite{Hartman:2020khs}, we neglect the (gravitational) area term contribution to $S_\mrm{gen}(R)$ coming from the boundary of $R$ because we will choose the latter to lie near $\mathcal{I}^+$ where $\phi \rightarrow +\infty$, which is our proxy for a non-gravitating region in any parametric regime.
Including this contribution would just shift $S_\mrm{gen}(R)$ by $(\phi + \phi_0)/4G_N$ evaluated at the endpoints of $R$. For any nontrivial island, $S_\mrm{gen}(R \cup I)$ would shift by the same amount, therefore an area term due to $\partial R$ would not affect the competition among extrema.

Motivated by the results of \cite{Hartman:2020khs}, next we search for a nontrivial island contained in the causal complement of $R$, whose endpoints are $(\sigma_I,-\varphi_I)$ and $(\sigma_I, \varphi_I)$; see \Fig{fig:dS2n-penrose}.
Since the CFT is in a pure vacuum state, we have that $S_\mrm{gen}(R\cup I)$ is equal to the generalized entropy evaluated for the complement, $S_\mrm{gen}( (R \cup I)^c )$, where $(R \cup I)^c$ denotes the complement of $R \cup I$ on any Cauchy slice that contains $R \cup I$.
$(R \cup I)^c$ is therefore a symmetric pair of intervals whose endpoints are $(\pm \varphi_I, \sigma_I)$ and $(\pm \varphi_R, \sigma_R)$, respectively. In the operator product expansion (OPE) limit, the disconnected components of $(R \cup I)^c$ are each small and spaced far apart, and so, the reduced state approximately factorizes across them. Thus, in the OPE limit, $S_\mrm{gen}((R \cup I)^c)$ is determined by the sum of the entropies of its two constituent intervals. Using \Eq{eq:2d-EE-no-island-3}, we get
\begin{equation}
S_{\text{gen}}((R\cup I)^c)=\frac{c}{3}\log\left[\frac{2n^2\left(\cos(\frac{\sigma_I-\sigma_R}{n})-\cos(\frac{\varphi_I-\varphi_R}{n})\right)}{\epsilon_{\mrm{rg}}\epsilon_{\mrm{uv}}\cos \sigma_I \cos \sigma_R}\right]+2\phi_r\frac{\cos \varphi_I}{\cos \sigma_I}+2\phi_0. \label{SgendS2n}
\end{equation}
Note that we again omit any area term contribution from $R$, but we include the area term  due to $\partial I$.
The latter has also the effect of renormalizing $\epsilon_{\mrm{uv}}$, as discussed in \Sec{sec:CFTentropy}.
The OPE limit approximation is checked in App.~\ref{subsec:two-interval}. Here and henceforth we set $4G_N = 1$.

In order for $I$ to be an entanglement island, the boundary of $I$ must be a quantum extremal surface. In other words, $S_\mrm{gen}(R \cup I)$ (or equivalently, $S_\mrm{gen}((R \cup I)^c)$) must be stationary with respect to variations of the endpoint coordinates $\sigma_I$ and $\varphi_I$.
The system of equations
\begin{equation}
    \begin{aligned}
    \frac{\partial}{\partial \sigma_I} S_{\text{gen}}((R\cup I)^c) &= 0 \\
    \frac{\partial}{\partial \varphi_I} S_{\text{gen}}((R\cup I)^c) &= 0
    \end{aligned}
    \label{eq:extremality-conditions}
\end{equation}
has no general closed-form solution that we could discern, but it can be solved in the limits $\phi_r \gg c$ and $\phi_r \ll c$, as well as numerically in other parametric regimes.
In all cases, we find a critical point, $(\sigma_{I*}, \varphi_{I*})$, located in the upper left corner of the crunching patch that is adjacent to $R$'s patch, as illustrated in \Fig{fig:dS2n-penrose}.
We remark that this critical point is actually a maximum with respect to variations of both $\sigma_I$ and $\varphi_I$, but evaluation of the Hessian reveals that this point is still a saddle of $S_{\text{gen}}((R\cup I)^c)$ as a function of $\sigma_I$ and $\varphi_I$.\footnote{Using the local hyperbolic coordinates $X$ and $T$ introduced in Eq.~(7.5) of \Ref{Hartman:2020khs} instead, it is possible to show that the saddle that we found is a maximum in $T$ and a minimum in $X$. The critical point that we identify here coincides with that found in \Ref{Hartman:2020khs} when we set $n=1$. Further note that the result of App.~B of \Ref{Hartman:2020khs} only guarantees that the critical point of $S_\mrm{gen}$ is a timelike maximum and makes no statement about the spacelike direction.}
We can then evaluate the generalized entropy \Eq{SgendS2n} at this critical point to obtain $S_\mrm{island}(R)$, which we denote as such to distinguish it from the (non-extremized) ansatz \eqref{SgendS2n}.

Having in mind that $\sigma_R$, $\varphi_R$, $\sigma_I$, and $\varphi_I$ are all close to the corners of their respective patches, let us write
\begin{equation} \label{eq:approx_Sgen_start}
\begin{aligned}
    \sigma_R &= \frac{\pi}{2} - \delta \sigma_R \qquad  \,\sigma_I = \frac{\pi}{2} - \delta \sigma_I \\
    \varphi_R &= \frac{\pi}{2} - \delta \varphi_R \qquad \varphi_I = \frac{\pi}{2} + \delta \varphi_I,
\end{aligned}
\end{equation}
where $\delta \sigma_R$, $\delta \varphi_R$, $\delta \sigma_I$, and $\delta \varphi_I$ are all positive and small.
Making these substitutions in \Eq{SgendS2n}, we get
\begin{equation}
    S_{\text{gen}}((R\cup I)^c) \approx \frac{c}{3} \log\left[\frac{2n^2\left(\cos(\frac{\delta \sigma_I-\delta \sigma_R}{n})-\cos(\frac{\delta \varphi_I+\delta \varphi_R}{n})\right)}{\epsilon_{\mrm{rg}}\epsilon_{\mrm{uv}} \delta \sigma_I \delta \sigma_R}\right] - 2\phi_r\frac{\delta \varphi_I}{\delta \sigma_I}+2\phi_0.
\end{equation}
Next, let us also assume that the sum $\delta \varphi_I + \delta \varphi_R$ and the difference $\delta \sigma_I - \delta \sigma_R$ are small, giving
\begin{equation} \label{eq:approx_Sgen}
    S_{\text{gen}}((R\cup I)^c) \approx \frac{c}{3} \log\left[\frac{ (\delta \varphi_I + \delta \varphi_R)^2 - (\delta\sigma_I - \delta\sigma_R)^2 }{\epsilon_{\mrm{rg}}\epsilon_{\mrm{uv}} \delta \sigma_I \delta \sigma_R}\right] - 2\phi_r\frac{\delta \varphi_I}{\delta \sigma_I}+2\phi_0.
\end{equation}
Notice that the generalized entropy is independent of $n$ to leading order.
Let us further assume that $\delta \sigma_I \gg \delta \sigma_R$, which we can justify later.
With that assumption, so that $(\delta \sigma_I - \delta \sigma_R) \approx \delta \sigma_I$ in the numerator above, the system of equations $\partial_{\delta \sigma_I} S_\mrm{gen} = 0$, $\partial_{\delta \varphi_I} S_\mrm{gen} = 0$ has a very simple solution:
\begin{equation} \label{eq:approx-crit}
    \delta \sigma_I = \frac{6\phi_r}{c} \delta \varphi_R,\qquad \delta \varphi_I = \sqrt{1+\frac{36\phi_r^2}{c^2}}\delta \varphi_R.
\end{equation}
Note that $\delta \varphi_I > \delta \sigma_I$, and so the endpoint of $I$ is in the crunching patch, as we initially required.
Plugging this solution back into \Eq{eq:approx_Sgen}, we get
\begin{equation}
    S_{\text{island}}(R) \approx \frac{c}{3} \log \left[ \frac{c}{3\phi_r \epsilon_\mrm{rg}\epsilon_\mrm{uv}} \left( 1 + \sqrt{1+\frac{36\phi_r^2}{c^2}}  \right) \frac{\delta \varphi_R}{\delta \sigma_R} \right] - \frac{c}{3}\sqrt{1+\frac{36\phi_r^2}{c^2}} + 2\phi_0.
\end{equation}

Now let us consider two separate parametric limits and choose the endpoint of $R$ accordingly.
First, suppose that $\phi_r \gg c$.
In this case, choose the endpoint of $R$ such that $\delta \sigma_R = \delta \varphi_R/N$, where $N$ is at least $O((\phi_r/c)^0)$.
In other words, we suppose that as we drag the right endpoint of $R$ toward the upper right corner of the expanding patch, we keep the ratio $\delta \sigma_R / \delta \varphi_R$ fixed.
It follows that $\delta \varphi_R > \delta \sigma_R$, so that the endpoint of $R$ is indeed in the expanding patch, and in this regime where $\phi_r \gg c$, the assumption $\delta \sigma_I \gg \delta \sigma_R$ is justified.
The endpoints of $R$ are also parametrically close to $\mathcal{I}^+$.
Making this choice and dropping subdominant terms, we arrive at
\begin{equation} \label{eq:Sisland-hierarchy1}
    S_{\text{island}}(R) \approx \frac{c}{3} \log \left[ \frac{2N}{ \epsilon_\mrm{rg}\epsilon_\mrm{uv}} \right] - 2\phi_r  + 2\phi_0.
\end{equation}
If we further drop the logarithmic correction, we have that $S_\mrm{island}(R) \approx 2(\phi_0 - \phi_r)$, which is just twice the value of the dilaton evaluated at the boundary of the crunching patch.
Either way, $S_\mrm{island}(R)$ is approximately constant, while $S_\mrm{CFT}(R)$ diverges as the endpoints of $R$ approach $(\pi/2, \pm \pi/2)$.
There is therefore a ``Page transition'' beyond which the nontrivial island entropy is the smaller extremum.

Let us compare $S_\mrm{island}(R)$ to $S_\mrm{CFT}(R)$ to determine the location of the Page transition.
Plugging \Eq{eq:approx_Sgen_start} into \Eq{eq:trivial-island}, we get
\begin{equation}
    S_\mrm{CFT}(R) \approx \frac{c}{3} \log \left[ \frac{f_n}{\epsilon_\mrm{uv} \delta \sigma_R} \right],
\end{equation}
where $f_n = 2n \sin(\pi/2n)$.
Equating $S_\mrm{CFT}(R)$ and $S_\mrm{island}(R)$, we find that the Page transition occurs at
\begin{equation} \label{eq:Page-approx}
    \delta \sigma_R^{\rm Page} = \frac{f_n \epsilon_\mrm{rg}}{2N} e^{-\frac{6}{c}(\phi_0 - \phi_r)}, \qquad \delta \varphi_R^{\rm Page} = \frac{f_n \epsilon_\mrm{rg}}{2} e^{-\frac{6}{c}(\phi_0 - \phi_r)}.
\end{equation}
We can also read off the mild dependence of the Page transition on $n$.
Because $f_n$ monotonically increases up to $\pi$ as $n \rightarrow +\infty$, we see that the size of $R$ at which the Page transition occurs correspondingly monotonically decreases to a finite size.

If we instead suppose that $\phi_r \ll c$, choose the endpoints of $R$ such that $\delta \sigma_R = (6\phi_r/Nc) \delta \varphi_R$ for the same consistency reasons as above, where $N$ is at least $O((c/\phi_r)^0)$.
It then follows that
\begin{equation} \label{eq:approx_Sgen_end}
    S_{\text{island}}(R) \approx \frac{c}{3}\left( \log \left[ \frac{N c^2}{9 \phi_r^2 \epsilon_\mrm{rg}\epsilon_\mrm{uv}} \right]-1\right) + 2\phi_0,
\end{equation}
which again remains constant while $S_\mrm{CFT}(R)$ diverges.
In this case, the Page transition occurs at
\begin{equation} \label{eq:Page-approx-2}
    \delta \sigma_R^{\rm Page} = \frac{f_n \epsilon_\mrm{rg}}{4N} \left(\frac{6 \phi_r}{c}\right)^2e^{-\frac{6}{c}(\phi_0-1)}, \qquad  \delta \varphi_R^{\rm Page} = \frac{f_n \epsilon_\mrm{rg}}{4} \left(\frac{6 \phi_r}{c}\right)e^{-\frac{6}{c}(\phi_0-1)}.
\end{equation}

\begin{figure}[t]
\centering
\includegraphics[width=0.7\textwidth]{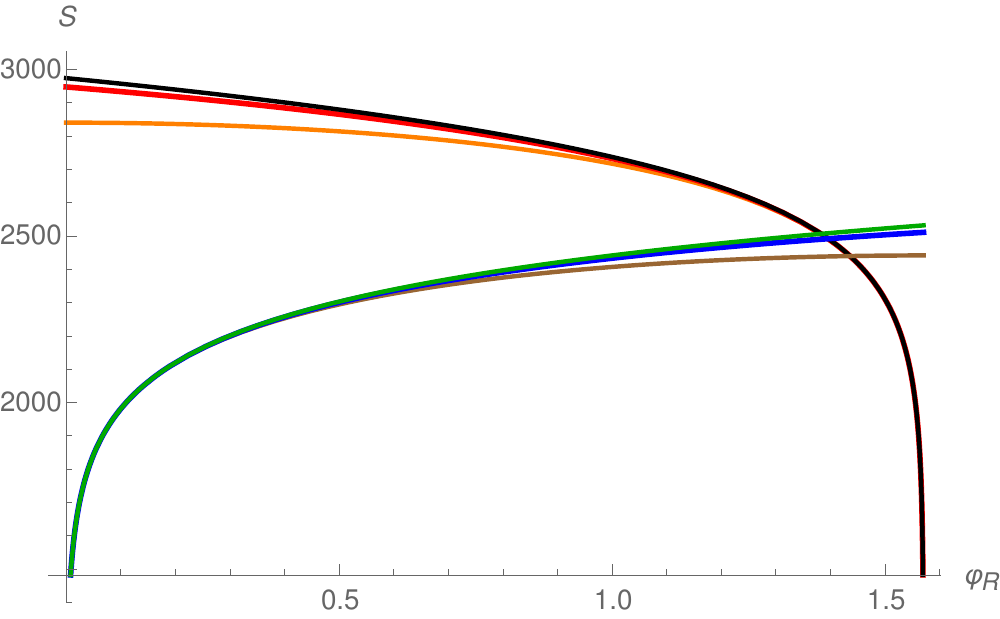}
\caption{$S_\mrm{CFT}(R)$ (brown, blue, dark green) versus $S_{\rm island}(R)$ (orange, red, black) in dS$^n_2$ for $n=1,\;2,\;+\infty$, respectively, with $\sigma_R =\pi/2-10^{-5}$ held fixed.  The size of $R$ beyond which the island contribution to generalized entropy becomes dominant slightly decreases as $n$ increases. Because $\sigma_R$ is held fixed in this plot, taking $\varphi_R$ all the way to $\pi/2$ moves the endpoint of $R$ outside of the expanding patch. In this limit, the endpoint of $I$ also moves outside of the crunching patch and $R \cup I$ tends to a full Cauchy slice on which the state is pure, resulting in vanishing entropy. The parameter values used for this plot are $c = 600$, $\phi_r =10$, $\phi_0 =0$, $\epsilon_\mrm{uv} = 1$, $\epsilon_\mrm{rg} =1$.}
\label{FigPagedS2n}
\end{figure}

\begin{figure}[t]
\centering
\includegraphics[width=0.8\textwidth]{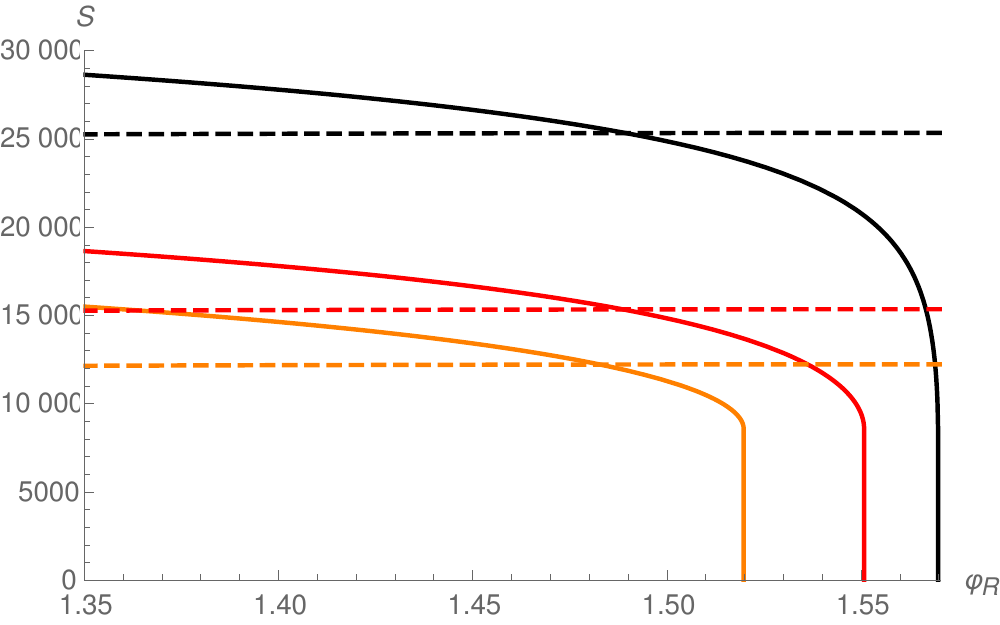}
\caption{$S_{\rm island}(R)$ (orange, red, black) versus $S_\mrm{CFT}(R)$ (dashed, same color scheme) in dS$_2$ (i.e. $n=1$) for $\sigma_R=\pi/2-0.001$, $\pi/2-0.02$, and $\pi/2-0.051$, respectively. As we decrease $\sigma_R$, the size of $R$ at which the Page transition occurs also decreases. The parameter values used for this plot are $c = 10000$, $\phi_r =100$, $\phi_0 =0$, $\epsilon_\mrm{uv} = 1$, $\epsilon_\mrm{rg} =1$. These exaggerated choices of parameters were made to clearly illustrate the shifts in entropy.}
\label{Figtime}
\end{figure}

For other parametric regimes, we must turn to numerics; see, for example, Figs.~\ref{FigPagedS2n} and \ref{Figtime}.\footnote{A Mathematica notebook that reproduces the plots in this manuscript is included as supplementary material.}
In all cases, however, we see the same basic physics at play: $S_\mrm{CFT}(R)$ gives the lesser entropy for small $R$, but there is a Page transition after which $S_\mrm{island}(R)$ is the lesser entropy for sufficiently large $R$.
The crossover point monotonically decreases as a function of $n$, and in numerical analyses, we can study the limiting behavior by taking the $n \rightarrow +\infty$ limit of \Eq{SgendS2n}, which gives
\begin{equation} \label{eq:dS-inf-n-island}
    \lim_{n \rightarrow \infty} S_{\text{gen}}((R\cup I)^c) = \frac{c}{3}\log\left[\frac{(\varphi_I-\varphi_R)^2-(\sigma_I-\sigma_R)^2}{\epsilon_{\mrm{rg}}\epsilon_{\mrm{uv}}\cos \sigma_I \cos \sigma_R}\right]+2\phi_r\frac{\cos \varphi_I}{\cos \sigma_I}+2\phi_0.
\end{equation}

\Fig{FigPagedS2n} Shows the competition between $S_\mrm{CFT}(R)$ and $S_\mrm{island}(R)$ for different $n$ as $\varphi_R$ varies with $\sigma_R$ held fixed.
These curves reproduce the same qualitative features that followed from the approximate analysis above.
We can also examine the competition between $S_\mrm{CFT}(R)$ and $S_\mrm{island}(R)$ as $\sigma_R$ is varied, as shown in \Fig{Figtime} for $n=1$. 
A Page transition still occurs as $\sigma_R$ is decreased, and the value of $\varphi_R$ at which the transition occurs also decreases.
We also find that moving the subregion $R$  back in time pushes the island forward in time toward ${\cal I}^+$. Below a limiting value $\sigma_R^\star$, the island is formally pushed beyond ${\cal I}^+$, outside the allowed range of the parameters of the crunching patch. A similar behavior persists for all values of $n$.

While we have found an island that extremizes $S_\mrm{gen}(R \cup I)$, one should ask whether there are other island configurations consisting of multiple disjoint components that give smaller values of $S_\mrm{gen}(R \cup I)$.
Heuristically, such islands are disfavored by the island formula.
One would expect that a single large island, such that $R \cup I$ covers as much of a Cauchy surface as possible, would be more efficient at purifying the state of $R$ compared to several smaller disjoint components, thus lowering the CFT entropy cost.
Moreover, the area of the boundary of every disjoint piece of an island contributes to the total generalized entropy.
Therefore (at least when $\phi_0 \gg \phi_r$) the geometric cost to form an island is larger for a greater number of disconnected components.
While we cannot prove that the single large island is the minimal extremum, we were able to verify that the extrema for which $I$ consists of two disconnected components result in a larger generalized entropy for a theory of $c\gg 1$ free Dirac fermions.
The details of our numerical analysis are elaborated in App.~\ref{subsec:three-interval}.
In particular, a plot of $S_\mrm{gen}(R \cup I)$ for these non-minimal extrema $I$ as a function of the size of $R$ is shown in \Fig{fig:2islands}.
This constitutes evidence that the single large island is indeed likely the minimal extremum.

We can also consider the case where $R$ is in a crunching patch.
However, if we look for an island whose endpoints lie in the surrounding expanding patches, we find that the extremality conditions \eqref{eq:extremality-conditions} cannot be satisfied.\footnote{This is consistent with the fact that the necessary conditions for island formation presented in \Ref{Hartman:2020khs} are not satisfied in the expanding patch.}
In other words, there are no quantum extremal surfaces% (or, more accurately, quantum extremal ``points'')
, and so no islands of this type form.

\subsection{Extended dS$_{2}$ with bubbles}\label{sec:withbubbles}

\begin{figure}[t]
\includegraphics[width=\textwidth]{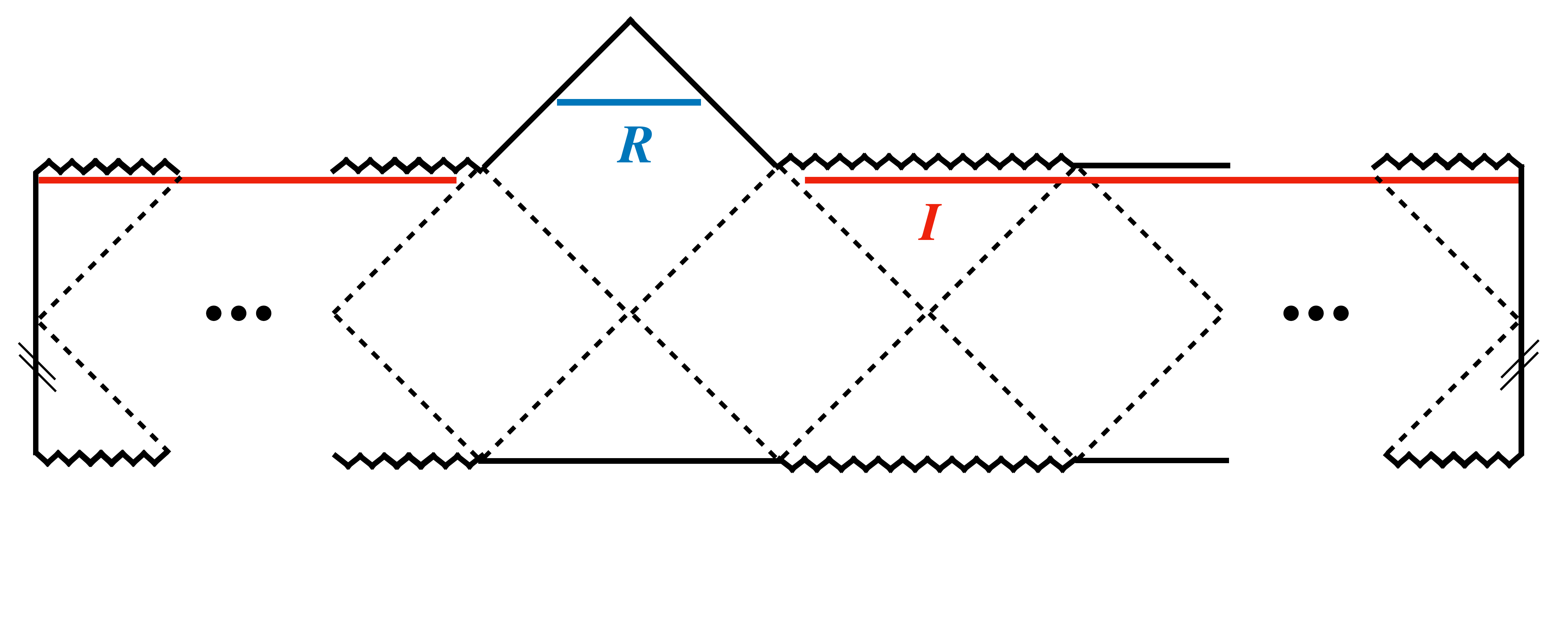}
\vspace{-17mm}
\caption{A single-component island $I$ in an extended JT multiverse for $R$ in a flat bubble.}\label{1isl}
\end{figure}

We now consider an $n$-fold extension of $\mrm{dS}_2$ where the expanding patch centred about $\varphi = 0$ has been replaced with a flat bubble with the line element \Eq{eq:mink2-line-element} and on which the dilaton is given by \Eq{eq:mink2-dilaton}; see \Fig{1isl}.
Let $R$ have endpoints $(\sigma_R, \pm \varphi_R)$ contained within this flat bubble; we will examine the entropy of $R$ as its size increases while keeping its endpoints close to $\mathcal{I}^+$ (i.e. $\varphi_R + \sigma_R \approx \pi$).

We again compute $S(\rho_R)$ using the island formula.
In this case, the trivial island gives
\begin{equation}
S_{\text{gen}}(R) \equiv S_\mrm{CFT}(R) =\frac{c}{3}\log\left[ \frac{2n\sin(\varphi_R/n)}{\epsilon_{\mrm{uv}} \cos(\tfrac{1}{2}(\sigma_R-\varphi_R))\cos(\tfrac{1}{2}(\sigma_R+\varphi_R))} \right],
\label{SnislBB}
\end{equation}
where we used \Eq{eq:2d-EE-no-island-3} with the flat Weyl factors for our chosen coordinates.
Next, we look for an island with endpoints $(\sigma_I, \pm \varphi_I)$ that lie in the crunching patches that are adjacent to the flat bubble, as depicted in \Fig{1isl}.
The generalized entropy for such an island is
\begin{equation}
S_{\text{gen}}((R\cup I)^c)=\frac{c}{3}\log\left[\frac{2n^2\left(\cos(\frac{\sigma_I-\sigma_R}{n})-\cos(\frac{\varphi_I-\varphi_R}{n})\right)}{\epsilon_{\mrm{rg}}\epsilon_{\mrm{uv}}\cos \sigma_I \cos(\tfrac{1}{2}(\sigma_R-\varphi_R))\cos(\tfrac{1}{2}(\sigma_R+\varphi_R))}\right]\!+2\phi_r\frac{\cos \varphi_I}{\cos \sigma_I}+2\phi_0
\label{Sgenbb}
\end{equation}
where we again compute the entropy of the complement and have invoked an OPE limit approximation.

As before, we can explicitly solve the extremality conditions \eqref{eq:extremality-conditions} in the limits $\phi_r \gg c$ and $\phi_r \ll c$, as well as numerically in other regimes.
In our current configuration, we set
\begin{equation}\label{eq:Ansatz}
\begin{aligned}
    \sigma_R &= \frac{\pi}{2} + \delta \sigma_R, \qquad  \,\sigma_I = \frac{\pi}{2} - \delta \sigma_I, \\
    \varphi_R &= \frac{\pi}{2} - \delta \varphi_R, \qquad \varphi_I = \frac{\pi}{2} + \delta \varphi_I,
\end{aligned}
\end{equation}
so that the endpoints of $R$ and $I$ are near $\mathcal{I}^+$.
With these definitions, \Eq{Sgenbb} approximately reduces to
\begin{equation} \label{eq:Sgenbb-approx}
    S_{\text{gen}}((R\cup I)^c) \approx \frac{c}{3}\log \left[ \frac{(\delta\varphi_I + \delta \varphi_R)^2 - (\delta \sigma_I + \delta \sigma_R)^2}{\epsilon_\mrm{uv} \epsilon_\mrm{rg} \delta \sigma_I (\delta\varphi_R - \delta \sigma_R)} \right] - 2 \phi_r \frac{\delta \varphi_I}{\delta \sigma_I} + 2 \phi_0.
\end{equation}
Again assuming that $\delta \sigma_I \gg \delta \sigma_R$, upon extremizing \Eq{eq:Sgenbb-approx} with respect to $\delta \sigma_I$ and $\delta \varphi_I$, we find the same critical point as \Eq{eq:approx-crit}.
This gives
\begin{equation}
    S_{\text{island}}(R) \approx \frac{c}{3} \log \left[ \frac{c}{3\phi_r \epsilon_\mrm{rg}\epsilon_\mrm{uv}} \left( 1 + \sqrt{1+\frac{36\phi_r^2}{c^2}}  \right) \frac{\delta \varphi_R}{\delta \varphi_R - \delta \sigma_R} \right] - \frac{c}{3}\sqrt{1+\frac{36\phi_r^2}{c^2}} + 2\phi_0
\end{equation}
for the generalized entropy corresponding to $R\cup I$.
Similarly, the no-island entropy is
\begin{equation}
    S_\mrm{CFT}(R) \approx \frac{c}{3}\log \left[ \frac{f_n}{ \epsilon_\mrm{uv} (\delta \varphi_R-\delta \sigma_R)} \right] \approx \frac{c}{3}\log \left[ \frac{f_n}{ \epsilon_\mrm{uv} \delta \varphi_R} \right],
\end{equation}
where the second step follows because we will always choose $\delta \sigma_R$ to be much less than $\delta \varphi_R$.

In the limit where $\phi_r \gg c$, we again choose $\delta \sigma_R = \delta \varphi_R / N$.
This gives
\begin{equation} \label{eq:Sisland-flat1}
    S_\mrm{island}(R) \approx \frac{c}{3} \log\left[ \frac{2}{(1-N^{-1}) \epsilon_\mrm{uv} \epsilon_\mrm{rg}}  \right] - 2 \phi_r + 2 \phi_0,
\end{equation}
and results in a Page transition at the same location as in \Eq{eq:Page-approx}.
In the limit where $\phi_r \ll c$, we choose $\delta \sigma_R = (6\phi_r/Nc)\delta \varphi_R$ as in the previous section, which gives
\begin{equation} \label{eq:Sisland-flat2}
    S_\mrm{island}(R) \approx \frac{c}{3}\left(\log\left[ \frac{2 c}{3\phi_r\epsilon_\mrm{uv}\epsilon_\mrm{rg}} \right] - 1\right) + 2\phi_0.
\end{equation}
The Page transition in this case happens as in \Eq{eq:Page-approx-2}.

\begin{figure}[t]
    \centering
    \includegraphics{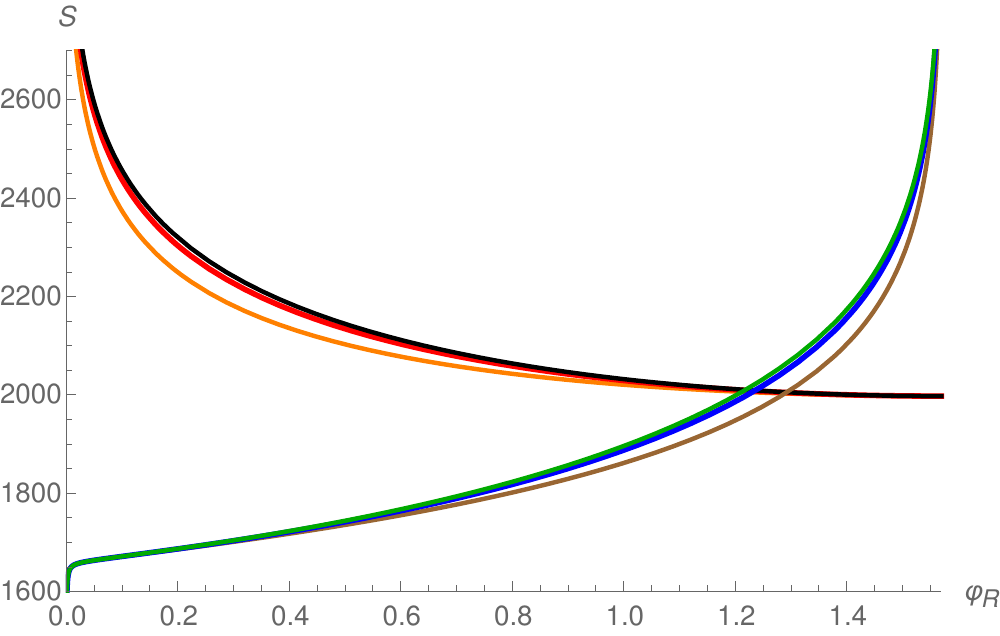}
    \caption{$S(\rho_R)$ evaluated for no islands (brown, blue, dark green ascending lines) and a single large island for $n=1, 2, +\infty$ (orange, red, black curves, respectively, from bottom to top). For a given $\varphi_R$, we set $\sigma_R = -(2/\pi)(\pi/2-10^{-3})\varphi_R + \pi-10^{-3}$ so that $R$ grows large while its endpoints remain near $\mathcal{I}^+$. Other parameter values are $c=600$, $\phi_r = 10$, $\phi_0 = 0$, $\epsilon_{\mrm{uv}}=1$, $\epsilon_{\mrm{rg}}=1$.}
    \label{fig:flat-bubble-Page}
\end{figure}

A plot of $S_\mrm{CFT}(R)$ and $S_\mrm{island}(R)$ outside of the regimes discussed above for different values of $n$ is shown in \Fig{fig:flat-bubble-Page}.
The $n \rightarrow +\infty$ limit is computed in the same way as \Eq{eq:dS-inf-n-island}, but with $\cos \sigma_R$ replaced with the Weyl factor for flat spacetime, $\tfrac{1}{2}(\cos \sigma_R + \cos \varphi_R)$.
The endpoints of $R$ are initially chosen such that $\sigma_R > \pi/2$ and are dragged toward $\sigma_R = \varphi_R = \pi/2$ to increase the size of $R$.
For these parameter choices, the Page transition is visually very clear.

As in our analysis of $\mrm{dS}_2^n$, we can also consider the case where $R$ is in a flat bubble that is surrounded by two expanding patches, e.g., a flat bubble that replaces the crunching patch centred about $\varphi = \pi$.
Once again, we find that the extremality conditions \eqref{eq:extremality-conditions} cannot be satisfied for an island whose endpoints are in the adjacent expanding patches.

\begin{figure}[t]
    \centering
    \includegraphics{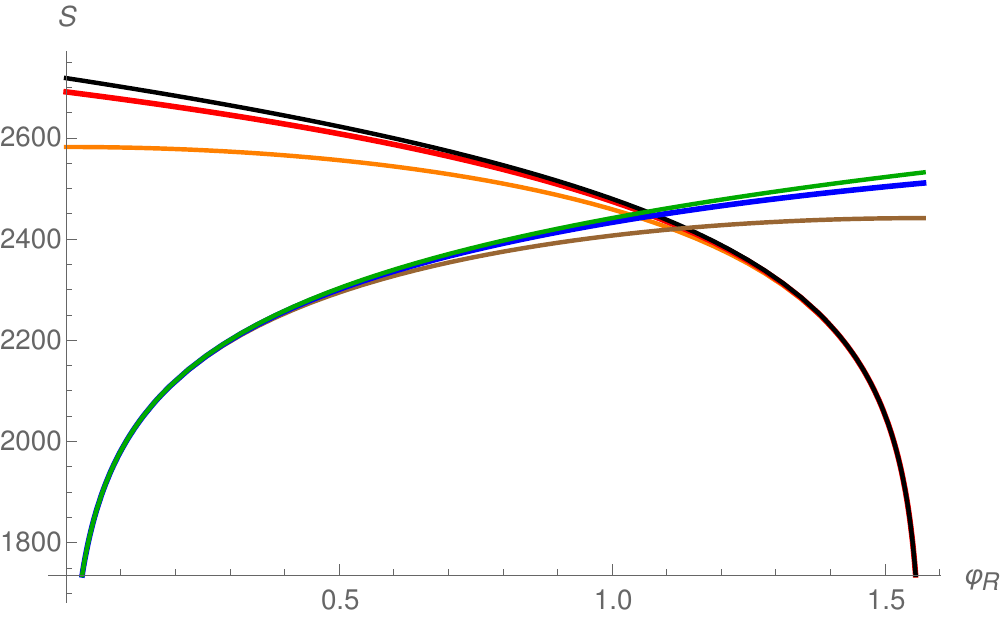}
    \caption{$S(\rho_R)$ evaluated for no islands (brown, blue, dark green ascending lines) with $R$ located in an expanding patch and a single large island for $n=1, 2, +\infty$ (orange, red, black curves, respectively, from bottom to top) whose endpoints lie in adjacent $\mrm{AdS}_2$ bubbles. $\sigma_R$ is fixed to $\pi/2 - 10^{-5}$. Other parameter values are $c=600$, $\phi_r = 10$, $\phi_0 = 0$, $\epsilon_{\mrm{uv}}=1$, $\epsilon_{\mrm{rg}}=1$.}
    \label{fig:AdS-bubble-Page}
\end{figure}

Finally, we note that islands continue to develop when the crunching $\mrm{dS}_2$ patches are replaced with bubbles of $\mrm{AdS}_2$.
To see this, we return to the case where $R$ has endpoints $(\sigma_R, \pm \varphi_R)$ that lie in a central $\mrm{dS}_2$ expanding patch, but we now replace the adjacent crunching patches with $\mrm{AdS}_2$ bubbles, in which the line element and dilaton are given by \Eqs{eq:AdS2-line-element}{eq:AdS2-dilaton}, respectively.
For an island with endpoints $(\sigma_I, \pm \varphi_I)$ with $\pi/2 < \varphi_I < 3\pi/2$, we now have
\begin{equation} \label{eq:SgenAdS}
S_{\text{gen}}((R\cup I)^c)=\frac{c}{3}\log\left[\frac{2n^2\left(\cos(\frac{\sigma_I-\sigma_R}{n})-\cos(\frac{\varphi_I-\varphi_R}{n})\right)}{\epsilon_{\mrm{rg}}\epsilon_{\mrm{uv}}\cos(\varphi_I-\pi) \cos \sigma_R}\right]-2\phi_r\frac{\cos \sigma_I}{\cos (\varphi_I-\pi)}+2\phi_0.
\end{equation}
The behavior of the dilaton near the corners of an AdS$_2$ bubble is very different compared to its behavior in dS$_2$ crunching patches.
In the latter case, $\phi\rightarrow-\infty$ near a crunching patch's corners at $\mathcal{I}^+$, whereas $\phi \rightarrow 0$ near the corners of an AdS$_2$ bubble.
This ultimately locates the endpoints of $I$ away from the corners of the AdS$_2$ bubbles surrounding $R$, and so a perturbative expansion like \Eq{eq:Ansatz} is no longer useful.
Nevertheless, we still find nontrivial islands numerically and we see that an island produces the minimal generalized entropy past a critical value of $\varphi_R$, which we show in \Fig{fig:AdS-bubble-Page}.

\subsection{Features of island formation}
\label{subsec:features}

To briefly summarize the last two subsections, we applied the island formula to compute the fine-grained entropy associated to a region $R$ that is confined to either an expanding patch of $\mrm{dS}_2^n$, or a flat bubble embedded in $\mrm{dS}_2^n$.
In both cases, once $R$'s endpoints are close enough to $\mathcal{I}^+$, once $R$ exceeds a certain size, and provided that $R$'s patch is surrounded by crunching patches or $\mrm{AdS}_2$ bubbles on either side, an island forms outside of $R$'s patch that covers most of the external universe.
Exactly where this transition happens depends mildly on $n$ (i.e., the size of the universe), but it monotonically decreases toward a limiting value $\varphi^{\rm Page}_R(n\rightarrow\infty) < \pi/2$.  Moreover, changing position of $R$ in time, we observe that an island may appear only for sufficiently large $\sigma_R$. As $\sigma_R$ approaches $\mathcal{I}^+$, the islands' endpoints move back in time toward a limiting location.

There is a clear interpretation for the island entropy, at least when $\phi_0 \gg \phi_r$ and $\phi_0\gg c$.
Examining Eqs.~\eqref{eq:Sisland-hierarchy1} and \eqref{eq:approx_Sgen_end} or Eqs.~\eqref{eq:Sisland-flat1} and \eqref{eq:Sisland-flat2}, we see that $S_\mrm{island}(R)$ is equal to $2\phi_0$ up to $O(\phi_r)$ or $O(c \log c)$ corrections depending on whether $\phi_r \gg c$ or $\phi_r \ll c$, respectively.
That is, in the regime of parametrically large $\phi_0$, $S_\mrm{island}(R) \approx 2\phi_0$ is the two-dimensional de Sitter horizon entropy.
Therefore, the Page transition in this cosmological setting occurs when $R$ grows so large that its matter entropy would exceed the de Sitter entropy.
If we view a maximum entropy as a bound on Hilbert space dimension, it is tempting to speculate that the appearance of islands is a semiclassical signal that the dimension of the Hilbert space for putative fine-grained degrees of freedom associated with $R$ is bounded by the de Sitter entropy.
Moreover, this bound persists regardless of how much spacetime lies outside of $R$'s patch.

In particular, one might have thought that the island entropy would depend on the pattern of patches and bubbles outside of $R$'s patch, but this is not the case.
The value of $S_\mrm{CFT}(R)$ depends only on the size of $R$ and on $n$ and, to leading order, $S_\mrm{island}(R)$ depends only on the former.
In other words, $S_\mrm{island}(R)$ is the same irrespective of the spacetime that lies beyond $R$'s patch.

The fact that $S_\mrm{island}(R)$ depends only on the endpoints of $R$ has further consequences for when an island forms and gives the minimal extremum of $S_\mrm{gen}(R \cup I)$.
As long as the endpoints of $R$ are sufficiently close to the corners of expanding patches or flat bubbles, an island will form with endpoints lying in the adjacent patches to those containing $R$'s endpoints provided they are crunching patches or $\mrm{AdS}_2$ bubbles.
Examples of such configurations are illustrated in Figs.~\ref{fig:crunching0a} and \ref{fig:crunching0b}.
Under these conditions, $(R \cup I)^c$ is locally identical to the cases that we examined in the previous two subsections, and so a Page transition occurs for sufficiently large $R$.

Much as crunching patches or $\mrm{AdS}_2$ regions are necessary to form islands, we also observed that islands do not form when $R$ is surrounded by expanding patches, or by flat bubbles; the extremality conditions \eqref{eq:extremality-conditions} cannot be satisfied in expanding $\mrm{dS}_2$ patches, or flat bubbles.
One might wonder, then, whether islands whose endpoints lie in the \textit{nearest} crunching or $\mrm{AdS}_2$ regions can form and whether they can give a lower generalized entropy than the absence of islands.
Examples of such configurations are illustrated in Figs.~\ref{fig:crunchinga}, \ref{fig:crunchingb}, and \ref{fig:crunchingc}.
Though these configurations lie outside of the OPE limit, we can compute their associated entropies for a specific choice of CFT. 
Taking the CFT to be a theory of $c\gg 1$ free Dirac fermions, we find that islands do indeed form, in the sense that extrema of $S_\mrm{gen}(R\cup I)$ with nontrivial $I$ exist, but that the configurations shown in Figs.~\ref{fig:crunchinga} and \ref{fig:crunchingb} always result in an entropy that is larger than $S_\mrm{CFT}(R)$.
Therefore, it appears that surrounding $R$ with expanding patches can ``screen'' the rest of the universe from $R$ to a certain extent.
However, this is not a hard and fast rule since, for example, the configuration shown in \Fig{fig:crunchingc} still exhibits a Page transition for sufficiently large $R$.
Computational details and the associated Page curves are elaborated in \App{subsec:two-interval}.

\section{False vacuum inflation in quantum cosmology}
\label{Sec:quantum-cosmology}

In the previous section, we have seen islands appear in the calculation of the Von Neumann entropy associated with a spacelike interval $R$ confined to a patch of the global spacetime in various two-dimensional toy-model multiverses (provided $R$ is taken sufficiently large). The formation of an island suggests that, if we had been working in the framework of semiclassical quantum cosmology, an additional saddle point geometry of the gravitational path integral would have come into play in the calculation of the Von Neumann entropy, and perhaps also of ``observables'' with a sufficiently rich information content.

In this section, we elaborate on this point by examining a similar, but more conventional toy-model multiverse in four dimensions, one that is often associated with the decay of an inflating false vacuum through bubble nucleation.
To be precise, we consider the Hartle-Hawking quantum state for universes that contain a scalar field whose potential possesses false and true vacua.
We will compare the calculation of Von Neumann entropy in the two-dimensional model to the calculation of probabilistic predictions for local cosmological observables using the Hartle-Hawking state in this model, and we will discuss how each model informs the other.
%Crucially, this is a setting that admits a fully quantum treatment of the gravitational state.

In the current setting, different gravitational saddles contribute to the calculation of probabilities depending on the level of detail of the local observation in question.
This leads us to draw an analogy between the appearance of new saddles here, when the observational question is made sufficiently precise, and the appearance of a large island in the two-dimensional model, when $R$ is sufficiently large.
These saddles are saddle point geometries of the Hartle-Hawking wavefunction, and in particular, they involve an enormous coarse-graining over the external (with respect to the local observation) fine-grained multiverse structure.
This further resonates with the fact that the formation of an island is insensitive to almost all of the multiverse structure external to $R$.

Altogether, the comparisons drawn here are meant to exemplify how semiclassical QC appears to incorporate the huge reduction of degrees of freedom suggested by the islands program in cosmology, while retaining some information in terms of a multiplicity of pasts. Our discussion in this section closely follows part of \cite{Hartle:2016tpo} albeit with a somewhat different emphasis.

\subsection{Multiverse Model}

\begin{figure}[t]
    \centering
    \includegraphics[width=0.5\textwidth]{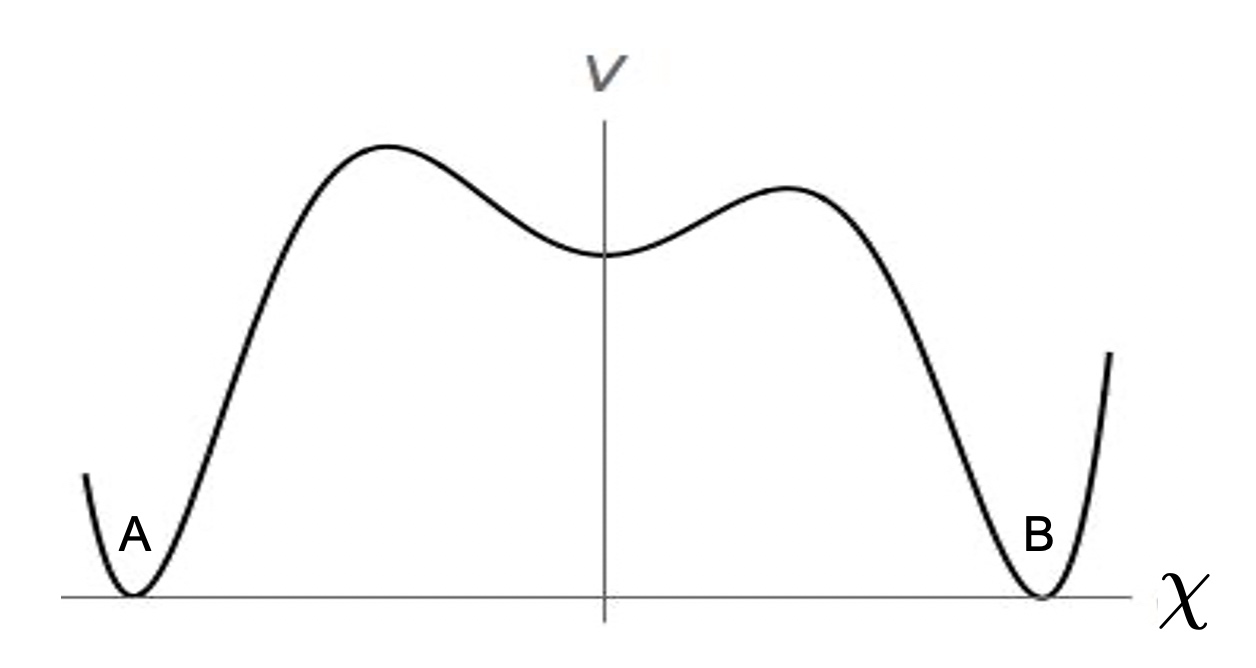}
    \caption{A potential with one false vacuum and two true vacua $A$ and $B$. The false vacuum is assumed to be separated from both true vacua by a barrier followed by a relatively flat patch where the slow roll conditions for inflation hold. The different shapes of the two barriers and of the potential in the two slow roll regimes leading on to the true vacua gives different false vacuum decay rates and different predictions for CMB related observables in universes ending up in either $A$ or $B$.}
    \label{fig:pot}
\end{figure}

We consider four-dimensional Einstein gravity coupled to a single scalar field $\chi$ moving in a positive potential. We take the potential to have a false vacuum $F$ with two quantum decay channels to two vacua $A$ and $B$ where the potential vanishes. \Fig{fig:pot} gives an example. Classically, this theory has an eternally inflating de Sitter solution with an effective cosmological constant given by the value of the potential in the false vacuum. Quantum mechanically, this solution decays through the nucleation of bubbles of true vacuum. The geometry inside these bubbles is that of an open universe which expands in the de Sitter background.

We allow for different decay rates of the false vacuum to $A$ and $B$. We further assume that the potential toward the vacua has flat patches where the slow roll conditions hold so that while the scalar slowly rolls down, the open universes\footnote{Whether the local geometry inside is open remains a matter of debate \cite{Cespedes:2020xpn}.} inside the bubbles undergo a period of inflation before the bubble universe reheats and standard cosmological evolution ensues. Finally, we assume the potential is such that detailed CMB-related observables, say the spectral tilt or the tensor to scalar ratio, enable observers inside one of the bubbles to determine whether they live in $A$ or $B$. 

The quantum mechanical nucleation of bubbles of type $A$ or $B$ in the false vacuum background is thought to give rise to a toy-model multiverse. These bubbles are the analog of the expanding or flat bubbles we patched in, in the two-dimensional toy-model multiverses in the preceding sections. The ``crunching,'' or strong gravity patches in the two-dimensional models correspond to the false vacuum background here. The CMB observables discriminating between $A$ and $B$ are the analog of the different dilaton behaviors in the patches containing $R$ in the two-dimensional models.

\begin{figure}[t]
    \centering
    \includegraphics[width=0.5\textwidth]{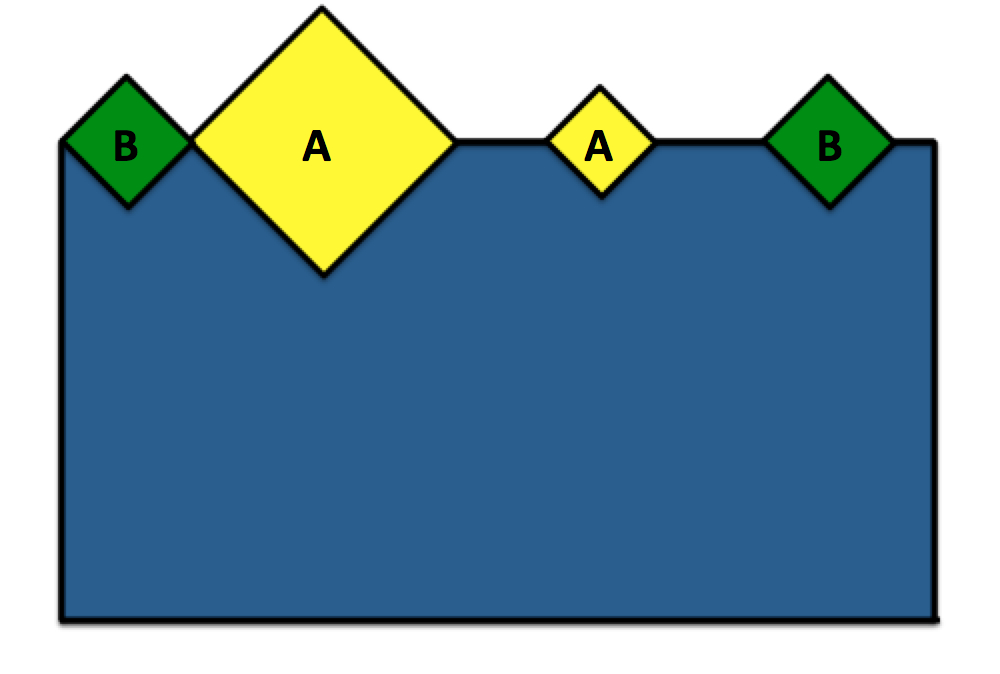}
    \caption{A conformal representation that evokes a fine-grained configuration of (possibly infinitely many) bubble universes in a false vacuum de Sitter background. The false vacuum is indicated in blue, regions inside bubbles of type A are in yellow, and regions inside bubbles of type B are in green. A quantum state of the universe $\Psi$ does not describe one specific such configuration, but an ensemble of possible ones.}
    \label{fig:bubble}
\end{figure}

A particular eternally inflating history consisting of a specific configuration of bubble universes in a false vacuum de Sitter background is illustrated in \Fig{fig:bubble}. In an essentially classical approach to eternal inflation, such a particular fine-grained global configuration is taken as a starting point for the calculation of predictions for local observations. In the absence of a quantum state, these predictions are derived by counting the number of Hubble volumes (or bubbles) in a global configuration where observables take different values. This requires an extraneous notion of typicality in sync with a prescription for regulating infinities because a fine-grained history typically follows an infinite number of bubbles, each of which is itself infinite. This prescription, which specifies a measure, supplements the theory. It consists, e.g., of specifying a spacelike three-surface beyond which one no longer counts instances of observations. It is well known, and hardly surprising, that the resulting predictions are highly regulator-dependent. This is known as the measure problem of eternal inflation. It is essentially a problem of information; the theory is underdetermined and that gives rise to a breakdown of predictivity. 

In the previous section, we have taken the analog of such a particular fine-grained configuration as a starting point for a semiclassical calculation of the Von Neumann entropy of subregions. It is tempting to interpret the formation of an island in that context as a signature that the idea of a normal definite global spacetime may be questionable. Here we take the complementary viewpoint and follow up on this reasoning with a conventional semiclassical quantum cosmology treatment of this false vacuum model. In the next Section, we substantiate the resonances between both analyses.

\subsection{Local predictions from coarse-grained saddle point geometries}

We consider the Hartle-Hawking no-boundary wavefunction (NBWF) in the model above on a closed spacelike three-surface $\Sigma$.
Schematically, we have
\begin{equation*}
\Psi_{HH}=\Psi[h(\vec x),\chi (\vec x), \zeta (\vec x)],
\end{equation*}
where $\zeta$ represents linear scalar perturbations around the background saddles, $h(\vec{x})$ is the induced metric on $\Sigma$ and $\chi$ is the scalar field. 

In the semiclassical approximation, the NBWF is  given by a sum of saddle points, each contributing a term of the form \cite{Hartle:1983ai}
\begin{equation}
\Psi[h,\chi,\zeta] \sim \exp(-I/\hbar) =  \exp\{(-I_R[h,\chi,\zeta] +i S[h,\chi,\zeta])/\hbar\} .
\label{semiclass}
\end{equation}
Here, $I_R[h,\chi,\zeta]$ and $-S[h,\chi,\zeta]$ are the real and imaginary parts of the Euclidean action $I$, evaluated on a saddle point solution of the field equations that matches $(h,\chi,\zeta)$ on its only boundary $\Sigma$ and is otherwise regular. In regions of superspace $\{h,\chi,\zeta \}$ where $S$ varies sufficiently rapidly, the semiclassical wavefunction \eqref{semiclass} describes a family of locally classical Lorentzian cosmologies that are the integral curves of $S$ and have amplitudes to leading order in $\hbar$ that are proportional to $\exp[-I_R(h,\chi,\zeta)]/\hbar]$, which is constant along the integral curve \cite{Hartle:2008ng}.

\begin{figure}[t]
    \centering
    \includegraphics[width=0.9\textwidth]{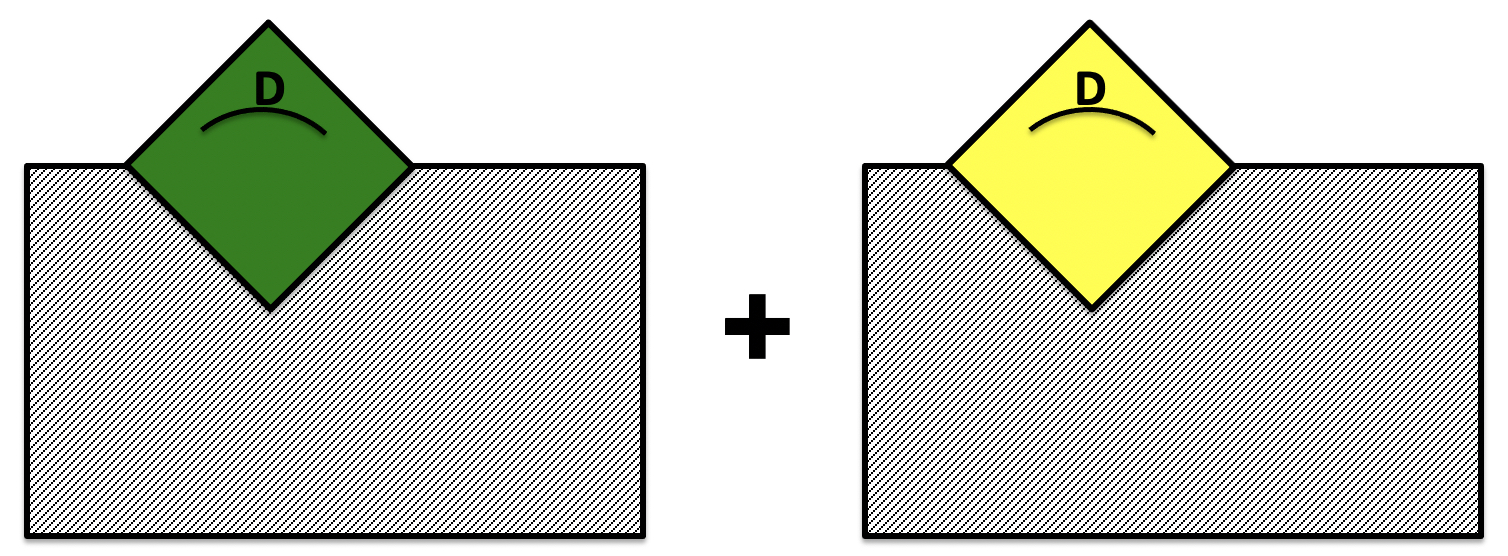}
    \caption{The semiclassical prediction of local observations in the false vacuum eternal inflation model of Fig \ref{fig:pot} involves two distinct saddle point geometries. Each of these corresponds to a coarse-grained history describing a different possible past of a given ``observer'' $D$, in which she evolves either toward the true vacuum $A$ or $B$. The saddle geometries average over any multi-bubble configuration outside---represented by gray---but enter as a superposition of pasts of $D$.}
    \label{fig:saddles}
\end{figure}

In the model we consider, the wavefunction comprises two distinct sets of cosmological backgrounds. First, there is a one-parameter family of saddle points, labeled by the absolute value $\chi_0$ of the scalar field at their ``South Pole'', with $\chi_0$ somewhere on the slow roll slope near one of the true vacua $A$ or $B$. Each of these saddles corresponds to a closed inflationary Friedmann-Lema\^itre-Robertson-Walker (FLRW) background, without eternal inflation, ending up in either $A$ or $B$ \cite{Hartle:2008ng}. Second, there are two isolated saddle geometries describing the expansion of a bubble, either of type $A$ or $B$, embedded in the false vacuum background at $\chi=0$. These saddles are no-boundary versions of the well-known Coleman-De Luccia (CDL) or Hawking-Moss (HM) instantons (which one dominates depends on the shape of the barrier). As regular compact solutions of the Euclidean field equations, these instantons are valid saddle points of the NBWF. In this interpretation, CDL instantons are associated with histories following a single bubble evolving toward a true vacuum that expands in a false vacuum background. The nucleation point of the bubble lies at the throat of the de Sitter background. Indeed the saddle describes the creation of both bubble and background. Crucially, the saddle does not keep track of other bubbles which may or may not be nucleating at various other locations in the false vacuum background. Instead it averages over everything happening outside one bubble \cite{Hartle:2016tpo}. This is illustrated in \Fig{fig:saddles}.

In the language of decoherent histories quantum mechanics, one says that CDL instantons, as NBWF saddle points, correspond to {\it coarse-grained} histories \cite{Hartle:1992as}. Quantum mechanical coarse-graining amounts to some sort of averaging whereby one bundles together detailed histories in coarser grained sets retaining less information. Taking the semiclassical wavefunction at face value, therefore, we see that this includes a huge amount of coarse-graining over possible multi-bubble configurations. Nevertheless, the semiclassical theory repackages some information contained in that putative fine-grained structure in terms of a limited set of distinct coarse-grained saddle geometries. While these saddles are consistent with there being no independent degrees of freedom in those far-flung regions at all, they do not necessarily imply this, although the dynamics of eternal inflation does suggest that at the very least, the wavefunction is very much spread out over a wide range of configurations on the largest scales. 

Decoherent histories quantum cosmology speaks of different descriptions of quantum systems at different levels of coarse-graining. Which level is the appropriate one depends on the correlation of interest \cite{Hartle:1992as}. The crux of previous semiclassical quantum cosmology calculations in models of this kind \cite{Hartle:2016tpo,Hartle:2010dq,Hertog:2013mra} is that the coarse-graining inherent in the semiclassical description is the relevant and appropriate one for the calculation of probabilities for local observations. 

To adopt an intrinsic perspective of a local observer, it is convenient to specify an observational situation in terms of data $D$ treated as part of the quantum system. Consider thus the correlation $p(\cF|\aoi)$ between a feature $\cF$ of the local universe, e.g., a statistical feature of the CMB that we seek to predict, and at least one instance of a set of local data $D$ (denoted by $\aoi$), which we take also to select a surface of homogeneity of a given density. One can think of $D$ as a specific local configuration of matter fields and geometry inside one Hubble volume that has ultimately evolved from the primordial perturbations $\zeta$ and whose probability to occur in any Hubble volume can therefore in principle be computed from the Gaussian fluctuation wavefunctions. All one knows from local observations is that there is at least one instance of $D$. For any kind of realistic data, this probability is of course exceedingly small. 

The condition on $\aoi$ in the correlation $p(\cF|\aoi)$ suppresses the first class of no-boundary saddle points labeled by $\chi_0$, i.e. those not associated with the false vacuum and the nucleation of a bubble of kind $A$ or $B$. The reason is that the surfaces of homogeneity in the former class are relatively small, rendering $p(\aoi) \ll 1$. On the other hand, bubbles have extremely large or even infinite surfaces of homogeneity, rendering the condition on $D$ trivial, i.e. $p(\aoi) \approx 1$ \cite{Hartle:2016tpo}. This selection of bubble saddles by taking $D$ sufficiently precise is the analog of taking $R$ sufficiently large in the island calculations above. Adding more bubbles does not change anything, for the condition on $\Daoi$ obviously remains moot. Assuming the probability of bubble collisions is negligible, a coarse-graining that follows what happens inside one bubble and ignores what goes on outside thus appears to be adequate for evaluating the quantum mechanical probabilities $p(\cF|\aoi)$. If the feature $\cF$ distinguishes between a bubble of type $A$ or $B$, then the relevant coarse-grained set of histories consists of the two remaining saddle point geometries, each following only one bubble (ours) but distinguished by whether this is of type $A$ or $B$. The picture one might have in mind here is one that is familiar from holography---that these two saddle geometries corresponds to two distinct ways of filling in the past (say, the bulk) leading to a given set of data $D$ on a homogeneous boundary surface of given density and subject to a no-boundary condition deep in the interior.

Evidently the relative probabilities for the outcome $\cF_A$ or $\cF_B$ will be specified by the action $I$ of the dominant saddle mediating the decay of the false vacuum toward resp. $A$ or $B$. Thus for the relative probabilities that we, systems characterized by data $D$, observe the physical properties of bubble $A$ or $B$, we get
\begin{equation}
\label{ans}
\frac{p(\cF_A \vert \Daoi)}{p(\cF_B \vert \Daoi)} = e^{2I_R^B - 2I_R^A}
\end{equation}
where $I_R^A$ and $I_R^B$ are the real parts of the Euclidean no-boundary actions of the (CDL or HM) saddle point geometries. This includes a weighting of the false vacuum background because no-boundary saddles describe the nucleation of both a bubble and the background, in contrast with the use of CDL or HM instantons in tunneling transitions where one assumes a pre-existing vacuum state. 

For a broad barrier where the Hawking-Moss saddle gives the dominant decay channel,
the no-boundary weighting in \Eq{ans} is given by
\begin{equation}
-I_R = 24\pi^2 \left(\frac{1}{V(\chi_{max})} -\frac{1}{V(0)} \right) + \frac{24\pi^2}{V(0)} \ ,
\label{HM}
\end{equation}
where $\chi_{max}$ denotes the value of $\chi$ at the maximum of $V$ and we have added and subtracted the false vacuum weighting. The terms inside the brackets are simply equal to the difference in entropy of both de Sitter backgrounds and combine to form the bubble nucleation rate, and the third term represents the no-boundary weighting of the false vacuum background.

For a narrow barrier, there is a CDL saddle that straddles the maximum. CDL instantons are slightly more complicated saddle points in which the field $\chi$ varies from an initial value $\chi_i$ near the false vacuum to a final value $\chi_{f}$ on the other side of the barrier. In the limit $V_{,\chi \chi}/H^2 (\chi_{max}) \rightarrow-4$, the CDL solution tends to HM and so does its action. By contrast, in the limit where the barrier is narrow and sharp and hence $\vert \chi_{f} - \chi_0 \vert \ll 1$, the decay rate implied by the CDL action tends to the well-known thin-wall result, giving for the no-boundary weighting in \Eq{ans},
\begin{equation}
-I_R = \left(\frac{27\pi^2 T^4}{2 V(0)^3} - \frac{24\pi^2}{V(0)}\right)+ \frac{24\pi^2}{V(0)} \ ,
\label{CDLthin}
\end{equation}
where $T$ is the tension of the narrow barrier separating $F$ from the true vacuum, 
\begin{equation}
T = \int_{\chi_i}^{\chi_{f}} d\chi \sqrt{2V(\chi)}.
\end{equation}

The third term in \Eq{HM} and \Eq{CDLthin} is just the de Sitter entropy of the false vacuum. Thus the no-boundary weighting of the saddles has formally a very similar structure to the island entropy we obtained above, given by the dS entropy plus a small correction. In relative probabilities, the weighting of the background cancels, of course, so we end up in \Eq{ans} with a difference between two relatively small corrections to the entropy of the false vacuum. 

To summarize, correlations of the kind $p(\cF|\aoi)$ that capture predictions for local cosmological observations in the Hartle-Hawking state in false vacuum models are specified in terms of a superposition of saddle point geometries, each representing a highly coarse-grained configuration. 

\subsection{Comparative summary of the models}

In \Sec{Sec:JT}, we built toy model multiverses out of solutions of two-dimensional JT gravity theories, and we used the islands prescription to compute the Von Neumann entropies associated to regions $R$.
In this section, we examined a more familiar four-dimensional multiverse model from quantum cosmology, and we explained how to compute probabilities for cosmological observations using saddles of the Hartle-Hawking wavefunction.
We end this section with a short recap of the similarities, differences, and resonances between the two models and the two calculations.

The two models are similar in the sense that they aim to model the physics of an inflationary multiverse with multiple vacua.
Roughly speaking, bubbles of the terminal $A$ and $B$ vacua in the QC model correspond to the weakly-gravitating patches of a JT multiverse where we situated $R$, and the false vacuum $F$ corresponds to strongly-gravitating regions.

When following the islands prescription, the appearance of a nontrivial island generally signals the appearance of a new saddle in an underlying Euclidean gravitational path integral.
While we currently lack a Euclidean description of the JT multiverses, the selection of particular gravitational saddles is clearly exhibited in the calculation of probabilities in the QC model.
Furthermore, this calculation is a true quantum gravitational calculation, albeit in the semiclassical approximation, in the sense that it explicitly tracks the state of the gravitational sector.
The QC calculation is of course not a calculation of Von Neumann entropy. %therefore its saddles and the putative saddles of a Euclidean JT multiverse calculation are somewhat different in nature.
Nevertheless, the selection of saddles in the QC model and the formation of islands in the JT multiverse model are analogous in that both are a consequence of having asked a sufficiently detailed question: Islands form when calculating the fine-grained entropy of regions that nearly span an entire cosmological patch in a JT multiverse, and sufficiently detailed observational conditions select saddles associated with false vacuum decay in the Hartle-Hawking wavefunction.

Finally, both models suggest a huge redundancy in the global picture of an eternally inflating spacetime.
The saddles in the QC model explicitly coarse-grain over scales larger than those of the observational condition in question.
In the JT multiverse models, an island always forms for sufficiently large $R$, independent of the global structure beyond $R$ and despite the fact that $R$ is locally the same for any such structure.
Furthermore, the fine-grained degrees of freedom of $R$ are supposed to encode the island itself, at least at a semiclassical level.
(We will further comment on this point in the Discussion.)
Given that the geometry of the JT multiverses reflect a more traditional view of the spacetime produced by eternal inflation, it is tempting to speculate that the formation of these islands is a semiclassical hint that the global view of an operationally well-defined, eternally inflating spacetime eventually breaks down, as is manifest in the QC model.
In particular, the fragmentation of the global description into a number of disconnected, separate saddle points for each type of bubble in the QC model suggests that something similar will happen in a quantum gravitational analysis of the two-dimensional model, at least when the latter is considered in (perhaps some appropriate generalization of) the Hartle-Hawking state.

\section{Discussion}\label{Sec:conclusion}

We have considered toy multiverse models inspired by false vacuum eternal inflation. We found that in the semiclassical calculation of the Von Neumann entropy associated with a sufficiently large spacelike interval $R$ in two-dimensional models, an island $I$ develops covering most of the rest of the multiverse. This further substantiates the quantum cosmology treatment of models of this kind in which predictions for local cosmological observables are specified by saddle points that discriminate between different pasts of $R$ but otherwise coarse-grain, or ``average,'' over any large-scale multiverse structure outside one bubble.

The two-dimensional multiverse geometries we considered contain bubbles of zero or negative curvature within an analytic extension of $\mrm{dS}_2$ that is a solution of the de Sitter version of JT gravity.
Within these geometries, we used the behavior of the dilaton to label regions of weak gravity and of strong gravity.
Coupling a CFT to these geometries as a matter model, we then calculated the generalized entropy $S_\mrm{gen}(R \cup I)$  associated to a spacelike interval $R$ and additional putative islands $I$. This let us implement the islands program to compute the Von Neumann entropy $S(\rho_R)$: given a spacelike interval $R$ and the generalized entropy  for different island configurations $S_\mrm{gen}(R \cup I)$, we looked for islands $I$ that extremize $S_\mrm{gen}(R \cup I)$, and then we took the minimum of these extrema.
For sufficiently large subregions $R$ in regions of weak gravity, an island forms with endpoints in surrounding regions of strong gravity.
The island covers most of the multiverse to the exterior of $R$, suggesting that the global spacetime does not capture fundamental degrees of freedom independent from those in $R$.

In our analysis of false vacuum eternal inflation, we considered observables localized within bubbles that have exited from eternal inflation where gravity can be said to be relatively weak. The averaging entering in the semiclassical QC setting amounts to a coarse-graining over the regime of eternal inflation surrounding the observer's patch. Stretching the analogy with our two-dimensional toy models as far as we can, we view the ``crunching'' patches, the proxies for a strong gravity regime, as the toy-model analog of the regime of eternal inflation. In establishing this connection, our analysis has implicitly assumed the validity of the islands program in a setting where $R$ and $I$ are part of the same spacetime.\footnote{When $R$ and $I$ are (parts of) disjoint spacetimes associated to a manifestly bipartite Hilbert space, such as in the study of entanglement between disjoint closed universes \cite{Balasubramanian:2020coy,Balasubramanian:2020xqf,Balasubramanian:2021wgd,Fallows:2021sge}, the development of entanglement islands can be thought of as a consequence of monogamy of entanglement. In such situations, one considers entangling a collection of non-gravitating degrees of freedom, $A$, with a disjoint collection of gravitating degrees of freedom, $B$. The latter are entangled with degrees of freedom in the gravitational sector due to gravitational interactions. A consequence of monogamy of entanglement is that the structure of entanglement between $B$ and the gravitational sector cannot be perfectly preserved as one increases the entanglement between $A$ and $B$. Therefore, the Von Neumann entropy of $A$ eventually becomes sensitive to geometric effects in $B$, which is manifested by the formation of entanglement islands.} Regardless, $R$ is always entangled with its complement, and taking $R$ larger thus increases the amount of this entanglement. It is thus plausible that the fine-grained Von Neumann entropy $S(\rho_R)$ becomes sensitive to geometric effects in the form of entanglement islands, and that is what we found.

We now comment on some loose ends and open questions in our two-dimensional multiverse models. In addition to the configurations that we considered, another possibility would have been to locate the subregion $R$ within one static patch of our JT multiverse models, in the spirit of \Ref{Sybesma:2020fxg}. In that work, given a region $R$ in a static patch of dS$_3$, an island develops in the opposite wedge, which leads to a Page transition.
Yet another avenue would be to consider island configurations which are timelike separated from the region $R$, as explored in \Ref{Chen:2020tes} for the simple case of a ${\rm dS}_2$ solution of JT gravity with positive cosmological constant. 
Similarly to the analysis of \Ref{Balasubramanian:2020xqf}, it should also be possible to entangle our multiverse configurations with a disjoint non-gravitating system, with the auxiliary system playing the role of $R$.

While islands and saddles are manifestly nonpertubative objects, an interesting question to ask is whether the perturbative dependence of $I$ on $R$ could be understood using the theory of bulk operator reconstruction.
In the AdS/CFT correspondence, bulk reconstruction gives a prescription for how to represent bulk operators that lie in the entanglement wedge of a given boundary subregion as CFT operators supported on that subregion \cite{Almheiri:2014lwa,Dong:2016eik,Cotler:2017erl,Chen:2019gbt}.
In the case of an AdS black hole coupled to an external reservoir in which an island develops in the black hole interior, bulk operators supported on the island are then represented through this prescription as operators in the reservoir \cite{Penington:2019kki,Almheiri:2019qdq}.
In both cases, the bulk operator is ``represented'' in the sense that both it and its reconstruction's expectation values agree on a restricted set of perturbatively close states known as the code subspace.
In the present cosmological setting, it would be interesting to investigate whether operators supported on $I$ can be represented as operators supported on $R$ for an appropriate code subspace and given a mapping between effective degrees of freedom on $R\cup I$ and fine-grained degrees of freedom on $R$.

A feature of the two-dimensional models that we considered is that they offer precise quantum control over the CFT matter model and its contribution to generalized entropy.
We did not focus on quantum aspects of the spacetime in these models; nevertheless, a Euclidean construction of JT multiverses and an analysis based on the gravitational path integral are interesting avenues for future inquiry.
These are not straightforward tasks, however, as there are subtleties involved in defining a quantum state for the gravitational sector.
Suppose, for example, that we wished to define a Hartle-Hawking-like state for dS${}_2^n$ (see \Fig{fig:dS2n-penrose}) by continuing the manifold into the Euclidean past at the $\sigma=0$ slice.
The resulting manifold possesses a conical excess, and so it cannot be a solution of the JT theory, which has $R = +2$ everywhere.
In principle, one would therefore have to modify the theory at the level of its action so that it could support a conical excess.\footnote{Similarly, as discussed in \Sec{Sec:JT}, additional degrees of freedom are required to support the discontinuities in the dilaton's first derivative in the bubble spacetimes.}
That said, it appears that a sensible gravitational path integral can still be defined, at least for pure dS${}_2^n$.
Ref.~\cite{Cotler:2019nbi} constructs a gravitational path integral that prepares a state at $\mathcal{I}^+$ via a double analytic continuation in time and of the Hubble length.
The problem is then mapped onto a path integral for Euclidean AdS${}_2$, which may be a possible starting point for a Euclidean calculation of Von Neumann entropies in JT multiverses.

In the language of semiclassical gravitational path integrals, the formation of entanglement islands signals that one or more new saddles comes into play. While we did not pursue a path integral analysis of our two-dimensional toy multiverses in de Sitter JT gravity, we pointed out that this observation is very much in line with existing results in quantum cosmology in models of this kind in four dimensions. As an illustration, we gave a brief discussion of the probabilistic predictions for local cosmological observables such as, say, the CMB temperature anisotropies, in a false vacuum model of eternal inflation with two distinct decay channels and in the Hartle-Hawking state. In these and other models of inflation, probabilities for observations in quantum cosmology typically involve a superposition of saddle point geometries that include an averaging over any multiverse structure on the largest scales. This built-in coarse-graining was one of the key elements behind the semiclassical resolution of the measure problem.  

The upshot of the quantum cosmology analysis of these models appears to be that the global spacetime breaks up into a sum of a small number of distinct saddle point geometries, each of which involving a huge coarse-graining over much, if not all, of the bubble exterior. The fundamentally classical picture of a global spacetime is thus basically replaced in semiclassical quantum cosmology by a multiplicity of a few past histories of $R$, combined with much ``uncertainty'' on super-bubble scales. That is, contrary to appearances, the saddle geometries would {\it not} specify a global classical state, but rather delineate the limitations of classical spacetime, a point much emphasized by Hartle et al.\cite{Hartle:2010dq,Hartle:2016tpo}. This may not be entirely inconsistent with the ideas behind bulk reconstruction that one can represent operators on $I$ as operators on $R$. Imagine one were interested in constructing, or better still, measuring some heavy operator in $R$ that contains a significant amount of information about a distant patch. One expects that such extraordinary and complex measurements would result in a backreaction on the spacetime in $R$ to the extent that the measurement amounts to selecting the saddle point corresponding to the patch in question.

The picture that emerges from the confluence of these analyses is one in which, fundamentally, a definite spacetime geometry in cosmology comes about in an ``inside out'' way. In a sense, the entire multiverse would be reduced to an oasis consisting of a patch of classical spacetime around us surrounded on all sides by quantum fuzziness. That was the essence of the ``top-down'' approach to (quantum) cosmology advocated by Hawking \cite{Hartle:2010dq,Hawking:2006ur}.

\vspace{5mm}
\begin{center} 
{\bf Acknowledgments}
\end{center}
\noindent 
We thank Ning Bao, Arjun Kar, Jason Pollack, Jacopo Sisti, James Sully, and Mark Van Raamsdonk for helpful discussions during the preparation of this manuscript, as well as Kristan Jensen, Edgar Shaghoulian, and Edward Witten for useful comments. We also thank the anonymous referee for their contributions to the review and publication process. S.E.A.G., T.H. and B.R. are supported in part by the KU Leuven research grant C16/16/005. S.E.A.G. also thanks Uppsala University for its hospitality during part of this work.
A.C.D. was supported for a portion of this work as a postdoctoral fellow (Fundamental Research) of the National Research Foundation -- Flanders (FWO), Belgium.
A.C.D. acknowledges the support of the Natural Sciences and Engineering Research Council of Canada (NSERC), [funding reference number PDF-545750-2020]. / La contribution d'A.C.D. {\`a} cette recherche a {\'e}t{\'e} financ{\'e}e en partie par le Conseil de recherches en sciences naturelles et en g{\'e}nie du Canada (CRSNG), [num{\'e}ro de r{\'e}f{\'e}rence PDF-545750-2020].
N.P.F. is supported by the European
Commission through the Marie Sk{\l}odowska-Curie Action UniCHydro (grant agreement ID: 886540).

\appendix

\section{Exact island entropies in a theory of free fermions} \label{app:approx-checks}

The different multiverse configurations in \Sec{Sec:islands} involve a region $(R\cup I)^c$ which consists of two disjoint intervals.
When the size of these intervals is small compared to their separation, the two-interval entropy is approximately given by the sum of the single interval components. This regime is sometimes called the OPE limit.

The single interval entropy is a universal quantity, i.e. valid for any CFT with a given central charge, up to a scheme-dependent constant. However, the Von Neumann entropy of the reduced state of a CFT on disconnected intervals is in general not universal.
In this appendix, we compute exact multi-interval entropies for free massless Dirac fermions, and we compare the exact result to the OPE limit approximation in order to check the latter's accuracy.
We also use the exact result to investigate island configurations in JT multiverses for which the OPE limit is not valid.
Finally, we compute $S_\mrm{gen}(R \cup I)$ directly for an island that consists of two disjoint intervals in order to give evidence that an island consisting of a single large interval is the extremum that gives the smallest value of $S_\mrm{gen}(R \cup I)$.

\subsection{Multi-interval entanglement entropy}

Consider $p$ disjoint intervals in $\mathbb{R}^2$ whose endpoints we label by $(u_i,v_i)$, with $i=1,\dots, p$.
The corresponding Von Neumann entropy for a two-dimensional Euclidean CFT consisting of $c$ free massless Dirac fermions was computed in \Ref{Casini:2005rm} and is given by
\begin{equation}\label{eq:disjoint}
S_{\rm CFT}^{(p)}=\frac{c}{3}\left(\sum_{i,j}\log |u_i-v_j|-\sum_{i<j} \log |u_i-u_j|-\sum_{i<j}\log |v_i-v_j| -p \,\log \epsilon_{\rm uv}\right),
\end{equation}
where $\epsilon_{\rm uv}$ is a small UV regulator.

Let us first consider the case where $p=2$, and suppose that we place the theory on a manifold with the line element  $ds^2=dzd\bar{z}/\Omega(z,\bar{z})^2$.
Labelling the two intervals' endpoints by $(\mbf{z}_1, \mbf{z}_2)$ and $(\mbf{z}_3,\mbf{z}_4)$, where $\mbf{z}_i \equiv (z_i, \bar{z}_i)$, the two-interval entropy is
\begin{equation}
\label{eq:S2intz}
S_{\rm CFT}^{(2)}=\frac{c}{6}{\rm log}\bigg[\frac{|z_{12}|^2|z_{23}|^2|z_{34}|^2|z_{14}|^2}{\epsilon_{\rm uv}^4|z_{13}|^2|z_{24}|^2 \Omega_1\Omega_2\Omega_3\Omega_4}\bigg]\,,
\end{equation}
where $z_{ij}:=z_i-z_j$ and  $\Omega_i=\Omega(z_i,\bar{z}_i)$. For simplicity, in writing \Eq{eq:S2intz}, we have dropped a scheme-dependent constant.

To calculate $S_{\rm CFT}^{(2)}$ using the replica trick is equivalent to inserting twist operators at the endpoints of the intervals \cite{Calabrese:2009qy}, meaning that with four endpoints, the computation reduces to evaluating a four-point function. In the limit where the cross ratios
\begin{equation}\label{eq:limit}
z_{13}z_{24}/(z_{23}z_{14})\rightarrow 1 \qquad \text{and} \qquad \bar{z}_{13}\bar{z}_{24}/(\bar{z}_{23}\bar{z}_{14})\rightarrow 1\,,
\end{equation}
the two contributions to $S_{\rm CFT}^{(2)}$ decouple, and the two-interval entropy written in \Eq{eq:S2intz} reduces to the sum of the single interval entropies of the form in \Eq{eq:2d-EE-no-island-2}. The limit \Eq{eq:limit} thus corresponds to evaluating this four-point function in the OPE limit.

For $p=3$, the three-interval entropy is given by
\begin{equation}
\label{eq:3int}
S_{\rm CFT}^{(3)}=    \frac{c}{6}{\rm log}\bigg[\frac{|z_{12}|^2|z_{14}|^2|z_{16}|^2|z_{23}|^2|z_{34}|^2|z_{36}|^2|z_{25}|^2|z_{45}|^2|z_{56}|^2}{\epsilon_{\rm uv}^6|z_{13}|^2|z_{24}|^2|z_{15}|^2|z_{35}|^2|z_{26}|^2|z_{46}|^2 \Omega_1\Omega_2\Omega_3\Omega_4\Omega_5\Omega_6}\bigg],
\end{equation}
where the three intervals' endpoints are $(\mbf{z}_1, \mbf{z}_2)$, $(\mbf{z}_3, \mbf{z}_4)$, and $(\mbf{z}_5, \mbf{z}_6)$.
We will utilize Eqs.~\eqref{eq:S2intz} and \eqref{eq:3int} in the rest of this appendix.

\subsection{Single component island}
\label{subsec:two-interval}
We start by focusing on configurations with a single island $I$ extending throughout the multiverse to be able to examine several cases of interest. First, we reproduce our results in the main text for dS$^n_2$ and for the case of flat bubbles in dS$^n_2$ with the exact free fermion entropy.  Subsequently, with the aim of testing how close the island must be to the radiation region in order to produce a Page transition,  we consider additional JT multiverse configurations where bubbles are inserted asymmetrically and where the crunching regions are not adjacent to the expanding patch. In all the following cases we set $\varphi_1=-\varphi_4=\varphi_I$, $\sigma_1=\sigma_4=\sigma_I$, $\varphi_2=-\varphi_3=\varphi_R$, and $\sigma_2=\sigma_3=\sigma_R$.

For the models studied in this work, namely pure dS$_2^n$ and dS$_2^n$ with bubbles, the two-interval entropy of free fermions can be expressed in global coordinates via \Eq{xxbar} as
\begin{eqnarray}
\label{eq:S2int}
    S_{\rm CFT}^{(2)}&=&\frac{c}{6}{\rm log}\bigg[4n^4\left(\cos\left(\frac{\sigma_{12}}{n}\right)-\cos\left(\frac{\varphi_{12}}{n}\right)\right)\left(\cos\left(\frac{\sigma_{23}}{n}\right)-\cos\left(\frac{\varphi_{23}}{n}\right)\right)\times\nonumber\\
    &&\qquad \qquad\times\left(\cos\left(\frac{\sigma_{14}}{n}\right)-\cos\left(\frac{\varphi_{14}}{n}\right)\right)\left(\cos\left(\frac{\sigma_{34}}{n}\right)-\cos\left(\frac{\varphi_{34}}{n}\right)\right)\bigg]\\
&& - \frac{c}{6}{\rm log}\bigg[  \left(\cos\left(\frac{\sigma_{31}}{n}\right)-\cos\left(\frac{\varphi_{31}}{n}\right)\right)\left(\cos\left(\frac{\sigma_{42}}{n}\right)-\cos\left(\frac{\varphi_{42}}{n}\right)\right)\epsilon_{\rm uv}^4\prod_{i=1}^4{\omega_i}\bigg],\nonumber
\end{eqnarray}
with $\sigma_{ij}=\sigma_i-\sigma_j$, $\varphi_{ij}=\varphi_i-\varphi_j$ and $\omega_i=\omega(\sigma_i,\varphi_i)$. We will use this expression shortly to evaluate the generalized entropy of interest.

\subsubsection{Extended dS$_2$}
\begin{figure}[t]
    \centering
    \includegraphics[width=0.7\textwidth]{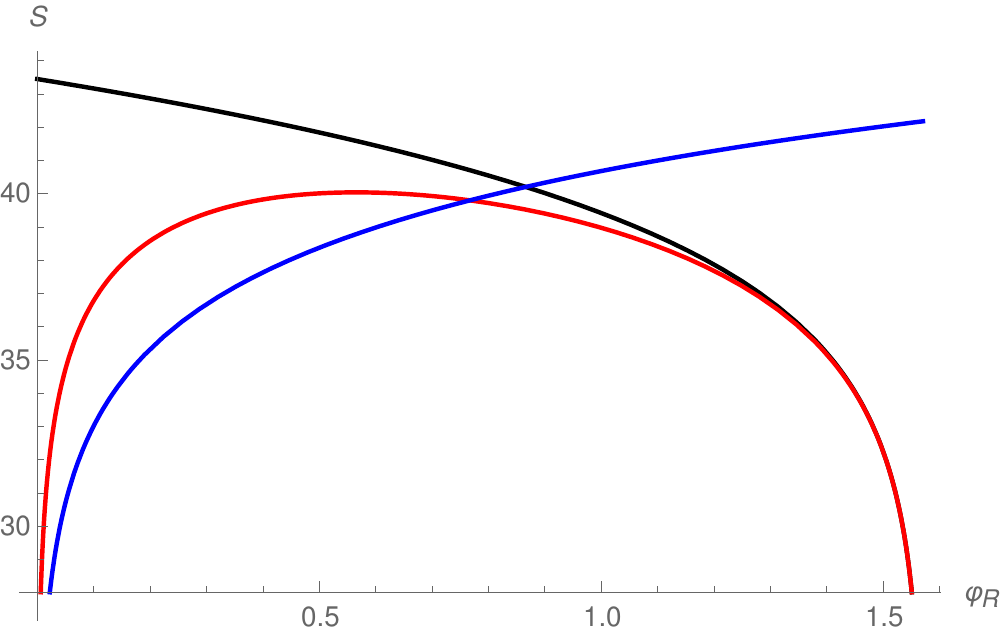}
    \caption{$S_{\text{gen}}(R \cup I)$ as derived using an exact two-interval formula for the generalized entropy in the context of free fermions (in red), the OPE limit (black), and the corresponding no-island entropy (blue) for a choice of parameters $n=10$, $\sigma_R=\frac{\pi}{2}-10^{-5}$, $c=100$, $\phi_r=10$, $\phi_0=0$, and $\epsilon_{\rm uv}=1$, $\epsilon_{\rm rg}=1$. The results coincide only for large spatial extension of the region $R$. The comparison shows that the matter entropy evaluated in the OPE limit gives a good approximation when the separation between the disjoint intervals is much greater than their proper lengths, which might occur before or after the Page transition depending on the parameters of the theory.}
    \label{fig:FFdS2n}
\end{figure}

Consider the symmetric configuration  $R\cup I$ shown in \Fig{fig:dS2n-penrose}. The conformal factors at the endpoints for this disjoint interval are given as $\omega_1=\omega_4=\cos{\sigma_I}$ and $\omega_2=\omega_3=\cos{\sigma_R}$. 
Plugging these factors into \Eq{eq:S2int}, we compute the corresponding generalized entropy \Eq{eq:gen} and extremize with respect to $\varphi_I$ and $\sigma_I$. We find numerically the entropy plots shown in \Fig{fig:FFdS2n}. Notice that, as guaranteed by subadditivity of Von Neumann entropy, the OPE approximation that we used in the main text provides an upper bound to the exact generalized entropy with a non-trivial island. While the qualitative behavior of the Page curve is unchanged, the precise value of $\varphi_R$ at which the Page transition occurs depends on whether we consider the free fermion model or the OPE limit of the twist operators.  Nevertheless, as explained in the introduction of this Appendix, both entropies agree  for large enough $\varphi_R$.

\subsubsection{Extended dS$_2$ with flat bubbles}
\begin{figure}[ht]
    \centering
    \includegraphics[width=0.7\textwidth]{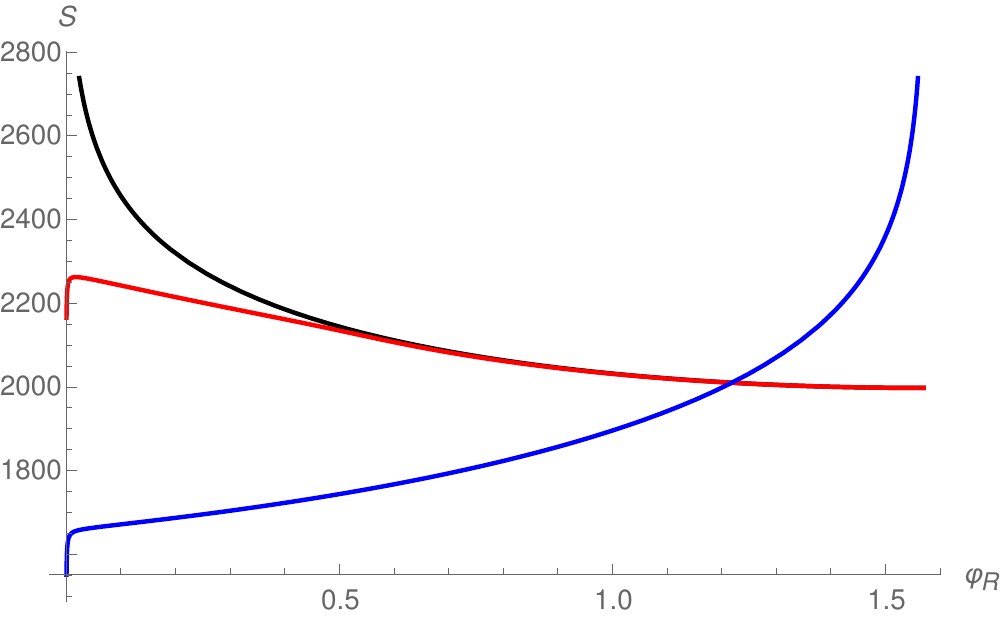}
    \caption{Comparison between generalized entropy with non-trivial island for a free fermion theory (red), the OPE limit of twist operators (black), and the no-island entropy (blue), illustrated when the endpoints of R satisfy the relation $\sigma_R=-\frac{2}{\pi}\left(\frac{\pi}{2}-10^{-3}\right)\varphi_R+\pi-10^{-3}$. The parameters of the theory are chosen as $n=10$, $c=600$, $\phi_r=10$, $\phi_0=0$, and $\epsilon_{\rm uv}=1$, $\epsilon_{\rm rg}=1$. The free fermion and the OPE generalized entropies coincide once $R$ is large enough. }\label{fig:FFBB}
\end{figure}

Next, we compare the exact generalized entropy for a theory of free fermions to the results in \Sec{sec:withbubbles} for the configuration illustrated in \Fig{1isl}.
The conformal factors at the endpoints of $R\cup I$ are $\omega_1=\omega_4=\frac{1}{2}\left(\cos{\sigma_R}+\cos{\varphi_R}\right)$ and $\omega_2=\omega_3=\cos{\sigma_I}$.
The plot showing this comparison is displayed in \Fig{fig:FFBB}. Again, for sufficiently large $R$, we find good agreement between the exact and the approximated generalized entropy.

\subsubsection{Alternative island configurations}
\begin{figure}[t]
  \centering
  \begin{subfigure}{\textwidth}
   \centering
   \begin{subfigure}{.66\textwidth}
   \includegraphics[width=\linewidth]{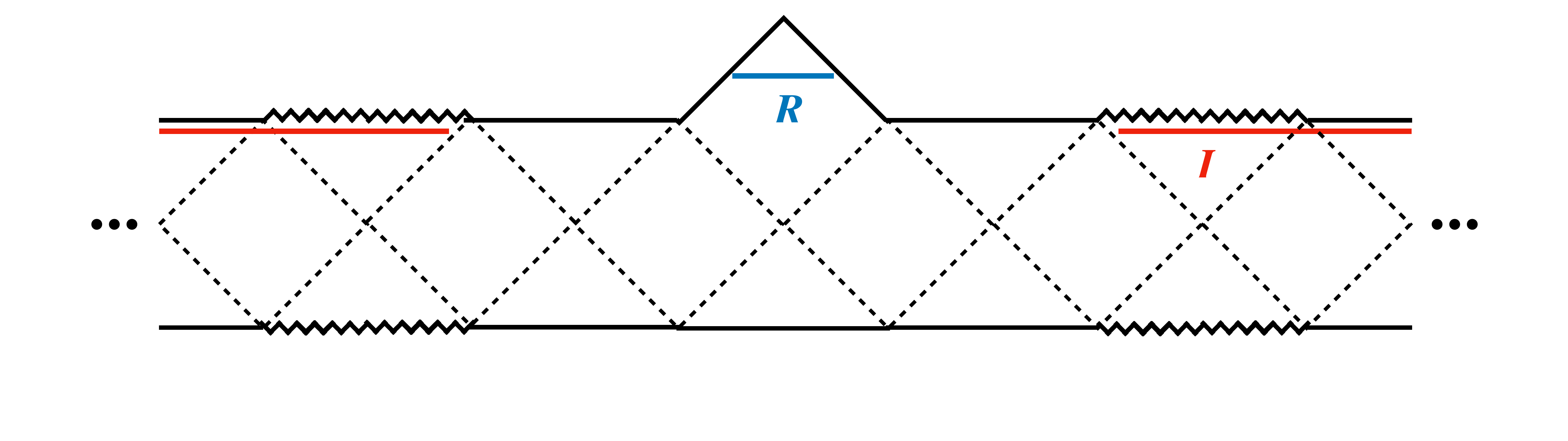}
  \end{subfigure}\begin{subfigure}{.33\textwidth}
   \includegraphics[width=\linewidth]{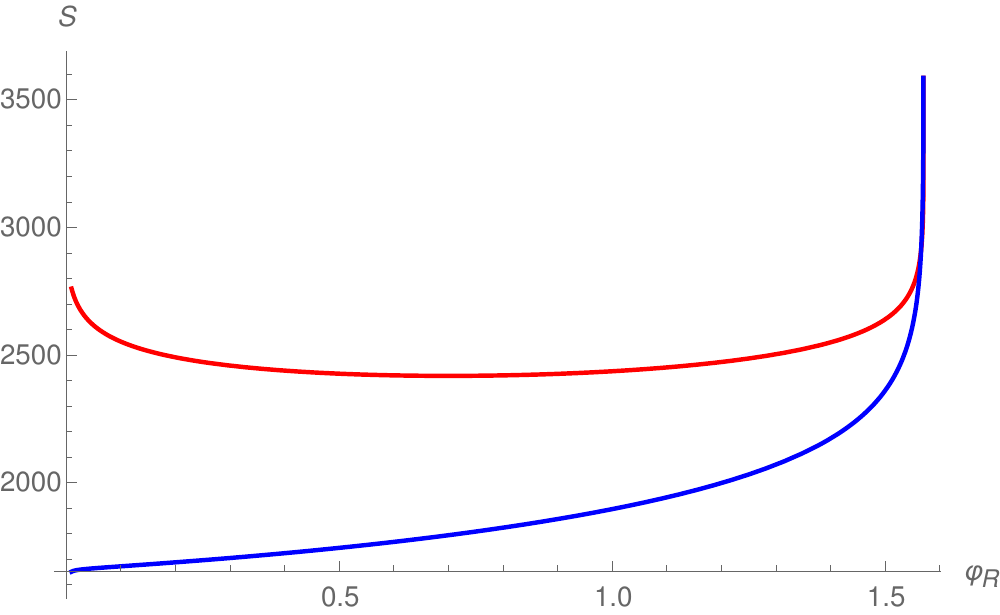}
   \end{subfigure}
   \caption{$ $}
   \label{fig:crunchinga}
    \end{subfigure}
    \begin{subfigure}{\textwidth}
   \centering
     \begin{subfigure}{.66\textwidth}
    \includegraphics[width=\linewidth]{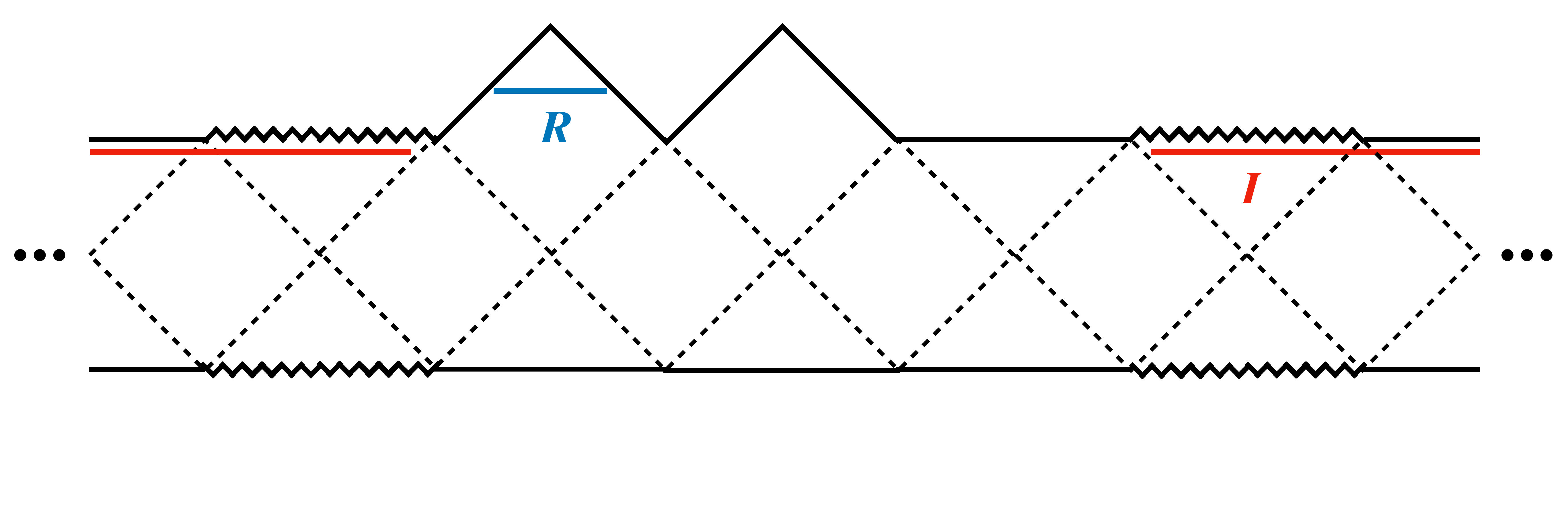}
   \end{subfigure}\begin{subfigure}{.33\textwidth}
   \includegraphics[width=\linewidth]{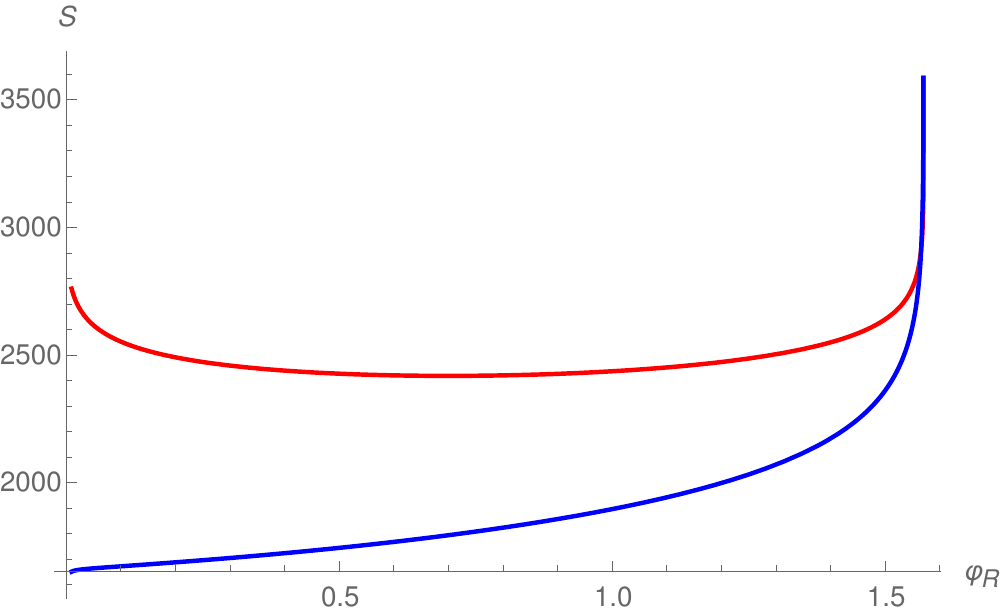}
    \end{subfigure}
   \caption{$ $}
   \label{fig:crunchingb}
   \end{subfigure}
    \begin{subfigure}{\textwidth}
    \centering
     \begin{subfigure}{.66\textwidth}
    \includegraphics[width=\linewidth]{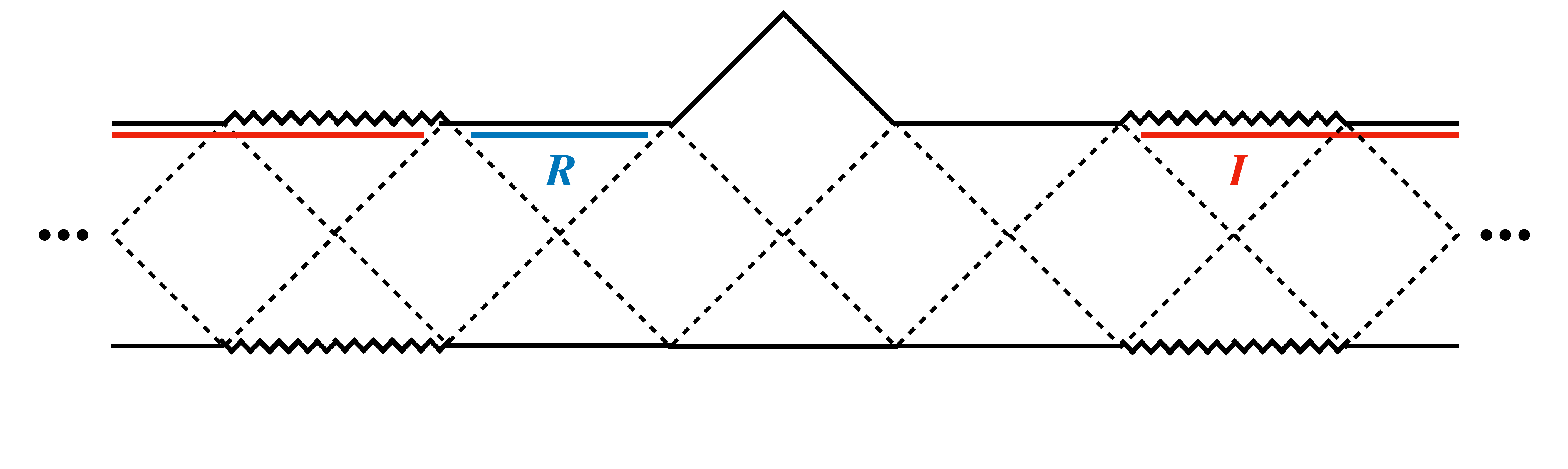}
    \end{subfigure}\begin{subfigure}{.33\textwidth}
    \includegraphics[width=\linewidth]{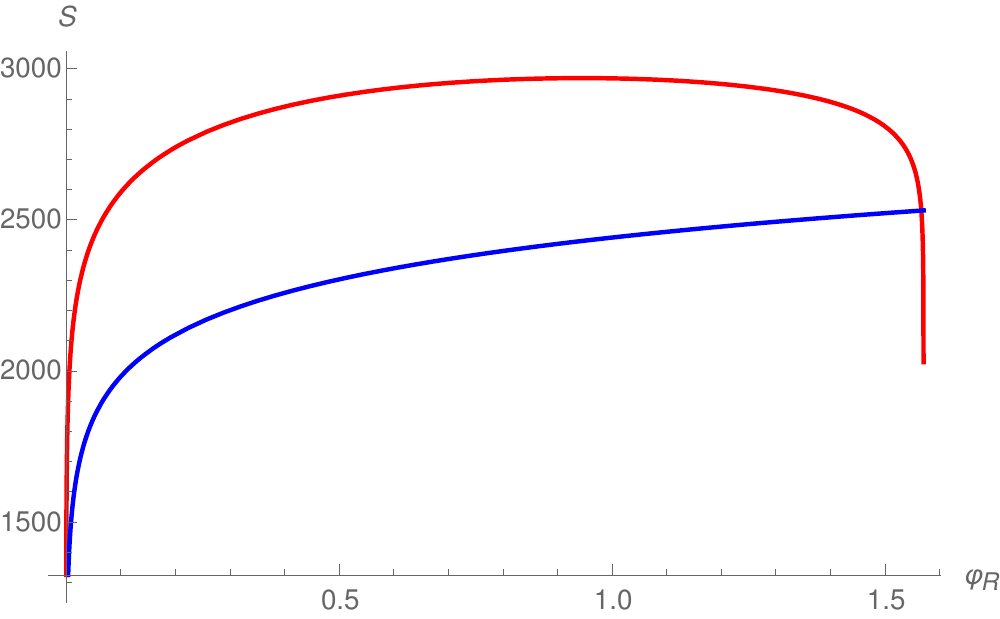}
    \end{subfigure}
    \caption{$ $}
   \label{fig:crunchingc}
   \end{subfigure}\vspace{0.5cm}
    \caption{(a)-(c) Penrose diagrams and the corresponding generalized entropy  for different $R\cup I$ configurations. The red curves indicate the generalized entropy including islands, while the blue ones indicate the no-island entropy. The endpoints of the non-trivial islands are inside crunching patches and they do not have to be located adjacent to the patch where $R$ resides. We observe, however, that the generalized entropy for the case (a) and (b) is minimal for the no-island configuration. The endpoints of $R$ are held at $\sigma_R=-\frac{2}{\pi}\left(\frac{\pi}{2}-10^{-3}\right)\varphi_R+\pi-10^{-3}$ for configurations (a)-(b), and $\sigma_R=\frac{\pi}{2}-10^{-5}$ for (c). Other parameters used in the plots are $n=10$, $c=600$, $\phi_r=10$, $\phi_0=0$, $\epsilon_{\rm rg}=1$, and $\epsilon_{\rm uv}=1$.}
    \label{fig:crunching}
\end{figure}

\begin{figure}[h]
    \centering
    \begin{subfigure}{\linewidth}
    \centering
     \begin{subfigure}{.66\textwidth}
    \includegraphics[width=\linewidth]{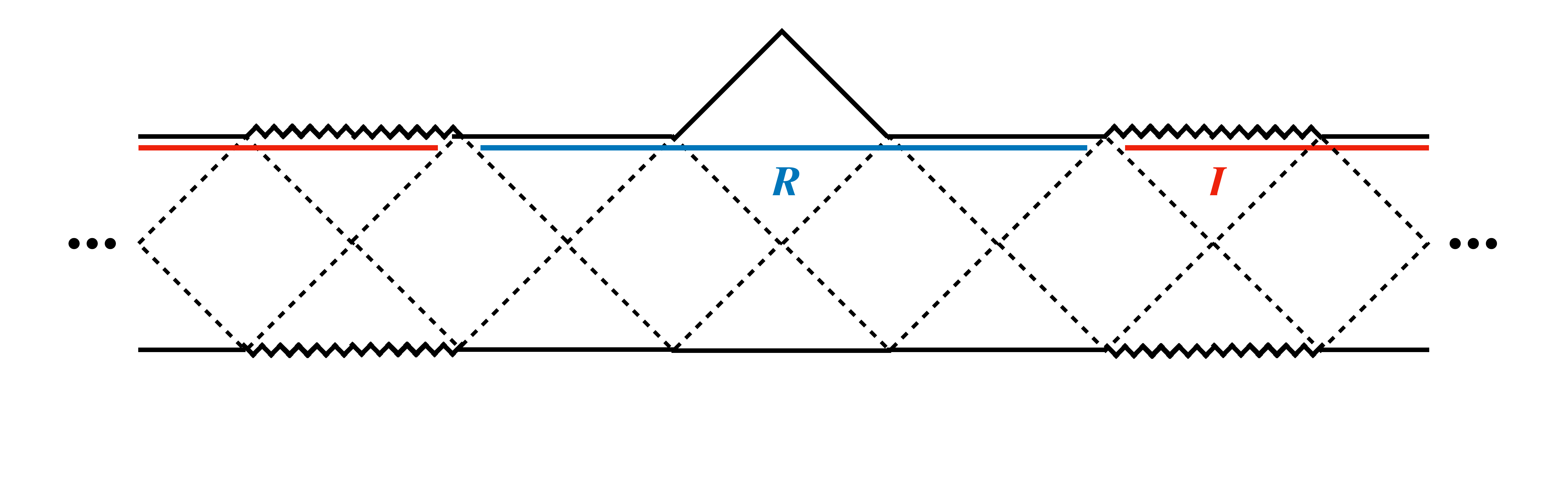}
    \end{subfigure}\begin{subfigure}{.33\textwidth}
    \includegraphics[width=\linewidth]{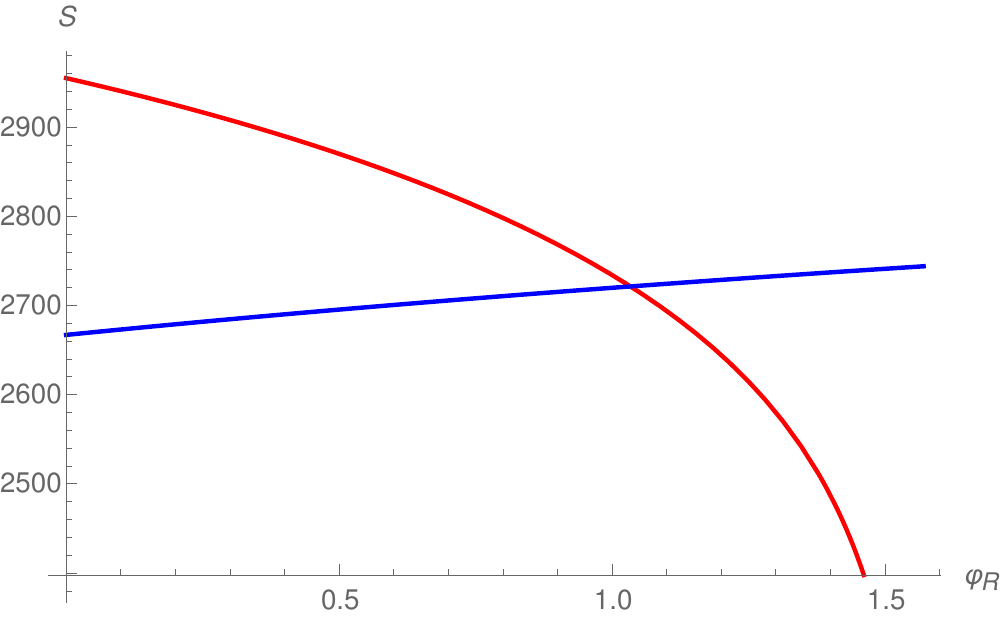}
    \end{subfigure}
    \caption{$ $}
    \label{fig:crunching0a}
    \end{subfigure}
    \begin{subfigure}{\linewidth}
    \centering
     \begin{subfigure}{.66\textwidth}
    \includegraphics[width=\linewidth]{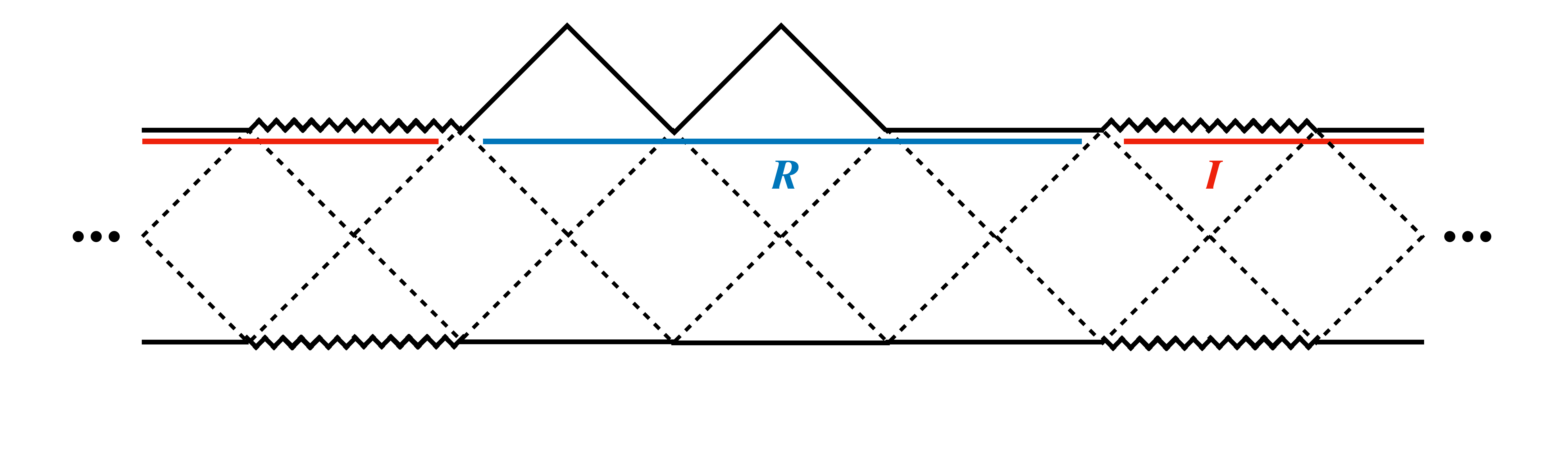}
    \end{subfigure}\begin{subfigure}{.33\textwidth}
    \includegraphics[width=\linewidth]{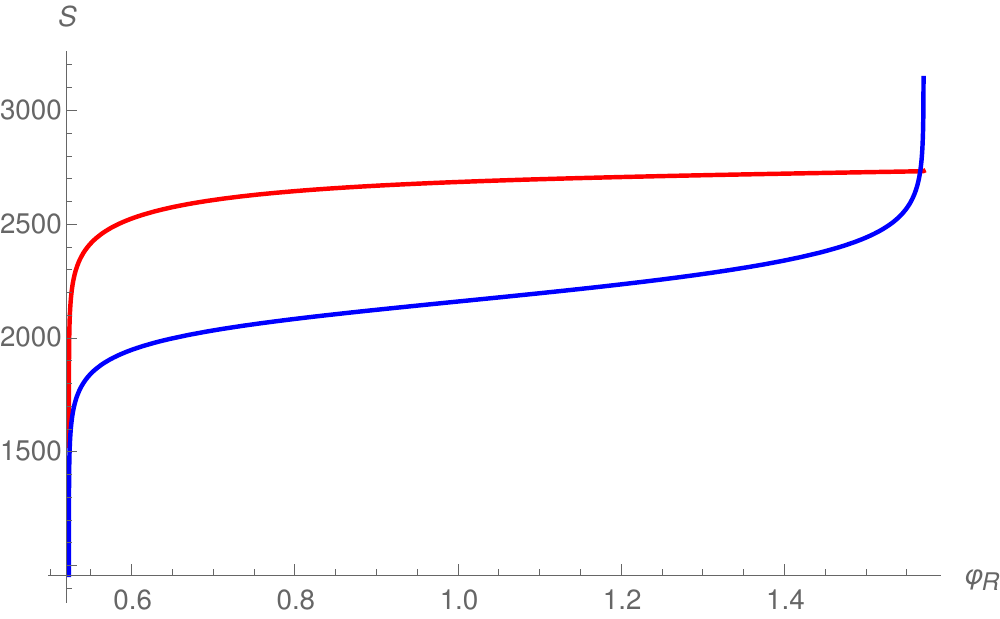}
    \end{subfigure}
    \caption{$ $}
    \label{fig:crunching0b}
    \end{subfigure}
    \caption{(a)-(b) Penrose diagrams and the corresponding Page curves, in which we relocate region $R$ to be as close to $I$ as possible to produce higher purification of the quantum state with respect to Figs. \ref{fig:crunchinga}-\ref{fig:crunchingb}. Red and blue denote the entropy with or without islands, and the constants are chosen as $n=10$, $c=600$, $\phi_r=10$, $\phi_0=0$, $\epsilon_{\rm rg}=1$, and $\epsilon_{\rm uv}=1$. Both endpoints of $R$ are located at $\sigma_R=\frac{\pi}{2}-10^{-5}$ in (a), while for (b) one endpoint is at $\sigma_{R1}=-\frac{2}{\pi}\left(\frac{\pi}{2}-10^{-3}\right)\varphi_R+\pi-10^{-3}$ and the other one at $\sigma_{R2}=\frac{\pi}{2}-10^{-5}$.}
    \label{fig:crunching0}
\end{figure}

Here, let us consider an alternative set of island configurations as displayed in Figs. \ref{fig:crunching} and \ref{fig:crunching0}. We are interested in investigating under what circumstances islands arise when the endpoints of $I$ lie in crunching regions that surround $R$.\footnote{There have been settings in which islands appear in expanding patches, e.g. \cite{Aalsma:2021bit}, albeit with a different assumption imposed on the CFT state.}

In particular, in  Fig. \ref{fig:crunchinga}-\ref{fig:crunchingc}, we consider configurations where the region $R$ is confined to one flat or expanding patch, which has at most one adjacent crunching patch.  In Figs.~\ref{fig:crunchinga}-\ref{fig:crunchingb}, we observe that islands may form, but they are somewhat screened by the intermediate additional flat and expanding bubbles, in the sense that they never have lower generalized entropy than the configuration with no island. In contrast, we find that  the configuration depicted in  \Fig{fig:crunchingc} exhibits a Page transition, albeit for a very large value of $\varphi_R$, when R is close enough of $\mathcal{I}^+$.

On the other hand, if we extend $R$ so that it spans multiple patches and so that its endpoints lie very close to crunching patches, we see that a Page transition occurs, as illustrated in Figs.
\ref{fig:crunching0a}-\ref{fig:crunching0b}.
These numerics are fully consistent with the analytic arguments from \Sec{subsec:features}.
Intuitively, the entropy of  $R \cup I$ is once again almost that of a pure state in these configurations.

\subsection{Two-component islands}
\label{subsec:three-interval}
\begin{figure}[t]
    \centering
    \includegraphics[width=0.75\textwidth]{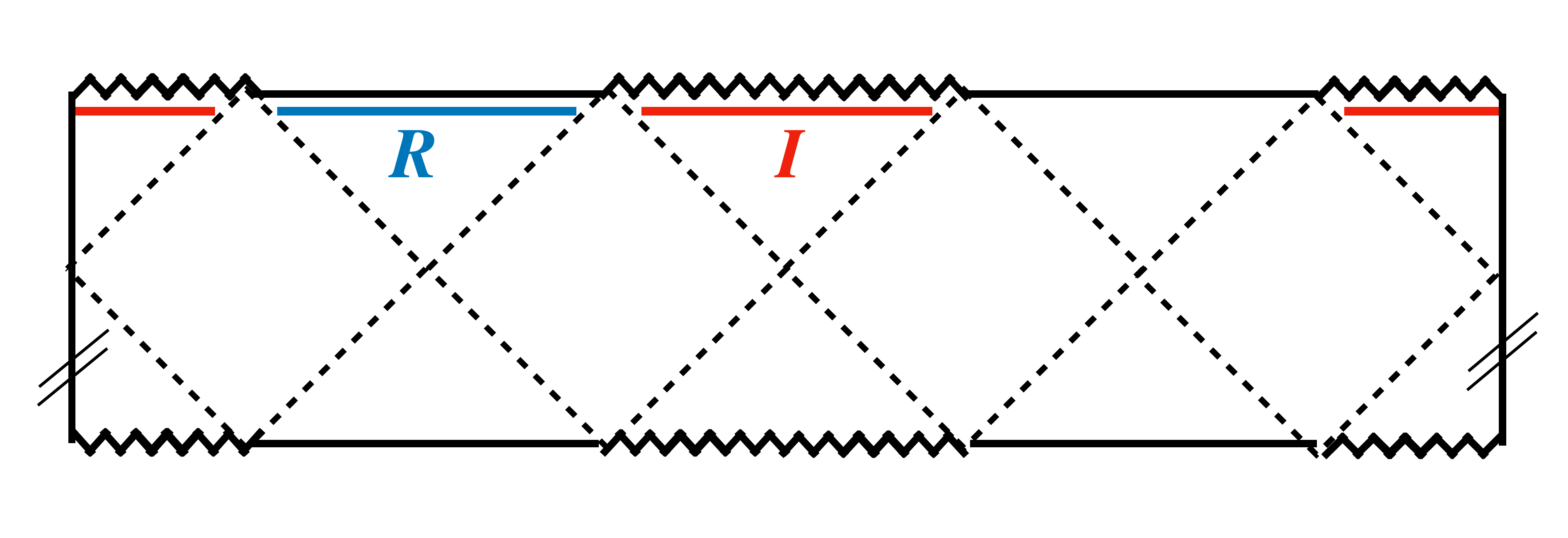}
    \caption{An island with two disconnected components in dS$_2^2$.}
    \label{fig:2isl}
\end{figure}

In this section, we consider the possibility of an island in ${\rm dS}_2^n$ that consists of two disconnected components, as depicted in \Fig{fig:2isl}.
To compute the generalized entropy $S_{\rm gen}(R\cup I)$, we once again consider the complement $(R\cup I)^c$, which is now comprised of three intervals. As already emphasized, such an entropy depends on the particular model under consideration. Here, we use the three-interval formula \Eq{eq:3int}, valid for the case of $c$ free massless Dirac fermions, and we adapt it to the global coordinates defined in Eqs.~\eqref{eq:dS2-line-element} and \eqref{eq:dS2-dilaton} through \Eq{xxbar}. 

\begin{figure}[ht]
    \centering
    \includegraphics[width=0.7\textwidth]{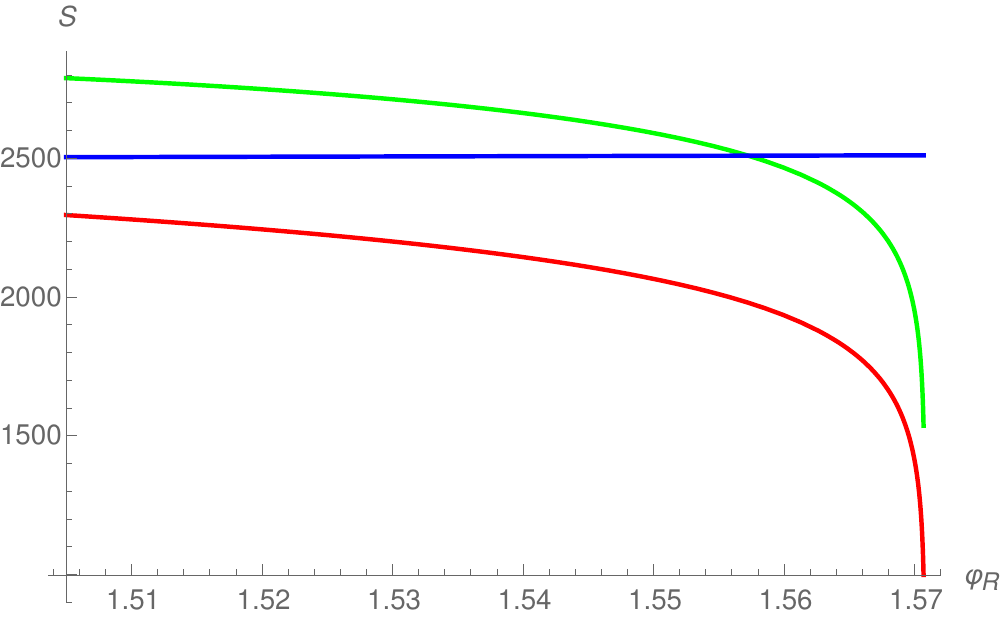}
    \caption{Two-component island (green curve) versus single-component island (red curve) versus no island (blue curve) for $n=2$. We plot the regime of $\varphi_R$ for which a two-island configuration exists as an extremum of $S_\mrm{gen}(R\cup I)$ and we observe that such a configuration is never dominant. We choose the endpoints of $R$ to be fixed at $\sigma_{R}=\frac{\pi}{2}-10^{-5}$, and the parameters of the theory are chosen as $n=2$, $c=600$, $\phi_r=10$, $\phi_0=0$, $\epsilon_{\rm rg}=1$, and $\epsilon_{\rm uv}=1$.}
    \label{fig:2islands}
\end{figure}

The result of our numerical computation for $n=2$ is shown in \Fig{fig:2islands}. We find that a two-component island configuration only appears for sufficiently large $\varphi_R$. The two components are symmetric about $R$, and the endpoints of the right component lie in the range $(\pi/2,\pi)$ within the crunching region to the right of $R$. For larger values of $n$, the value of $\varphi_R$ beyond which these island configurations appear increases. In all circumstances, the corresponding generalized entropy is always greater than the case of no islands or a single component island, and so a two-component island never dominates.

\clearpage

\section{Addendum}
\label{app:addendum}

In \Sec{subsec:JT}, we considered vacuum solutions of Jackiw-Teitelboim (JT) gravity with positive, zero, and negative cosmological constant, and in \Sec{sec:CFTentropy}, we deformed these theories by coupling the background metric to a two-dimensional CFT with field content collectively denoted by $\psi$.

We can describe each of these three deformed theories in a unified way with the action
\begin{equation}
\begin{aligned}
    I[g_{\mu\nu},\phi,\psi] &= \frac{\phi_0}{16\pi G_N}\int \dee^2x\sqrt{-g}\mathcal{R} - \frac{1}{16\pi G_N}\int \dee^2x\sqrt{-g} (\phi \mathcal{R}-V(\phi)) \\[2mm]
    &+ I_{GHY}[g_{\mu\nu},\phi] +I_{CFT}[g_{\mu\nu},\psi],
\end{aligned}
\end{equation}
where $V(\phi) = 2\phi$ results in a de Sitter (dS) solution, $V(\phi) = 2$ results in a flat solution, and $V(\phi) = -2\phi$ results in an anti-de Sitter (AdS) solution.
Varying this action with respect to the metric and the dilaton field $\phi$ produces the following equations of motion in the semiclassical limit:
\begin{align}
    \nabla_\mu\nabla_\nu\phi-g_{\mu\nu}\nabla^2\phi-\frac{1}{2}g_{\mu\nu}V(\phi)&=-8\pi G_N \expval{T_{\mu\nu}} \label{eq:EOMdil}\,,\\
    \mathcal{R} &=V'(\phi)\,, \label{eq:EOMmet}
\end{align}
where $\langle T_{\mu\nu}\rangle$ is the expectation value of the covariant stress-energy tensor of the CFT. The vacuum solution corresponds to setting $\expval{T_{\mu\nu}} = 0$.

Upon including the coupling to the CFT, we showed that a contribution to $\expval{T_{\mu\nu}}$ coming from a trace anomaly can be removed by a suitable field redefinition, and we subsequently proceeded with the vacuum solution for $\phi$ in \Sec{sec:CFTentropy} and beyond. However, we did not account for contributions from the Weyl anomaly and the Casimir energy when we examined an $n$-fold extension of dS${}_2$.
These contributions cancel when $n=1$.

Begin with the line element defined in Eqs.~\eqref{eq:generic-extended-metric-1} and \eqref{eq:conformal-factors} and let $x^\pm = \sigma \pm \varphi$.
As before, we can remove $\expval{T_{+-}}$ by a suitable redefinition of the constant $\phi_0$ (as it arises from the conformal anomaly).
The Weyl anomaly and Casimir energy \cite{Balasubramanian:2020xqf} combine to give
\begin{equation} \label{eq:source}
    \expval{T_{\pm\pm}} = \frac{c}{48\pi}\left(1-\frac{1}{n^2}\right),
\end{equation}
where we recall that the spatial coordinate $\varphi$ is $2\pi n$-periodic, and we have chosen the state in the $z$, $\bar{z}$ coordinates of \eqref{xxbar} to be in vacuum. Therefore, the sourceless solutions for the dilaton will acquire a supplementary additive term due to the source on the right-hand-side of \Eq{eq:EOMdil} when $n > 1$.

Our starting point was the case $\mathcal{R} = 2$, i.e. dS${}_2^n$.
In this case, the solution of \Eq{eq:EOMdil} with $V(\phi) = 2\phi$ and the source \eqref{eq:source} is
\begin{equation}
    \phi=\phi_r\frac{\cos\varphi}{\cos\sigma}-\frac{cG_N}{3}\qty(1-\frac{1}{n^2})(\sigma\tan\sigma+1), \label{eq:dildS}
\end{equation}
cf. \Eq{eq:dS2-dilaton}. Notice, however, that when $\phi_r/G_N \gg c$, the vacuum contribution dominates over the additive correction, and so we can safely neglect the correction in this limit.

In particular, we can still attempt to build up a JT multiverse using vacuum solutions as follows.
Starting with pure dS${}_2^n$ with $\phi_r/G_N \gg c$, we drop the correction due to a CFT stress-energy source and work with the vacuum dS dilaton.
If we want to include bubbles, the only choice is to patch in bubbles in which the dilaton obeys vacuum equations of motion; otherwise, the gluing would result in a dilaton that is not continuous across bubble interfaces.
However, the CFT then cannot be in the Minkowski vacuum of \eqref{xxbar}, for which there would have to be nonzero $\langle T_{\pm \pm} \rangle$ sources in flat and AdS regions per \Eq{eq:source}.
Instead, the CFT is in some state such that $\langle T_{\pm \pm} \rangle$ vanishes everywhere.
It is unclear whether such a CFT state is well defined and whether its entropy is close to that of the Minkowski vacuum of \eqref{xxbar}, so that \eqref{eq:2d-EE-no-island-3} continues to hold.
But, with these caveats, the existing analysis goes forward.

\begin{figure}[h]
    \centering
    \includegraphics[scale=0.2]{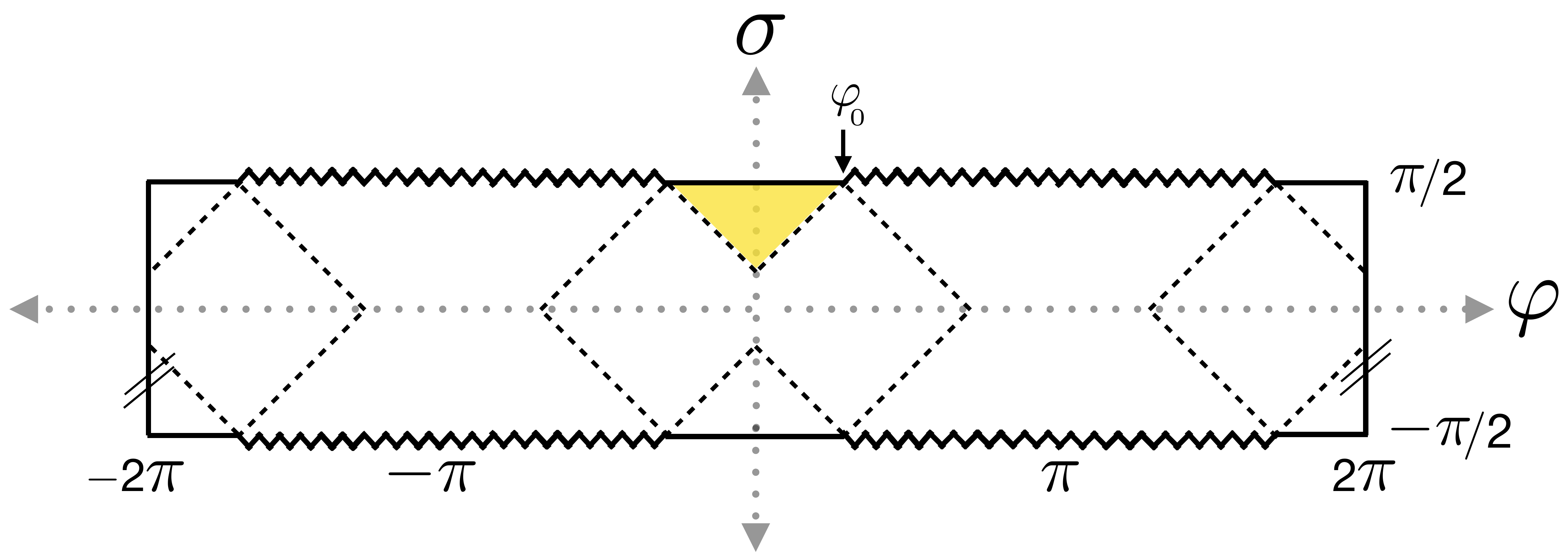}
    \caption{Penrose diagram for dS${}_2^n$ with $n=2$. An expanding patch is shaded in yellow, and it intersects $\mathcal{I}^+$ on the interval $(-\varphi_0,\varphi_0)$ where $\varphi_0 < \pi/2$.}
    \label{fig:Addendum-Sketch-1}
\end{figure}

Alternatively, if we do not neglect the source \eqref{eq:source}, we can solve for a backreacted dilaton on the manifolds that we specified in Eqs.~\eqref{eq:generic-extended-metric-1} and \eqref{eq:conformal-factors}.
Let us start with dS${}_2^n$ in the absence of any bubbles.
Compared to the vacuum solution, the modified dilaton solution \eqref{eq:dildS} causes the parts of $\mathcal{I}^+$ on which $\phi \rightarrow +\infty$ (i.e., the future boundary of the expanding patches) to shrink.
By inspection, these are the parts of $\mathcal{I}^+$ for which
\begin{equation}
    \cos \varphi > \frac{\pi c G_N}{6\phi_r}\left(1-\frac{1}{n^2}\right)\,.
\end{equation}
For such regions of $\mathcal{I}^+$ to exist, one must have that
\begin{equation}\label{eq:criterion}
    \frac{\phi_r}{G_N} > \frac{\pi c}{6} \left(1-\frac{1}{n^2}\right)\,,
\end{equation}
which we assume here.
This is illustrated in \Fig{fig:Addendum-Sketch-1}.

Our analysis of quantum extremal islands then proceeds essentially verbatim.
We again consider a region $R$ whose endpoints lie near the corners of the expanding patch centred about $\varphi = 0$, and we posit an island whose endpoints lie just beyond the patch's corners.
Because the locations of the corners are now shifted relative to the vacuum case, instead of the ansatz \eqref{eq:approx_Sgen_start}, we write
\begin{equation}
\begin{aligned}
    \sigma_R &= \frac{\pi}{2} - \delta \sigma_R\,, \qquad  \,\sigma_I = \frac{\pi}{2} - \delta \sigma_I\,, \\
    \varphi_R &= \varphi_0 - \delta \varphi_R\,, \qquad \varphi_I = \varphi_0 + \delta \varphi_I,
\end{aligned}
\end{equation}
where we have defined
\begin{equation}
    \varphi_0=\arccos{\frac{\pi cG_N}{6\phi_r}\qty(1-\frac{1}{n^2})}.\label{eq:varphi0}
\end{equation}
Making these substitutions in \eqref{SgendS2n}, we get
\begin{equation}
\begin{aligned}
    S_{\text{gen}}((R\cup I)^c) \approx &\frac{c}{3} \log\left[\frac{2n^2(\cos{} (\frac{\delta\sigma_I-\delta\sigma_R}{n}) -\cos{} (\frac{\delta\varphi_I+\delta\varphi_R}{n}))}{\epsilon_{\mrm{rg}}\epsilon_{\mrm{uv}} \delta \sigma_I \delta \sigma_R}\right]\\
    &\qquad + 2\phi_r\frac{\cos(\varphi_0+\delta \varphi_I)}{\delta \sigma_I}-\frac{\pi c}{12}\qty(1-\frac{1}{n^2})\frac{1}{\delta\sigma_I} +2\phi_0\,,
\end{aligned}
\end{equation}
where we have set $4G_N=1$.
Next, if we expand $\cos(\varphi_0+\delta \varphi_I)$ about $\varphi_0$ and assume that the sum $\delta \varphi_I + \delta \varphi_R$ and the difference $\delta \sigma_I - \delta \sigma_R$ are small, we arrive at
\begin{equation}
\begin{aligned}
    S_{\text{gen}}((R\cup I)^c) \approx &\frac{c}{3} \log\left[\frac{ (\delta \varphi_I + \delta \varphi_R)^2 - (\delta\sigma_I - \delta\sigma_R)^2 }{\epsilon_{\mrm{rg}}\epsilon_{\mrm{uv}} \delta \sigma_I \delta \sigma_R}\right]-2\phi'_r\frac{\delta\varphi_I}{\delta\sigma_I}+2\phi_0,
\end{aligned}
\end{equation}
where we have defined
\begin{equation}
    \phi'_r=\phi_r\sqrt{1-\qty(\frac{\pi c(1-n^{-2})}{24\phi_r})^2}.\label{eq:rescaling}
\end{equation}
This is identical in form to \Eq{eq:approx_Sgen}, and so the rest of the analysis proceeds as before, but with $\phi_r \rightarrow \phi_r'$.
While $\phi_r'$ now depends on $n$ explicitly, any $n$-dependence only enters at $O((c/\phi_r)^2)$.

Next, we consider multiverse models in the presence of flat or AdS${}_2$ bubbles, like the ones in \Sec{sec:withbubbles}.
The task is to show that the backreacted dilaton solutions continuously join up along bubble interfaces.
Let us first examine the flat case.
With a judicious choice of integration constants, the solution of \Eq{eq:EOMdil} with $V(\phi) = 2$ and the source \eqref{eq:source} for $\phi$ in a flat bubble centred about $\varphi=0$ reads (cf. \Eq{eq:mink2-dilaton})
\begin{equation} \label{eq:dil-flat-backrxn}
\phi = \phi_r - K_n + 2\frac{\cos \varphi - \cos \sigma}{\cos \varphi + \cos \sigma} - K_n \frac{\sigma \sin \sigma +\varphi \sin \varphi}{\cos \varphi + \cos \sigma},
\end{equation}
where we have defined
\begin{equation}
    K_n = \frac{cG_N}{3}\left( 1 - \frac{1}{n^2} \right).
\end{equation}
In particular, the flat solution \eqref{eq:dil-flat-backrxn} and the dS solution \eqref{eq:dildS} coincide along $\sigma = |\varphi|$, and so they can be continuously joined together.

\begin{figure}[t]
    \centering
    \includegraphics[scale=0.2]{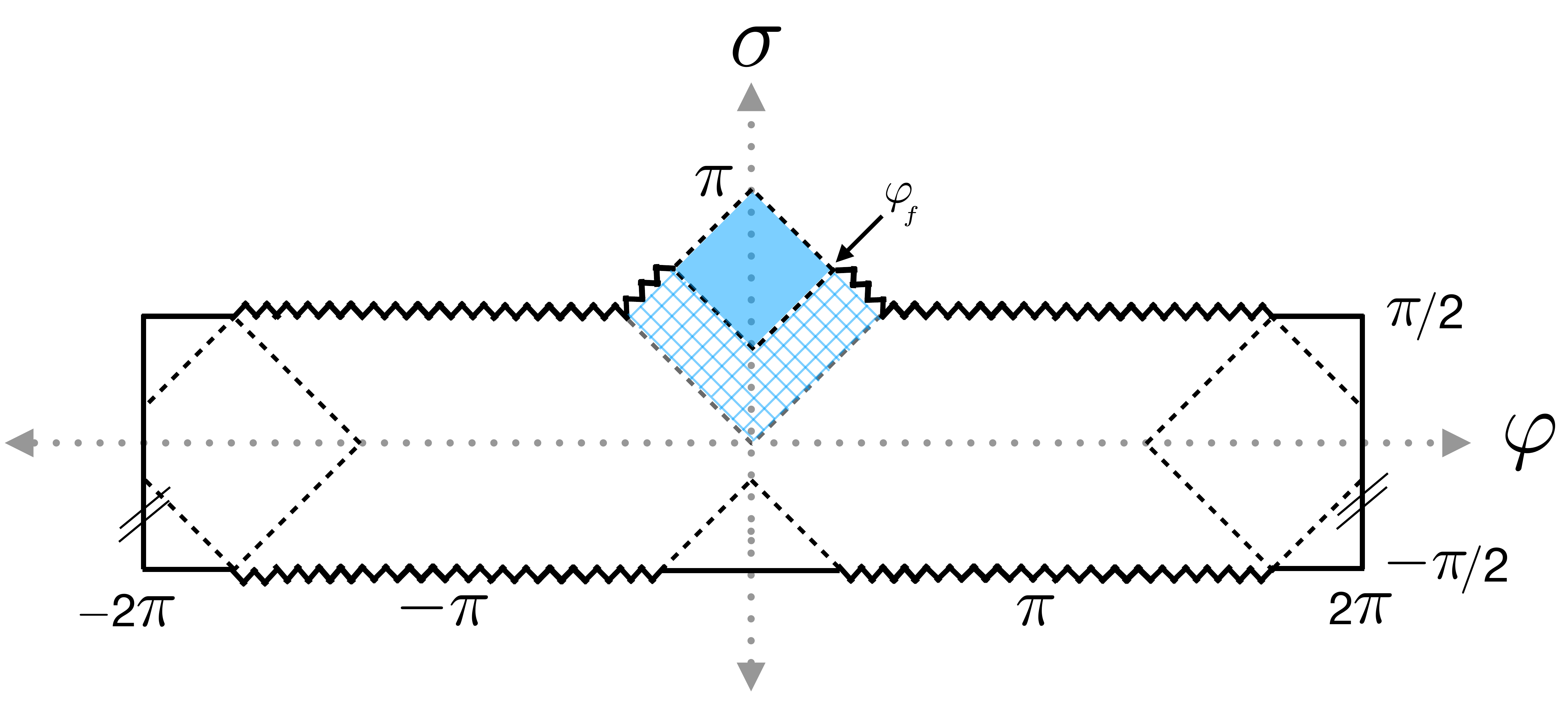}
    \caption{Penrose diagram for dS${}_2^n$ with $n=2$ and a flat bubble centred at $\varphi=0$. The domain of dependence of the part of $\mathcal{I}^+$ on which $\phi \rightarrow \infty$ is shaded in blue, but the metric is still flat in the blue hatched region.}
    \label{fig:Addendum-Sketch-2}
\end{figure}

A consequence of the backreaction is that $\mathcal{I}^+$ in a flat bubble develops segments where $\phi \rightarrow -\infty$.
By examining the behaviour of \eqref{eq:dil-flat-backrxn} as one approaches the lines $\sigma = \pi \pm \varphi$, one concludes that $\varphi < |\varphi_f|$ is the portion of $\mathcal{I}^+$ on which $\phi \rightarrow \infty$, where
\begin{equation}
    \tan \varphi_f = \frac{4}{\pi K_n}.
\end{equation}
This is illustrated in \Fig{fig:Addendum-Sketch-2}.
Therefore, should islands still develop, we expect that the endpoints of a region $R$ inside of a flat bubble should be placed just to the interior of $(\sigma,\varphi) = (\pi-\varphi_f,\pm\varphi_f)$.
We were unable to locate extrema of $S_\mrm{gen}((R \cup I)^c)$ by placing $(\sigma_I,\varphi_I)$ perturbatively to the past of $(\pi - \varphi_f, \varphi_f)$ and mirroring our earlier analysis, and so additional work would be needed to conclusively determine whether or not islands develop.

For an AdS${}_2$ bubble, the solution of \Eq{eq:EOMdil} with $V(\phi) = -2\phi$ and the source \eqref{eq:source} for $\phi$ reads (cf. \Eq{eq:AdS2-dilaton})
\begin{equation}
    \phi= -\phi_r\frac{\cos\sigma}{\cos\tilde\varphi} - K_n(\tilde\varphi\tan\tilde\varphi+1), \label{eq:dilAdSback}
\end{equation}
where $\tilde\varphi = 0$ corresponds to the centre of the bubble.
Its value along the line $\sigma = \tilde \varphi$  for $0 < \tilde \varphi < \pi/2$  coincides with the value of the dS dilaton \eqref{eq:dildS} along the line $\sigma = \pi + \varphi$ for a constant shift $\tilde \varphi = \varphi + \pi$.
Similarly, the AdS dilaton's value along the line $\sigma = -\tilde \varphi$ for $-\pi/2 < \tilde \varphi < 0$ coincides with that of \eqref{eq:dildS} along the line $\sigma = \pi - \varphi$ for a constant shift $\tilde \varphi = \varphi - \pi$.
Therefore, with the appropriate coordinate translations, the backreacted dS and AdS dilaton profiles continuously join up along AdS${}_2$ bubble walls.

In our previous analysis, we were only able to locate islands with endpoints in AdS${}_2$ bubbles numerically in the regime $\phi_r/G_N \ll c$.
Therefore, we cannot conclusively say whether or not islands develop when we take backreaction into account, which requires $\phi_r/G_N > c$ for $n > 1$ per \Eq{eq:criterion}.

\clearpage

\bibliographystyle{utphys-modified}
\bibliography{references.bib}

\providecommand{\href}[2]{#2}\begingroup\raggedright\begin{thebibliography}{10}

\bibitem{Linde:1993xx}
A.~D. Linde, D.~A. Linde, and A.~Mezhlumian, ``{From the Big Bang theory to the
  theory of a stationary universe},''
  \href{http://dx.doi.org/10.1103/PhysRevD.49.1783}{{\em Phys. Rev. D}
  {\bfseries 49} (1994) 1783--1826},
  \href{http://arxiv.org/abs/gr-qc/9306035}{{\ttfamily arXiv:gr-qc/9306035}}.

\bibitem{Hartle:2010dq}
J.~Hartle, S.~W. Hawking, and T.~Hertog, ``{Local Observation in Eternal
  inflation},'' \href{http://dx.doi.org/10.1103/PhysRevLett.106.141302}{{\em
  Phys. Rev. Lett.} {\bfseries 106} (2011) 141302},
  \href{http://arxiv.org/abs/1009.2525}{{\ttfamily arXiv:1009.2525 [hep-th]}}.

\bibitem{Penington:2019npb}
G.~Penington, ``{Entanglement Wedge Reconstruction and the Information
  Paradox},'' \href{http://dx.doi.org/10.1007/JHEP09(2020)002}{{\em JHEP}
  {\bfseries 09} (2020) 002}, \href{http://arxiv.org/abs/1905.08255}{{\ttfamily
  arXiv:1905.08255 [hep-th]}}.

\bibitem{Almheiri:2019psf}
A.~Almheiri, N.~Engelhardt, D.~Marolf, and H.~Maxfield, ``{The entropy of bulk
  quantum fields and the entanglement wedge of an evaporating black hole},''
  \href{http://dx.doi.org/10.1007/JHEP12(2019)063}{{\em JHEP} {\bfseries 12}
  (2019) 063}, \href{http://arxiv.org/abs/1905.08762}{{\ttfamily
  arXiv:1905.08762 [hep-th]}}.

\bibitem{Penington:2019kki}
G.~Penington, S.~H. Shenker, D.~Stanford, and Z.~Yang, ``{Replica wormholes and
  the black hole interior},'' \href{http://arxiv.org/abs/1911.11977}{{\ttfamily
  arXiv:1911.11977 [hep-th]}}.

\bibitem{Almheiri:2019qdq}
A.~Almheiri, T.~Hartman, J.~Maldacena, E.~Shaghoulian, and A.~Tajdini,
  ``{Replica Wormholes and the Entropy of Hawking Radiation},''
  \href{http://dx.doi.org/10.1007/JHEP05(2020)013}{{\em JHEP} {\bfseries 05}
  (2020) 013}, \href{http://arxiv.org/abs/1911.12333}{{\ttfamily
  arXiv:1911.12333 [hep-th]}}.

\bibitem{Marolf:2020xie}
D.~Marolf and H.~Maxfield, ``{Transcending the ensemble: baby universes,
  spacetime wormholes, and the order and disorder of black hole information},''
  \href{http://dx.doi.org/10.1007/JHEP08(2020)044}{{\em JHEP} {\bfseries 08}
  (2020) 044}, \href{http://arxiv.org/abs/2002.08950}{{\ttfamily
  arXiv:2002.08950 [hep-th]}}.

\bibitem{Almheiri:2020cfm}
A.~Almheiri, T.~Hartman, J.~Maldacena, E.~Shaghoulian, and A.~Tajdini, ``The
  entropy of hawking radiation,''
  \href{http://dx.doi.org/10.1103/RevModPhys.93.035002}{{\em Rev. Mod. Phys.}
  {\bfseries 93} (2021) 035002},
  \href{http://arxiv.org/abs/2006.06872}{{\ttfamily arXiv:2006.06872
  [hep-th]}}.

\bibitem{Marolf:2020rpm}
D.~Marolf and H.~Maxfield, ``{Observations of Hawking radiation: the Page curve
  and baby universes},'' \href{http://dx.doi.org/10.1007/JHEP04(2021)272}{{\em
  JHEP} {\bfseries 04} (2021) 272},
  \href{http://arxiv.org/abs/2010.06602}{{\ttfamily arXiv:2010.06602
  [hep-th]}}.

\bibitem{Page:1993wv}
D.~N. Page, ``{Information in black hole radiation},''
  \href{http://dx.doi.org/10.1103/PhysRevLett.71.3743}{{\em Phys. Rev. Lett.}
  {\bfseries 71} (1993) 3743--3746},
  \href{http://arxiv.org/abs/hep-th/9306083}{{\ttfamily arXiv:hep-th/9306083}}.

\bibitem{Page:2013dx}
D.~N. Page, ``{Time Dependence of Hawking Radiation Entropy},''
  \href{http://dx.doi.org/10.1088/1475-7516/2013/09/028}{{\em JCAP} {\bfseries
  09} (2013) 028}, \href{http://arxiv.org/abs/1301.4995}{{\ttfamily
  arXiv:1301.4995 [hep-th]}}.

\bibitem{Ryu_2006}
S.~Ryu and T.~Takayanagi, ``{Holographic derivation of entanglement entropy
  from AdS/CFT},'' \href{http://dx.doi.org/10.1103/PhysRevLett.96.181602}{{\em
  Phys. Rev. Lett.} {\bfseries 96} (2006) 181602},
  \href{http://arxiv.org/abs/hep-th/0603001}{{\ttfamily arXiv:hep-th/0603001}}.

\bibitem{Hubeny:2007xt}
V.~E. Hubeny, M.~Rangamani, and T.~Takayanagi, ``{A Covariant holographic
  entanglement entropy proposal},''
  \href{http://dx.doi.org/10.1088/1126-6708/2007/07/062}{{\em JHEP} {\bfseries
  07} (2007) 062}, \href{http://arxiv.org/abs/0705.0016}{{\ttfamily
  arXiv:0705.0016 [hep-th]}}.

\bibitem{Faulkner:2013ana}
T.~Faulkner, A.~Lewkowycz, and J.~Maldacena, ``{Quantum corrections to
  holographic entanglement entropy},''
  \href{http://dx.doi.org/10.1007/JHEP11(2013)074}{{\em JHEP} {\bfseries 11}
  (2013) 074}, \href{http://arxiv.org/abs/1307.2892}{{\ttfamily arXiv:1307.2892
  [hep-th]}}.

\bibitem{Engelhardt:2014gca}
N.~Engelhardt and A.~C. Wall, ``{Quantum Extremal Surfaces: Holographic
  Entanglement Entropy beyond the Classical Regime},''
  \href{http://dx.doi.org/10.1007/JHEP01(2015)073}{{\em JHEP} {\bfseries 01}
  (2015) 073}, \href{http://arxiv.org/abs/1408.3203}{{\ttfamily arXiv:1408.3203
  [hep-th]}}.

\bibitem{Teitelboim:1983ux}
C.~Teitelboim, ``{Gravitation and Hamiltonian Structure in Two Space-Time
  Dimensions},'' \href{http://dx.doi.org/10.1016/0370-2693(83)90012-6}{{\em
  Phys. Lett. B} {\bfseries 126} (1983) 41--45}.

\bibitem{Jackiw:1984je}
R.~Jackiw, ``{Lower Dimensional Gravity},''
  \href{http://dx.doi.org/10.1016/0550-3213(85)90448-1}{{\em Nucl. Phys. B}
  {\bfseries 252} (1985) 343--356}.

\bibitem{Almheiri:2014cka}
A.~Almheiri and J.~Polchinski, ``{Models of AdS$_{2}$ backreaction and
  holography},'' \href{http://dx.doi.org/10.1007/JHEP11(2015)014}{{\em JHEP}
  {\bfseries 11} (2015) 014}, \href{http://arxiv.org/abs/1402.6334}{{\ttfamily
  arXiv:1402.6334 [hep-th]}}.

\bibitem{Maldacena:2016upp}
J.~Maldacena, D.~Stanford, and Z.~Yang, ``{Conformal symmetry and its breaking
  in two dimensional Nearly Anti-de-Sitter space},''
  \href{http://dx.doi.org/10.1093/ptep/ptw124}{{\em PTEP} {\bfseries 2016}
  no.~12, (2016) 12C104}, \href{http://arxiv.org/abs/1606.01857}{{\ttfamily
  arXiv:1606.01857 [hep-th]}}.

\bibitem{Saad:2019lba}
P.~Saad, S.~H. Shenker, and D.~Stanford, ``{JT gravity as a matrix integral},''
  \href{http://arxiv.org/abs/1903.11115}{{\ttfamily arXiv:1903.11115
  [hep-th]}}.

\bibitem{Maldacena:2019cbz}
J.~Maldacena, G.~J. Turiaci, and Z.~Yang, ``{Two dimensional Nearly de Sitter
  gravity},'' \href{http://dx.doi.org/10.1007/JHEP01(2021)139}{{\em JHEP}
  {\bfseries 01} (2021) 139}, \href{http://arxiv.org/abs/1904.01911}{{\ttfamily
  arXiv:1904.01911 [hep-th]}}.

\bibitem{Cotler:2019nbi}
J.~Cotler, K.~Jensen, and A.~Maloney, ``{Low-dimensional de Sitter quantum
  gravity},'' \href{http://dx.doi.org/10.1007/JHEP06(2020)048}{{\em JHEP}
  {\bfseries 06} (2020) 048}, \href{http://arxiv.org/abs/1905.03780}{{\ttfamily
  arXiv:1905.03780 [hep-th]}}.

\bibitem{Chen:2020tes}
Y.~Chen, V.~Gorbenko, and J.~Maldacena, ``{Bra-ket wormholes in gravitationally
  prepared states},'' \href{http://dx.doi.org/10.1007/JHEP02(2021)009}{{\em
  JHEP} {\bfseries 02} (2021) 009},
  \href{http://arxiv.org/abs/2007.16091}{{\ttfamily arXiv:2007.16091
  [hep-th]}}.

\bibitem{Hartman:2020khs}
T.~Hartman, Y.~Jiang, and E.~Shaghoulian, ``{Islands in cosmology},''
  \href{http://dx.doi.org/10.1007/JHEP11(2020)111}{{\em JHEP} {\bfseries 11}
  (2020) 111}, \href{http://arxiv.org/abs/2008.01022}{{\ttfamily
  arXiv:2008.01022 [hep-th]}}.

\bibitem{Balasubramanian:2020xqf}
V.~Balasubramanian, A.~Kar, and T.~Ugajin, ``{Islands in de Sitter space},''
  \href{http://dx.doi.org/10.1007/JHEP02(2021)072}{{\em JHEP} {\bfseries 02}
  (2021) 072}, \href{http://arxiv.org/abs/2008.05275}{{\ttfamily
  arXiv:2008.05275 [hep-th]}}.

\bibitem{Sybesma:2020fxg}
W.~Sybesma, ``{Pure de Sitter space and the island moving back in time},''
  \href{http://dx.doi.org/10.1088/1361-6382/abff9a}{{\em Class. Quant. Grav.}
  {\bfseries 38} no.~14, (2021) 145012},
  \href{http://arxiv.org/abs/2008.07994}{{\ttfamily arXiv:2008.07994
  [hep-th]}}.

\bibitem{Aalsma:2021bit}
L.~Aalsma and W.~Sybesma, ``{The Price of Curiosity: Information Recovery in de
  Sitter Space},'' \href{http://dx.doi.org/10.1007/JHEP05(2021)291}{{\em JHEP}
  {\bfseries 05} (2021) 291}, \href{http://arxiv.org/abs/2104.00006}{{\ttfamily
  arXiv:2104.00006 [hep-th]}}.

\bibitem{Manu:2020tty}
A.~Manu, K.~Narayan, and P.~Paul, ``{Cosmological singularities, entanglement
  and quantum extremal surfaces},''
  \href{http://dx.doi.org/10.1007/JHEP04(2021)200}{{\em JHEP} {\bfseries 04}
  (2021) 200}, \href{http://arxiv.org/abs/2012.07351}{{\ttfamily
  arXiv:2012.07351 [hep-th]}}.

\bibitem{Langhoff:2021uct}
K.~Langhoff, C.~Murdia, and Y.~Nomura, ``{The Multiverse in an Inverted
  Island},'' \href{http://arxiv.org/abs/2106.05271}{{\ttfamily arXiv:2106.05271
  [hep-th]}}.

\bibitem{Bousso:2021sji}
R.~Bousso and A.~Shahbazi-Moghaddam, ``{Island Finder and Entropy Bound},''
  \href{http://dx.doi.org/10.1103/PhysRevD.103.106005}{{\em Phys. Rev. D}
  {\bfseries 103} no.~10, (2021) 106005},
  \href{http://arxiv.org/abs/2101.11648}{{\ttfamily arXiv:2101.11648
  [hep-th]}}.

\bibitem{Hartle:2016tpo}
J.~Hartle and T.~Hertog, ``{One Bubble to Rule Them All},''
  \href{http://dx.doi.org/10.1103/PhysRevD.95.123502}{{\em Phys. Rev. D}
  {\bfseries 95} no.~12, (2017) 123502},
  \href{http://arxiv.org/abs/1604.03580}{{\ttfamily arXiv:1604.03580
  [hep-th]}}.

\bibitem{Anninos:2017hhn}
D.~Anninos and D.~M. Hofman, ``{Infrared Realization of dS$_2$ in AdS$_2$},''
  \href{http://dx.doi.org/10.1088/1361-6382/aab143}{{\em Class. Quant. Grav.}
  {\bfseries 35} no.~8, (2018) 085003},
  \href{http://arxiv.org/abs/1703.04622}{{\ttfamily arXiv:1703.04622
  [hep-th]}}.

\bibitem{Anninos:2018svg}
D.~Anninos, D.~A. Galante, and D.~M. Hofman, ``{De Sitter horizons \&
  holographic liquids},'' \href{http://dx.doi.org/10.1007/JHEP07(2019)038}{{\em
  JHEP} {\bfseries 07} (2019) 038},
  \href{http://arxiv.org/abs/1811.08153}{{\ttfamily arXiv:1811.08153
  [hep-th]}}.

\bibitem{Sarosi:2017ykf}
G.~S\'arosi, ``{AdS$_{2}$ holography and the SYK model},''
  \href{http://dx.doi.org/10.22323/1.323.0001}{{\em PoS} {\bfseries Modave2017}
  (2018) 001}, \href{http://arxiv.org/abs/1711.08482}{{\ttfamily
  arXiv:1711.08482 [hep-th]}}.

\bibitem{Holzhey:1994we}
C.~Holzhey, F.~Larsen, and F.~Wilczek, ``{Geometric and renormalized entropy in
  conformal field theory},''
  \href{http://dx.doi.org/10.1016/0550-3213(94)90402-2}{{\em Nucl. Phys. B}
  {\bfseries 424} (1994) 443--467},
  \href{http://arxiv.org/abs/hep-th/9403108}{{\ttfamily arXiv:hep-th/9403108}}.

\bibitem{Calabrese:2004eu}
P.~Calabrese and J.~L. Cardy, ``{Entanglement entropy and quantum field
  theory},'' \href{http://dx.doi.org/10.1088/1742-5468/2004/06/P06002}{{\em J.
  Stat. Mech.} {\bfseries 0406} (2004) P06002},
  \href{http://arxiv.org/abs/hep-th/0405152}{{\ttfamily arXiv:hep-th/0405152}}.

\bibitem{Calabrese:2009qy}
P.~Calabrese and J.~Cardy, ``{Entanglement entropy and conformal field
  theory},'' \href{http://dx.doi.org/10.1088/1751-8113/42/50/504005}{{\em J.
  Phys. A} {\bfseries 42} (2009) 504005},
  \href{http://arxiv.org/abs/0905.4013}{{\ttfamily arXiv:0905.4013
  [cond-mat.stat-mech]}}.

\bibitem{Bousso:2015mna}
R.~Bousso, Z.~Fisher, S.~Leichenauer, and A.~C. Wall, ``{Quantum focusing
  conjecture},'' \href{http://dx.doi.org/10.1103/PhysRevD.93.064044}{{\em Phys.
  Rev. D} {\bfseries 93} no.~6, (2016) 064044},
  \href{http://arxiv.org/abs/1506.02669}{{\ttfamily arXiv:1506.02669
  [hep-th]}}.

\bibitem{Affleck:1991tk}
I.~Affleck and A.~W.~W. Ludwig, ``{Universal noninteger 'ground state
  degeneracy' in critical quantum systems},''
  \href{http://dx.doi.org/10.1103/PhysRevLett.67.161}{{\em Phys. Rev. Lett.}
  {\bfseries 67} (1991) 161--164}.

\bibitem{Cespedes:2020xpn}
S.~Cespedes, S.~P. de~Alwis, F.~Muia, and F.~Quevedo, ``{Lorentzian vacuum
  transitions: Open or closed universes?},''
  \href{http://dx.doi.org/10.1103/PhysRevD.104.026013}{{\em Phys. Rev. D}
  {\bfseries 104} no.~2, (2021) 026013},
  \href{http://arxiv.org/abs/2011.13936}{{\ttfamily arXiv:2011.13936
  [hep-th]}}.

\bibitem{Hartle:1983ai}
J.~B. Hartle and S.~W. Hawking, ``{Wave Function of the Universe},''
  \href{http://dx.doi.org/10.1103/PhysRevD.28.2960}{{\em Phys. Rev. D}
  {\bfseries 28} (1983) 2960--2975}.

\bibitem{Hartle:2008ng}
J.~B. Hartle, S.~W. Hawking, and T.~Hertog, ``{The Classical Universes of the
  No-Boundary Quantum State},''
  \href{http://dx.doi.org/10.1103/PhysRevD.77.123537}{{\em Phys. Rev. D}
  {\bfseries 77} (2008) 123537},
  \href{http://arxiv.org/abs/0803.1663}{{\ttfamily arXiv:0803.1663 [hep-th]}}.

\bibitem{Hartle:1992as}
J.~B. Hartle, ``{Space-time quantum mechanics and the quantum mechanics of
  space-time},'' in {\em {Les Houches Summer School on Gravitation and
  Quantizations, Session 57}}.
\newblock 7, 1992.
\newblock \href{http://arxiv.org/abs/gr-qc/9304006}{{\ttfamily
  arXiv:gr-qc/9304006}}.

\bibitem{Hertog:2013mra}
T.~Hertog, ``{Predicting a Prior for Planck},''
  \href{http://dx.doi.org/10.1088/1475-7516/2014/02/043}{{\em JCAP} {\bfseries
  02} (2014) 043}, \href{http://arxiv.org/abs/1305.6135}{{\ttfamily
  arXiv:1305.6135 [astro-ph.CO]}}.

\bibitem{Balasubramanian:2020coy}
V.~Balasubramanian, A.~Kar, and T.~Ugajin, ``{Entanglement between two disjoint
  universes},'' \href{http://dx.doi.org/10.1007/JHEP02(2021)136}{{\em JHEP}
  {\bfseries 02} (2021) 136}, \href{http://arxiv.org/abs/2008.05274}{{\ttfamily
  arXiv:2008.05274 [hep-th]}}.

\bibitem{Balasubramanian:2021wgd}
V.~Balasubramanian, A.~Kar, and T.~Ugajin, ``{Entanglement between two
  gravitating universes},'' \href{http://arxiv.org/abs/2104.13383}{{\ttfamily
  arXiv:2104.13383 [hep-th]}}.

\bibitem{Fallows:2021sge}
S.~Fallows and S.~F. Ross, ``{Islands and mixed states in closed universes},''
  \href{http://dx.doi.org/10.1007/JHEP07(2021)022}{{\em JHEP} {\bfseries 07}
  (2021) 022}, \href{http://arxiv.org/abs/2103.14364}{{\ttfamily
  arXiv:2103.14364 [hep-th]}}.

\bibitem{Almheiri:2014lwa}
A.~Almheiri, X.~Dong, and D.~Harlow, ``{Bulk Locality and Quantum Error
  Correction in AdS/CFT},''
  \href{http://dx.doi.org/10.1007/JHEP04(2015)163}{{\em JHEP} {\bfseries 04}
  (2015) 163}, \href{http://arxiv.org/abs/1411.7041}{{\ttfamily arXiv:1411.7041
  [hep-th]}}.

\bibitem{Dong:2016eik}
X.~Dong, D.~Harlow, and A.~C. Wall, ``{Reconstruction of Bulk Operators within
  the Entanglement Wedge in Gauge-Gravity Duality},''
  \href{http://dx.doi.org/10.1103/PhysRevLett.117.021601}{{\em Phys. Rev.
  Lett.} {\bfseries 117} no.~2, (2016) 021601},
  \href{http://arxiv.org/abs/1601.05416}{{\ttfamily arXiv:1601.05416
  [hep-th]}}.

\bibitem{Cotler:2017erl}
J.~Cotler, P.~Hayden, G.~Penington, G.~Salton, B.~Swingle, and M.~Walter,
  ``{Entanglement Wedge Reconstruction via Universal Recovery Channels},''
  \href{http://dx.doi.org/10.1103/PhysRevX.9.031011}{{\em Phys. Rev. X}
  {\bfseries 9} no.~3, (2019) 031011},
  \href{http://arxiv.org/abs/1704.05839}{{\ttfamily arXiv:1704.05839
  [hep-th]}}.

\bibitem{Chen:2019gbt}
C.-F. Chen, G.~Penington, and G.~Salton, ``{Entanglement Wedge Reconstruction
  using the Petz Map},'' \href{http://dx.doi.org/10.1007/JHEP01(2020)168}{{\em
  JHEP} {\bfseries 01} (2020) 168},
  \href{http://arxiv.org/abs/1902.02844}{{\ttfamily arXiv:1902.02844
  [hep-th]}}.

\bibitem{Hawking:2006ur}
S.~W. Hawking and T.~Hertog, ``{Populating the landscape: A Top down
  approach},'' \href{http://dx.doi.org/10.1103/PhysRevD.73.123527}{{\em Phys.
  Rev. D} {\bfseries 73} (2006) 123527},
  \href{http://arxiv.org/abs/hep-th/0602091}{{\ttfamily arXiv:hep-th/0602091}}.

\bibitem{Casini:2005rm}
H.~Casini, C.~D. Fosco, and M.~Huerta, ``{Entanglement and alpha entropies for
  a massive Dirac field in two dimensions},''
  \href{http://dx.doi.org/10.1088/1742-5468/2005/07/P07007}{{\em J. Stat.
  Mech.} {\bfseries 0507} (2005) P07007},
  \href{http://arxiv.org/abs/cond-mat/0505563}{{\ttfamily
  arXiv:cond-mat/0505563}}.

\end{thebibliography}\endgroup

\end{document}